%% file: SanchezGallegoetal-NGLS_Ha_arXiv.tex
\documentclass[useAMS,usenatbib]{mn2e}
\usepackage{graphicx}
\usepackage{amssymb}
\usepackage{amsmath}
\usepackage{natbib} 

\usepackage{subfig}
\usepackage{float}

\usepackage{longtable,lscape}

\DeclareMathAlphabet{\mathsc}{OT1}{cmr}{m}{sc}

\def\HI{\mbox{H\,{\sc i}}~}
\def\HInoSp{\mbox{H\,{\sc i}}}
\def\NII{\mbox{N\,{\sc ii}}}

\def\TabObs{1}
\def\TabResults{2}

\title[The JCMT Nearby Galaxies Legacy Survey. VII.]{The JCMT Nearby Galaxies Legacy Survey. VII. H$\alpha$ imaging and massive star formation properties}
\author[J.~R.~S\'{a}nchez-Gallego et al.]{J.~R.~S\'{a}nchez-Gallego,$^{1,2}$\thanks{E-mail:
jrsg@iac.es} J.~H~.~Knapen,$^{1,2}$ C.~D.~Wilson,$^{3}$ P.~Barmby,$^{4}$
\newauthor
M.~Azimlu$^{4,5}$ and S.~Courteau$^{6}$\\
$^{1}$Instituto de Astrof\'{i}sica de Canarias, E-38205 La Laguna, Tenerife, Spain\\
$^{2}$Departamento de Astrof\'{i}sica, Universidad de La Laguna, E-38200, La Laguna, Tenerife, Spain\\
$^{3}$Department of Physics \& Astronomy, McMaster University, Hamilton, Ontario, L8S 4M1, Canada\\
$^{4}$Physics and Astronomy Department, University of Western Ontario, 1151 Richmond Street, London, Ontario, N6A 3K7, Canada\\
$^{5}$Harvard-Smithsonian center for Astrophysics, 60 Garden St., Cambridge, MA, 02138, USA\\
$^{6}$Department of Physics, Engineering Physics and Astronomy, Queen's University, Kingston, ON K7L 3N6, Canada}
\begin{document}

\date{Accepted 1988 December 15. Received 1988 December 14; in original form 1988 October 11}

\pagerange{\pageref{firstpage}--\pageref{lastpage}} \pubyear{2011}

\maketitle

\label{firstpage}

\begin{abstract}
We present H$\alpha$ fluxes, star formation rates (SFRs) and equivalent widths (EWs) for a sample of 156 nearby galaxies observed in the $^{12}$CO~$J=3-2$ line as part of the James Clerk Maxwell Telescope Nearby Galaxies Legacy Survey. These are derived from images and values in the literature and from new H$\alpha$ images for 72 galaxies which we publish here. We describe the sample, observations and procedures to extract the H$\alpha$ fluxes and related quantities. We discuss the SFR properties of our sample and confirm the well-known correlation with galaxy luminosity, albeit with high dispersion. Our SFRs range from 0.1 to 11$\,M_\odot\rm\,yr^{-1}$ with a median SFR value for the complete sample of 0.2$\,M_\odot\rm\,yr^{-1}$. This median values is somewhat lower than similar published measurements, which we attribute, in part, to our sample being \HInoSp{}-selected and, thus, not biased towards high SFRs as has frequently been the case in previous studies. Additionally, we calculate internal absorptions for the H$\alpha$ line, $A({\rm H}\alpha)$, which are lower than many of those used in previous studies. Our derived EWs, which range from 1 to 880$\,\rm \AA$ with a median value of 27$\,\rm\AA$, show little dependence with luminosity but rise by a factor of five from early- to late-type galaxies. This paper is the first in a series aimed at comparing SFRs obtained from H$\alpha$ imaging of galaxies with information derived from other tracers of star formation and atomic and molecular gas.
\end{abstract}


%
%
%%%%%%%%%%%%%%%%%%  INTRODUCTION  %%%%%%%%%%%%%%%%%%%%%%%
%
%

\section{Introduction}
\label{sec:Intro}

Along with galaxy interactions and the presence of active nuclei, star formation efficiency is one of the most important factors that controls a galaxy's structure and evolution. Although the subject has been extensively researched both in the local Universe (e.g., \citealt{Kennicutt1983}; \citealt{Gallego1995}; \citealt{Alonso1999}; \citealt{James2004}; \citealt{Kennicutt2008}) and in the more distant Universe (e.g., \citealt{Madau1996}; \citealt{Hopkins2000}; \citealt{Lanzetta2002}; \citealt{Nagamine2006}), many fundamental questions in galactic and extragalactic research remain.

In the local Universe, many surveys have been carried out in different wavelengths (see, for example, \citealt{Kennicutt2008} for an extensive compilation of literature data) to probe the star formation rate. Among these, the H$\alpha$ line is one of the most widely used star formation rate (SFR) tracers, both photometrically and spectroscopically \citep{Kennicutt1998}. In contrast to, e.g., the Far Ultra-Violet (FUV), H$\alpha$ lies in the optical region for nearby galaxies, thus allowing easy ground-based observations. In addition, the spatial resolution which can be obtained with H$\alpha$ is superior to what is normally obtained at other wavelengths such as the Infrared (IR), which are also more complicated to observe with ground-based telescopes. Studies carried out throughout almost two decades have proved that the H$\alpha$ line is a solid tracer of star formation, especially if the measurements are complemented by the use of Far Infrared (FIR) data (e.g., \citealt{Calzetti2007, Calzetti2010, Leroy2012}).

Recent studies (see \citealt{Wilson2009} and references therein) state that the warm, dense molecular hydrogen is better correlated with star formation than the cold, more dispersed gas. Due to the lack of a permanent dipole moment, all the ground-based observations aimed to detect molecular hydrogen are made using alternative tracers. Although not the only method at hand (\citealt{Gautier1976}; \citealt{Neufeld2008}), the spectral lines produced by the different rotational transitions of the CO molecule stand as the best known and most widely used tracers of molecular hydrogen. However, the vast majority of surveys aiming to study the molecular gas in nearby galaxies have focused on the lower transitions ($J=1-0$, $J=2-1$) of CO, which do not discriminate between cold and warm gas. 

The James Clerk Maxwell Telescope (JCMT) on Mauna Kea, Hawai'i is currently hosting a series of Legacy Surveys, including the Nearby Galaxies Legacy Survey (NGLS) which aims to map 156 galaxies in the CO $J=3-2$ spectral line and in the 450 and 850$\,\mu\rm m$ continuum (C. Wilson et al., in preparation). All the CO $J=3-2$ observations have been completed and the continuum observations have begun. Further information about the scope and science goals of this survey can be found in \citet{Wilson2009} as well as in the subsequent series of papers which make use of the CO $J=3-2$ data (\citealt{Warren2010}; \citealt{Bendo2010}; \citealt{Irwin2011}; \citealt{Wilson2011}; \citealt{Sanchez-Gallego2011}). 

Because of the physical conditions needed to activate this transition, the $J=3-2$ line is perfect to trace and study the warm molecular gas which is most likely to form stars. This directly inspires the interest in observing this same sample using a well proven tracer of star formation such as H$\alpha$.

This paper is organized as follows. Section \ref{sec:Sample} introduces the sample and the criteria used for its selection; Sections \ref{sec:NewObservations} and \ref{sec:LiteratureData} describe the observations and the origin of the literature data used in this work. Section \ref{sec:Reduction} explains in detail the complete process of data analysis, from the most basic reduction to the continuum subtraction, flux measurements and SFR and EW calculations. In Section \ref{sec:Results} we analyze the SFR and equivalent width (EW) results obtained. They are compared with previous data from the literature in Section \ref{sec:OtherStudies}. Finally, Section \ref{sec:Conclusions} presents the main conclusions and outlines the future work we plan to do with these data.

%
%
%%%%%%%%%%%%%%%%%%  SAMPLE SELECTION  %%%%%%%%%%%%%%%%%%%%%%%
%
%

\section{Sample selection}
\label{sec:Sample}

This work presents H$\alpha$ photometry for a total of 156 nearby galaxies. The sample of galaxies has been mostly \HInoSp-flux selected in order to avoid a SFR-driven selection (as would have happened using a FIR flux criterion) while ensuring that the galaxies have a rich interstellar medium. All our galaxies are contained within a range between 2 and 25$\,\rm Mpc$ in distance ($v<1875\,\rm km\,s^{-1}$) and were selected to cover the complete morphological spectrum, from early-type spirals and ellipticals to late-type spirals and irregular galaxies. Both field and Virgo cluster galaxies are well represented in each morphological bin. Most of the Virgo galaxies have $D_{25}<4\rm\,arcmin$, which allowed us to use the jiggle map method during the CO $J=3-2$ observations \citep{Warren2010} carried out with the JCMT. C. Wilson et al. (2011, in prep.) give a more detailed explanation of the criteria and parameters used to construct the sample.

\begin{figure*}
  \centering
  \includegraphics[width=17cm]{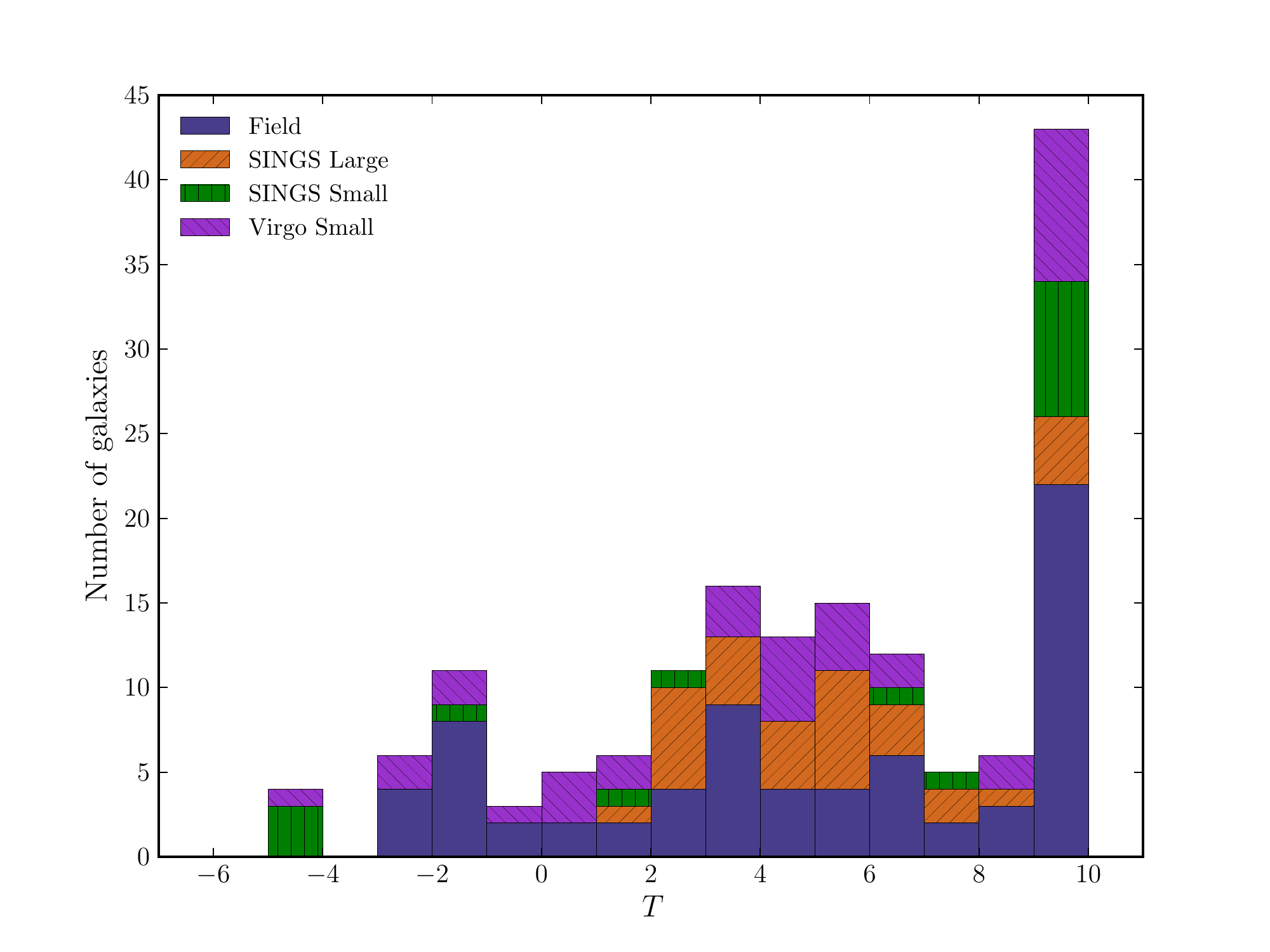}
  \caption{Histogram of the number of galaxies as a function of the Hubble stage, $T$, as listed in the Third Reference Catalogue of Bright Galaxies (RC3). The four different subsamples are shown with different patterns and colors.}
  \label{fig:Histogram}
\end{figure*}

The complete sample has been compiled from three different subsets: (1) 48 galaxies taken from the \emph{Spitzer} Infrared Nearby Galaxies Survey (SINGS; \citealt{Kennicutt2003}); (2) 36 galaxies chosen from a \HI flux-limited sample of galaxies from the Virgo cluster; and (3) 72 \HI flux limited field galaxies. The fact that a significant part of our list of targets matches the SINGS sample ensures that a large amount of multi-wavelength data is available for those galaxies, from the infrared imaging obtained by the SINGS team to \HI data (for example, from THINGS; \citealt{Walter2008}) and UV imaging from \emph{GALEX}. However, SINGS used FIR emission to constrain its sample, which leads to a certain bias towards spiral, star-forming galaxies. This bias also appears in our sample, as can be seen in Figure \ref{fig:Histogram}, which shows the distribution of the galaxies of our sample across the Hubble sequence and the contribution of each subsample. This bias will also become apparent when we discuss the SFRs (Section \ref{sec:Results}). The morphological breakdown is as follows: 4 elliptical galaxies $\rm(T\in[-6,-4))$, 20 lenticular galaxies $\rm(T\in[-4,0))$, 50 early spiral galaxies $\rm(T\in[0,4.5))$, 49 late spiral galaxies $\rm(T\in[4.5,9.5))$ and 33 irregular galaxies $\rm(T\in[9.5,10])$. We make a distinction between small ($D_{25}<4\rm\,arcmin$) and large galaxies in the SINGS subsample. The sample is given in Table \TabObs{}.

%
%
%%%%%%%%%%%%%%%%%%  NEW HA OBSERVATIONS  %%%%%%%%%%%%%%%%%%%%%%%
%
%

\section{New H$\alpha$ observations}
\label{sec:NewObservations}

We have imaged 72 out of the 156 sample galaxies in the H$\alpha$ line. While a small subset of these galaxies had H$\alpha$ data available in the literature, we decided that the images were either not deep or not detailed enough to serve our purpose and we reobserved them. 

For these observations, we made extensive use of several of the telescopes located on the Roque de los Muchachos Observatory (La Palma, Canary Islands). In particular, we used: (1) the Wide Field Camera (WFC) mounted on the 2.5 meter Isaac Newton Telescope (INT), (2) the Auxiliary port Camera (ACAM) and the auxport camera (AUX) on the 4.2 meter William Herschel Telescope (WHT), and (3) the Andalucia Faint Object Spectrograph and Camera (ALFOSC) on the 2.5 meter Nordic Optical Telescope (NOT). Additional images were taken using the MDM8K mosaic camera on the 2.4 meter Hiltner telescope at the Michigan-Dartmouth-MIT Observatory (Kitt Peak). Table \TabObs{} summarizes the dates, observing conditions, exposure times and filters used for each galaxy.

Although working with different telescopes and instruments, we kept a systematic observation strategy. Each galaxy was observed through a narrow-band filter redshifted to the position of the H$\alpha$ line for each galaxy. The narrow filters used had a width in the range $15-50\,\rm\AA$ with transmissions always above 50\% and, generally, higher than 70\%. We ensured that the H$\alpha$ line always fell in the region of maximum transmission of the filter. For each target we obtained at least three H$\alpha$ images in order to remove cosmic ray hits through the combination of the different images. Standard exposure times were $3\times300\,\rm s$.

The continuum subtraction was done using an \emph{R} broad-band image. Due to the low recession velocity of our galaxies, no other filters were necessary to obtain the continuum contribution in the H$\alpha$ region. We typically obtained $3\times180\,\rm s$ continuum exposures which were combined to produce a single \emph{R}-band image.

Sky flat frames were taken during both the evening and morning twilights. In some cases this was not possible due to adverse atmospheric conditions during one or both twilights. In some of these cases, dome flats were taken. A series of bias exposures was obtained at the beginning of the night and again at the end. Typically the bias was constant and uniform throughout the night.

The photometric calibration was carried out by observing at least one spectrophotometric standard star at the beginning of the night and at the end. \citet{Landolt1992} photometric stars where used for the \emph{R}-band calibration. While most of the images were obtained during photometric nights, some galaxies were observed in clear but not perfect conditions. In those cases we did not calibrate the images. Instead, we reobserved the galaxies during photometric nights, obtaining short exposures (typically 1/4 the original exposure times) which we used to calibrate the previous images.

%
%
%%%%%%%%%%%%%%%%%%  LITERATURE DATA  %%%%%%%%%%%%%%%%%%%%%%%
%
%

\section{Literature data}
\label{sec:LiteratureData}

The remaining 84 galaxies in our list had previously been observed, and photometrically calibrated H$\alpha$ imaging was available in the literature. These include: 25 galaxies from the H$\alpha$ Galaxy Survey (H$\alpha$GS; \citealt{James2004}); 33 galaxies from SINGS \citep{Kennicutt2003}; 4 galaxies from \citet{Knapen2004}; 4 galaxies from \citet{Kennicutt2008a}; and 18 galaxies from the GoldMine project (\citealt{Gavazzi2002, Gavazzi2006, Boselli2002}).

For these galaxies, the headers of the FITS files have been modified to match the global standards of this work (see Section \ref{sec:RedRegistration}) and the images have been divided by the exposure time to get an effective exposure time ($t\rm_{exp}^*$) equal to 1 second. If necessary, the astrometric solution in the images was improved.

When possible (i.e., when both the original H$\alpha$ and the continuum images were available) we did not use the H$\alpha$ continuum-subtracted image from the literature, but we performed our own continuum subtraction as explained in Section \ref{sec:RedContinuum}. We used the final images to measure our own flux values and derived properties, as described in Sections \ref{sec:RedFlux}-\ref{sec:Uncertainties}.

%
%
%%%%%%%%%%%%%%%%%%  DATA REDUCTION  %%%%%%%%%%%%%%%%%%%%%%%
%
%

\section{Data reduction}
\label{sec:Reduction}

\subsection{Basic reduction}
\label{sec:RedBasic}

The data reduction is quite straightforward and we refer to the corresponding procedures described in \citet{Knapen2004}. The reduction was performed using mainly packages and tasks from IRAF\footnote{http://iraf.noao.edu/}. Again, slightly different procedures were carried out depending on the instrument used to acquire the images. The main steps were, however, common to all the images regardless of their origin. The basic steps were as follows.

\begin{itemize}
    \item[a)] A median bias frame was produced combining all valid bias exposures. This median bias was then subtracted from each of the flat field and science images. Some images from the literature were bias-corrected using a constant value (namely the mean value of the median bias frame) rather than the master bias image (see \citealt{Knapen2004}). This is no problem as the structure of the individual bias frames was negligible in those cases.
    \item[b)] The flat field correction was applied using an adequate median sky flat obtained from the flat frames for each filter. Evening and morning sky flats were analyzed separately although, in the majority of cases, no important differences were found and all the flat frames were combined.
    \item[c)] The images were aligned, their background was subtracted and they were weighted by their exposure time and then combined by filter using the median. A sigclip rejection algorithm was applied to remove pixels with values above or below 3$\sigma$. This ensures that most of the cosmic rays were successfully removed.
    \item[d)] Although in most cases the images had a good astrometric calibration, a new astrometric solution was applied to each image using the Guide Star Catalogue II (GSC-II; \citealt{Lasker2008}). The final precision is better than one arcsec.
\end{itemize}

\subsection{Continuum subtraction}
\label{sec:RedContinuum}

Once both the H$\alpha$ on-line and the continuum images are completely reduced we must subtract the contribution of the continuum to the H$\alpha$ flux. The great majority of our continuum images were obtained using $R$-band filters and the rest of this subsection will deal with that situation. However, for a small number of cases (all from the literature) the continuum image is a narrow filter of width similar to that of the H$\alpha$ filter but shifted in wavelength so that only the continuum and not the line is measured. In these cases the procedure to subtract the contribution of the continuum is similar to when using $R$-band, but much more straightforward as the scale factor between the H$\alpha$ and the continuum images is always $\sim1$.

The method used to calculate the scale factor of the continuum image is well explained in \citet{Knapen2004} and we refer to that paper for further details. First, if there is an important difference between the full widths at half maximum (FWHMs) of the H$\alpha$ and continuum images, we smooth the images to the worst FWHM using a Gaussian convolution. Then, we plot the intensity of each pixel in the H$\alpha$ image, in counts per second, versus the intensity of the same pixel in the \emph{R}-band image. In general, a certain percentage of the brightest pixels (usually around 3\%) is removed to avoid the contribution of field stars. The resulting plot shows a strong linear correlation which can be fitted using a least squares regression line, the slope of which gives the scale factor between the two images. Although this method has been found to produce consistently good results, some galaxies (especially those with very low emission) require further processing to obtain the optimum scale factor value. Thus, a final small correction by hand was often made to ensure that the continuum subtraction was optimal by looking at different regions of the galaxy and ensuring that no area was oversubtracted. While it is not always trivial to find the most accurate scale factor, small differences in the value of this parameter do not affect the derived values (flux, SFR, EW) in a significant way.

It must be taken into account that the continuum image used is contaminated by the flux of the H$\alpha$ line. The contribution of the line has previously been estimated to be 3\% of the total integrated flux in the $R$-band \citep{James2004}. This effect does not occur when the filter used for the continuum subtraction is a narrow one, as in that case the H$\alpha$ line is not present at all. Integrating the transmission curves of the filters used, we calculated how much the H$\alpha$ line contributed to the continuum flux. We found that, for most sets of filters, the correction was in the range 2-4\% with some extreme cases as high as 6\%. The calibration was modified according to the value found for each galaxy. Table \TabResults{} shows the correction applied.

\subsection{Photometric calibration}
\label{sec:RedCalib}

The next stage in the reduction process was to calibrate the science frames photometrically. The \emph{R}-band calibration was done using \cite{Landolt1992} standard stars and taking the airmass of the objects into account. For the calibration of the H$\alpha$ images, following \cite{Knapen1992}, we considered the flux density, $f_\lambda$, given by
$$\log f_\lambda = -0.4 {\rm mag}_\lambda - 8.42,$$
with $f_\lambda$ in units of $\rm erg\,s^{-1}\,cm^{-2}\,\AA^{-1}$. To calculate the zero point factor we performed aperture photometry of the spectrophotometric standard stars selected. For these stars, the value of ${\rm mag}_\lambda$ is tabulated every few angstroms. All the stars used are from either \citet{Stone1977}, \citet{Massey1988} or \citet{Oke1990}. If $\chi$ is the total flux of the standard star in units of $\rm counts\,s^{-1}$, the zero point factor can be written as
$${\rm ZP} = \frac{\Delta\lambda f_\lambda}{\chi},$$
where $\Delta\lambda$ corresponds to the width of the H$\alpha$ filter used. ZP has units of $\rm erg\,cm^{-2}\,count^{-1}$. The flux for a certain region can be calculated as
$${\rm log_{10}}F = \rm ZP + log_{10}(counts),$$
in units of $\rm erg\,cm^{-2}\,s^{-1}$. In this expression we have taken into account that the effective exposure time of our images is one second.

\subsection{Final image preparation}
\label{sec:RedRegistration}

The final images for each galaxy (H$\alpha$ without continuum subtraction, \emph{R}-band and continuum-subtracted H$\alpha$) were rotated North-up and trimmed. The FITS headers of all images (including those obtained from the literature) have been modified in the following way: first, a custom header has been inserted including the relevant data (total exposure time, zero point, flux and luminosity per count, etc); second, the original header has been preserved after our custom header. The units in the final images are always counts per pixel (i.e., the exposure time has been set to one second). The original exposure time for each frame is stored in the header of the image under the field {\sc darktime}.

When working with literature data, the astrometric solution provided with most of the images was well below the FWHM of the frames and no correction was made. In some cases, however, the astrometric solution was wrong or nonexistent. Those images were astrometrically calibrated using the \emph{Guide Star Catalogue} \citep{Lasker1990} which provides an accuracy better than one arcsec. 

\subsection{Flux measurement}
\label{sec:RedFlux}

In order to determine the H$\alpha$ flux without the contribution of field stars, we masked all the stars near the galaxy while taking special care not to remove any H{\,\mbox{\sc ii}} regions. We used the masking utility integrated in Gaia \citep{Draper2009} which substitutes the region to mask based on an estimate of the values of the surrounding pixels. We applied a third order surface fit, which in all cases produced an excellent result. The stars were masked in the continuum band and, when necessary, also in the H$\alpha$ continuum-subtracted image.

To calculate the integrated H$\alpha$ flux of the galaxies we used the task \textsc{ellipse} in IRAF. On the continuum subtracted frame we ran \textsc{ellipse} to calculate the integrated flux for a series of concentric ellipses centred on the nucleus of the galaxy and with ellipticities ($\varepsilon$) and position angles (PA) as obtained from the Third Reference Catalogue of Bright Galaxies (RC3; \citealt{de-Vaucouleurs1991}). For a number of galaxies, especially in the case of irregular ones, these data were either not available or not reliable. In these cases circular apertures were used. Table 2 includes the values of $\varepsilon$ and PA we used in our measurements.  The separation between the different ellipses is a function of the pixel scale of the image used. Typically we calculated ellipses every 5 arcsec although this value was manually modified for special cases such as extremely large or small galaxies. The position of the nucleus was determined in each case from the continuum image, except for some irregular galaxies where the estimated centroid of the galaxy was used. The determination of the optimum semi-major axis of the ellipse was done using the same criterion as \citet{James2004}: we consider that the photometry has converged when the measured flux between two consecutive ellipses varies by less than 0.5\%. This method works efficiently for most of the galaxies in our sample. However, for a small number of galaxies (mainly irregular ones) the resulting aperture was clearly wrong. In those cases we used the D$_{25}$ value from the RC3. The aperture determined in this way was then used on every available band. The fluxes obtained were converted from counts to $\rm erg\,cm^{-2}\,s^{-1}$ (in the case of the H$\alpha$ band) or to magnitudes (case of $R$-band) using the previously calculated conversion factors.

Figures \ref{fig:ImagesField} to \ref{fig:ImagesVirgoSmall} show the final H$\alpha$ continuum-subtracted images. In each case, the elliptical aperture where we measured the H$\alpha$ flux is plotted on the image.

\subsection{[\NII] correction}
\label{sec:NII}

Although for the sake of clarity we have only discussed H$\alpha$ fluxes, a number of our H$\alpha$ images (mainly those observed using filters wider than 35-40$\,\rm\AA$) are contaminated with emission from the [\NII] lines at $\lambda\lambda6548,6584\,\rm\AA$. We estimated the contribution of the [\NII] lines using the expression from \citet{Kennicutt2008}
\begin{equation}
\label{eq:NII}
{\rm \log\left([\NII]/H\alpha \right)} = 
\left\lbrace
\begin{array}{ll}
-0.173\pm0.007\,M_B - & \\
\qquad -(3.903\pm0.137) & {\rm if}\ M_B>-21 \\
0.54 & {\rm if}\ M_B\le-21
\end{array}
\right.
\end{equation}

\citet{Helmboldt2004} provide a similar expression but using $M_R$ instead of $M_B$. We decided against using this latter reference because all but two of our galaxies have measured values of $M_B$, while many do not have reported $M_R$. For those cases where both $M_B$ and $M_R$ measurements were available we compared the $\rm [\NII]/H\alpha$ obtained using both methods, finding a good agreement. \citet{James2005} studied the $\rm [\NII]/H\alpha$ ratio for the H$\alpha$GS galaxies obtaining values which are compatible with ours.

Even when it is possible to know the relation between [\NII] and H$\alpha$ using Equation \ref{eq:NII}, we still have to deal with the problem of the different transmissions for the H$\alpha$ and the [\NII] lines. Except for the broadest filters, the H$\alpha$ line is located in the region of maximum transmission of the filter (see Section \ref{sec:NewObservations}) but the [\NII] lines at $\lambda\lambda6548,6584\,\rm\AA$ lie at a position where the transmission is much reduced or even marginal. Solving this problem is not easy as filter shapes are often different and difficult to model. Most of the filters we used are neither Gaussian nor rectangular but a convolution of those two shapes: i.e., the transmission remains approximately constant for most of the filter width and then decays exponentially. To estimate the real transmission of the [\NII] lines, we took a series of filters with different FWHM (from 36 to 90$\rm\AA$) for which a transmission curve had been accurately measured. We calculated the relation between the transmission of the filter for the position of the H$\alpha$ and the [\NII] lines, $T_{\rm [NII]/H\alpha}$. Figure \ref{fig:NII} shows the values obtained. We fitted the points to a sigmoid model obtaining that $T_{\rm [NII]/H\alpha}=1-2.670\,e^{\rm -FWHM/12.385}$. Using this expression we calculated the observed ratio, $\rm ([\NII]/H\alpha)_{obs}$, as  $\rm \xi_{obs}=([\NII]/[H\alpha])\times {\it T}_{[NII]/H\alpha}$, for each galaxy (see Table 2).

\subsection{Star formation rates and equivalent widths}
\label{sec:SFRandEW}

To determine the star formation rate (SFR) from the measured fluxes we applied the following conversion from \citet{Kennicutt2009}:
$$
{\rm SFR}(M{\rm _\odot \, yr^{-1}}) = 5.5\,\times\,10^{-42}\,L({\rm H\alpha}),
$$
\noindent where $L({\rm H\alpha})$ is the luminosity, calculated as
$$
L({\rm H\alpha)[\rm erg\,s^{-1}]} = 4\pi D^2 (3.086\,\times\,10^{24})^2 F^*_{\rm H\alpha},
$$
\noindent with $D$ the distance to the galaxy in Mpc and $F^*_{\rm H\alpha}$ the absorption-corrected flux. This determination of the SFR conversion factor is based on a ``Kroupa'' initial mass function \citep{Kroupa2003}. The SFRs based on this estimation are lower than others calculated using older relations (for example, \citealt{Kennicutt1994} or \citealt{Kennicutt1998}) by approximately a factor of 1.5 \citep{Kennicutt2009}. The distances used here (see Table \TabResults{}) are the same as those indicated in C. Wilson et al. (2011, in prep.).

To correct for Galactic absorption we used, for each galaxy, the total absorption value for the $R$ band, $A(R)$ given by NED\footnote{The NASA/IPAC Extragalactic Database; http://nedwww.ipac.caltech.edu/} (see Table \TabResults{}), which is based on the dust maps published by \citet{Schlegel1998}. To correct for the internal absorption, several solutions have been proposed in the literature. A number of studies over the years have adopted a constant value $A(H\alpha)=1.1\,\rm mag$ \citep{Kennicutt1983}, regardless of the morphological type or inclination of the galaxy (see, among others, \citealt{Kennicutt1984,Kennicutt1998,James2004}). \citet{Niklas1997}, however, reported that the value for $A(H\alpha)$ might be as low as $\rm 0.8\, mag$ and \citet{James2005} found values as low as $\rm0.4\,mag$ for some early-type galaxies. On the other hand, \citet{Helmboldt2004} proposed an expression to calculate $A(\rm H\alpha)$ in function of the absolute $R$ magnitude of the galaxy, $M_R$. This expression takes the form
\begin{equation}
\label{eq:AHa}
\log A({\rm H\alpha})=(-0.12\pm0.048)\,M_R + (-2.5\pm0.96).
\end{equation}

We used this equation to calculate the internal absorption of the galaxies in our sample. Unfortunately, not all of our galaxies have been accurately measured in the $R$-band. For those galaxies we overcame this problem by selecting different values of $A(H\alpha)$ according with the type of galaxy studied. Specifically, we used
\begin{equation*}
{A({\rm H\alpha})}=
\left\lbrace
\begin{array}{ll}
0.3 & {\rm if}\ M_B\le-16, \\
1.1 & {\rm otherwise,}
\end{array}
\right.
\end{equation*}

\noindent which are in good agreement with the values yielded by Equation \ref{eq:AHa}. Our method provides absorption values which are somewhat lower than those often used in previous studies of the same type. This explains that the mean SFR values we obtain here are lower than those frequently found in the literature (see Section \ref{sec:SFR}).

The equivalent width (EW) of the H$\alpha$ line is a direct and widely used method to estimate the intensity of the line as compared to the emission of the underlying continuum. As is usual in the literature, we calculate the EW by forming a rectangle with the height of the continuum under the line and finding the width such that the area of the rectangle matches the area of the spectral line. In practice, we calculated the EW of our galaxies applying two different methods. First, we used the scale factor obtained during the continuum subtraction process, together with the FWHM of the H$\alpha$ narrow-band filter. The equivalent width can then be determined as
\begin{equation}
\label{eq:EW_Method1}
{\rm EW} = \frac{\rm CPS_{H\alpha}}{k\,({\rm CPS_{cont}}-{\rm CPS_{H\alpha}})/{\rm FWHM_{H\alpha}}},
\end{equation}
\noindent where $k$ is the scale factor and CPS$\rm _x$ are the counts per second of the H$\alpha$ and continuum bands,  with CPS$\rm_{H\alpha}$ already corrected for [\NII] contamination and internal absorption. We found that this method yielded very accurate results, and our EW measurements are determined mainly using this system. Unfortunately, some of the literature data sets included neither the continuum scale factor, nor the original H$\alpha$ image, which is necessary to perform our own continuum subtraction. In these cases, we used an alternative method. The EW can be simply calculated as the ratio between the H$\alpha$ continuum subtracted flux and the flux density in the continuum band. As our continuum images were obtaining using standard $R$-band filters, we converted from $R$ magnitudes to flux densities using the parameters which define the system (we mostly used Cousins $R$ filters with an effective wavelength of 6400$\,\rm\AA$ and a zero point of 3080$\,\rm Jy$; \citealt{Cox2000}). As this method relies on physical measurements (with the uncertainties associated with every calibration) and not all the $R$-band filters are identical, we found that the EWs determined in this way are less accurate.

Finally, we used the EW values from the literature where available and perform the calculations described above in the rest of the cases. Section \ref{sec:OtherStudies} compares our EW values with those in the literature to study the quality of our measurements. Table \TabResults{} indicates the final SFR and EW measurements for each galaxy and, for the latter, the method used for the calculation or the source of the value.

\subsection{Uncertainties}
\label{sec:Uncertainties}

Several factors contribute to the final uncertainties of the SFR and EW measurements we present here. First, we estimated the uncertainty associated with the flat-fielding correction. Although this value changes depending on the instrument and even with the specific conditions of the night, we find the typical uncertainty to be well in the range $1-2\%$ with some values as high as $5\%$. 

Uncertainties in the zero point calibrations during photometric nights have typical values of $2\%$. We used several galaxies observed both in photometric and non-photometric conditions to estimate uncertainties associated with the latter. We find that, even in the worst cases, the uncertainty in the calibration is never higher than $5\%$. Similarly, the uncertainty in the flux measurement is typically below $1\%$, with some extreme cases as high as $5\%$ for a few very weak galaxies.

For the uncertainties associated with the distances, we found that the values given in NED were frequently unreliable and underestimated the real uncertainty. Except for a few galaxies with quality distance measurements, we adopted a typical uncertainty of 20\% in the estimated distance to the galaxy.

Along with the distance, the main contribution to the uncertainty for the vast majority of our galaxies is the continuum subtraction. As a galaxy is an extended object there is often not a single scale factor applicable to the whole image. This implies that any method aimed to measure the scale factor used in the continuum subtraction is affected by a certain amount of subjectivity. To measure the uncertainty associated with the removal of the continuum we selected 10 galaxies of different morphological types, SFRs and EWs. For each galaxy we produced new continuum subtracted frames using both the first scale factor which led to a clearly under-subtracted image and the first producing an over-subtracted one. Using these extreme values as a reference, we estimated that the uncertainty due to the scale factor is in the range from 5 to 15\% for most galaxies. Unlike some previous studies (\citealt{James2004}; \citealt{Kennicutt2008}) which reported a correlation between EW and the uncertainty of the scale factor, we find no trend whatsoever in this sense.

Without considering the problem of distance (i.e., taking into account only those uncertainties intrinsic to the quality of the images) we find that the median uncertainty in flux for all our galaxies is $\sim18\%$. This values is in very good agreement with recent surveys using similar techniques. For instance, \citet{Kennicutt2008} obtained a median uncertainty (also using the flux measurements) of $\sim16\%$ while \citet{James2004} reported $\sim18\%$.

\subsection{Image availability}
\label{sec:ImagesAvailability}

With the publication of this article, the H$\alpha$ and \emph{R}-band images for all the galaxies in our sample are made publicly available through the Strasbourg astronomical Data Center (CDS).

%
%
%%%%%%%%%%%%%%%%%%  RESULTS  %%%%%%%%%%%%%%%%%%%%%%%
%
%

\section{Results}
\label{sec:Results}

Table 2 shows the main results of the photometry of our 156 galaxies as well as their associated uncertainties. The column description is as follows.

Column (1): the name of the galaxy in the designation used in this work. 

Column (2): the morphological class of the galaxy as catalogued in NED. 

Column (3): the recession velocity used to calculate the Doppler shift of the H$\alpha$ line, obtained from NED. 

Column (4): the distance to the galaxy considering $\rm H_0=70.5\,km\,s^{-1}\,Mpc^{-1}$, $\rm \Omega_M=0.27$ and $\rm \Omega_{vac}=0.73$ and using a model that takes into account the Virgo infall velocity \citep{Mould2000}. For the galaxies belonging to the Virgo subsample we have adopted a distance of $16.7\pm1.1\,\rm Mpc$ \citep{Mei2007}. Additionally, we used the distances from \citet{Freedman2001} for NGC\,925, NGC\,2403, NGC\,3031, NGC\,3198, NGC\,3351, NGC\,3627 and NGC\,4725; \citet{Karachentsev2002} for NGC\,2976, NGC\,4236, UGC\,04305 and UGC\,08201; \citet{Tonry2001} for  NGC\,584, NGC\,855, NGC\,4594, NGC\,4736, NGC\,4826 and NGC\,5194; \citet{Karachentsev2004} for NGC\,628; \citet{Dalcanton2009} for IC\,2574; \citet{Macri2001} for NGC\,2841; \citet{Leonard2002} for NGC\,3184; and \citet{Seth2005} for NGC\,4631. For NGC\,3034, PGC\,023521, UGC\,05139, UGC\,05336 and UGC\,05423 we used the same distance as NGC\,3031 \citep{Freedman2001}.

Columns (5): the major and minor axis of the galaxy in arcmin. 

Column (6): the \emph{B}-band absolute magnitude derived from the apparent magnitude in the RC3 catalogue and the redshift of each galaxy. The luminosity distance, necessary to derive the absolute magnitude, was calculated using a model which takes $\rm H_0=70.5$, $\rm \Omega_M=0.27$ and $\rm \Omega_{vac}=0.73$. In a few cases (marked in the table) no $B$-band magnitude was available in RC3 and we use the closest match from the photometric data points gathered in NED for each galaxy. For PGC\,140287 we used the $B$-band magnitude from LEDA \citep{Paturel2003}.

Column (7): the scale factor applied to the continuum image during the continuum subtraction (see Section \ref{sec:RedContinuum}).

Column (8): the estimated contribution of the H$\alpha$ line to the flux measured in the continuum filter.

Column (9): the total flux of the galaxy in the H$\alpha$ line in units of $\rm erg\,cm^{-2}\,s^{-1}$. The area used to perform the photometry has been selected as indicated in Section \ref{sec:RedFlux}. This flux has not been corrected for any kind of absorption or [\NII] contribution.

Column (10): the total absorption for the galaxy, $A_{\rm T}$, in mag. This value comprises the Galactic foreground absorption in the $R$ band, $A(R)$, for the position of the galaxy taken from \citet{Schlegel1998}, and the internal absorption, $A(H\alpha)$ calculated as explained in Section \ref{sec:SFRandEW}.

Column (11): the estimated ratio $\rm \xi_{obs}=([\NII]/H\alpha)_{obs}$ calculated as explained in Section \ref{sec:NII}.

Column (12): the H$\alpha$ luminosity of the galaxy, in $\rm erg\,s^{-1}$. This value has been corrected for absorption using $A_{\rm T}$ and for the contribution of the [\NII] lines.

Column (13): the star formation rate for the galaxy in units of $M_{\odot}\rm\,yr^{-1}$ (corrected for absorption and [\NII]). 

Column (14): the equivalent width, in \AA{}ngstr\"{o}m, calculated as described in Section \ref{sec:SFRandEW}. The method used for the calculation is indicated in parenthesis meaning (1) that the equivalent width is derived from the scale factor used for the continuum subtraction as in Equation (\ref{eq:EW_Method1}); (2) that the equivalent width is derived from the fluxes of the images; and (3) that we used the value from the literature. In the latter case, the reference is always the same as the one listed in Table \TabObs{} for the corresponding galaxy.

\subsection{Star formation rate measurements}
\label{sec:SFR}

Figure \ref{fig:SFRvsMB} shows the correlation between the SFR (in units of solar masses per year) and the \emph{B}-band absolute magnitude, $M(B)$. Our four subsamples (field, small Virgo, and large and small SINGS galaxies) are plotted separately in Figure \ref{fig:SFRvsMB_Panels} to consider the individual properties and possible biases of these subsets.

We find an expected trend, with brighter galaxies hosting a higher amount of star formation. This behavior has been previously reported by several authors such as \citet{Tresse1998}, who used spectroscopic H$\alpha$ data from a sample of galaxies at $\rm z\sim0.2$, and more recently \citet{Knapen2009} (their Figure 2), using the sample of the H$\alpha$GS survey (see also Section \ref{sec:OtherStudies}). Figure 6 in \citet{Lee2009a}, who use data of $\sim300$ galaxies from 11HUGS \citep{Kennicutt2008a}, shows a similar trend for galaxies with ${\rm M}(B)\le-15$ (for very faint galaxies, ${\rm M}(B)>-15$,  the SFR seems to  drop suddenly to values below $10^{-4}\,M_\odot\rm\,yr^{-1}$). This same result is found by \citet{Karachentsev2010} using 435 galaxies in a volume-limited region of $10\,\rm Mpc$. The slope of our least squares fit is -0.29, somewhat higher than the value of -0.41 found by \citet{Lee2009a}.

The top right corner in Figure \ref{fig:SFRvsMB} is exclusively occupied by galaxies from the SINGS large subsample. The galaxy with the highest SFR in our sample (11.31$\,M_\odot\rm\,yr^{-1}$) is NGC$\,$4254 (Messier 99). In the other extreme of the plot we find two small irregular galaxies (PGC$\,$23521 and IC$\,$3105), both with a SFR below $10^{-3}\,M_\odot\rm\,yr^{-1}$.

It is interesting to consider the behavior of the four different subsamples which compose our sample. Field galaxies in the NGLS sample are distributed uniformly along a wide range of $M(B)$, which is expected as they have been selected based on their \HI emission and not on their optical luminosity. The same can be said for the galaxies contained in the SINGS subsamples. The Virgo cluster subsample, however, contains galaxies with $M(B)$ well constrained in the range $[-17,-22]$ magnitudes.

We calculate the median\footnote{The uncertainties are calculated by selecting a random subsample from the complete set of SFR values and calculating its median. We repeated this process a thousand times and selected the maximum and minimum median values obtained as the upper and lower uncertainties.} SFR for all the galaxies in our sample as $0.21^{+0.37}_{-0.12}\,M_\odot\rm\ yr^{-1}$. This value is consistent with those of the different subsamples ($0.22^{+0.46}_{-0.12}\,M_\odot\rm\ yr^{-1}$ for the field galaxies, $0.11^{+0.55}_{-0.01}$ for the SINGS small galaxies, and $0.09^{+0.21}_{-0.05}\,M_\odot\rm\ yr^{-1}$ in the case of Virgo galaxies). Not unexpectedly, the SINGS large subsample shows a higher median SFR ($1.04^{+1.77}_{-0.31}\,M_\odot\rm\ yr^{-1}$) as it is mainly populated by spiral galaxies (see Figure \ref{fig:Histogram}) with high FIR luminosities and, thus, large SFR (see, for example, \citealt{Calzetti2009}). On the other hand, the large uncertainty in the SINGS median value shows that this subsample contains galaxies with a large range of SFRs. The median value for the whole sample is lower than those of other studies in the literature (see Section \ref{sec:OtherStudies}). We attribute this difference to the fact that the internal absorptions we calculate here (see Section \ref{sec:SFRandEW}) are, in general, lower than those previously used. For example, if we adopt a typical, constant, $A(\rm H\alpha)=1.1\,mag$ value for the internal absorption, the median SFR becomes $0.38^{+0.60}_{-0.24}\,M_\odot\rm\ yr^{-1}$, more consistent with other values found in the literature.

To study the range of SFRs in the different subsamples, we binned the data into 0.5 magnitude bins and calculated the range (the difference between the maximum and minimum values) for each bin. We find maximum ranges between 2.0 and 2.9 dex for the field, SINGS large and Virgo subsamples. In every instance, the peak of maximum range is located around $\rm M({\it B})=20.0\,magnitudes$. On the other hand, the range of values in the SINGS small sample, as can easily be appreciated from Figure \ref{fig:SFRvsMB_Panels}, is much lower, with a maximum value of 1.3 dex. The same results, for the whole sample, are shown in the lower plot of Figure \ref{fig:SFRvsMB} as the normalized root mean square distance of each data point to the best fit. The largest dispersion occurs between ${\rm M}(B)\sim-17$ and ${\rm M}(B)\sim-20$.

Figure \ref{fig:SFRvsT} presents the measured SFR as a function of the Hubble stage of each galaxy. We find that most of the star forming members of our sample are spiral galaxies with stage $T\sim4$ (Sb/Sbc). A secondary peak is found for late-type spirals ($T\sim7$). These results are in good agreement with previous studies (see, for example, \citealt{James2004}). While there is an expected decline in the SFR toward late-type and irregular galaxies the same is not true for very early ellipticals. A slight increase in SFR appears to exist for early-type galaxies $(T\le-2)$. This behavior seems to contradict the widely accepted theory that elliptical galaxies have exhausted most of their gas reservoirs and do not form stars. A series of recent studies (see, e.g., \citealt{Sarzi2006, Sarzi2010}) has shown that these galaxies may contain more ionized gas than traditionally thought. However, this may be mostly due to internal shocks and not necessarily to the star formation processes occurring in the galaxy. The number of early-type galaxies in our sample is, unfortunately, not large enough to make a definitive statement in this debate.

Figure \ref{fig:SigmavsSFR} shows $\Sigma$ (the SFR surface density, in units of $M_\odot\rm\,yr^{-1}\,kpc^{-2}$) of each galaxy plotted against the corresponding SFR. To compute the size of the galaxy we used the optical size $D_{25}$ from RC3 and converted it to kpc using the distance from Table 2. We find that galaxies with higher $\Sigma$ tend to host larger amounts of star formation, with a regression fit of $\rm log(\Sigma)=0.37\,log(SFR)-2.28$. \citet{Knapen2009} (their Figure 4) published a similar plot for the galaxies in the H$\alpha$GS sample. In their case, however, $\Sigma$ remains constant for SFRs lower than $\sim1\,M_\odot\rm\,yr^{-1}$. 

As in the case of the SFRs, the result for $\Sigma$ shows a wide dispersion, with values ranging across three orders of magnitude. This can be said for each one of the subgroups which make up our sample.

We also compared our SFRs with the $M(K)$ magnitudes of the galaxies in our sample, following the same process as explained above. We find that the general behavior of the plotted data does not change when using $M(K)$ instead of $M(B)$ for the SINGS Large and Small and for the Virgo subsamples. In the case of the Field subsample, the data show a higher dispersion when using $M(K)$.

\subsection{Equivalent widths}
\label{sec:EW}

Figures \ref{fig:EWvsMB} and \ref{fig:EWvsMB_Panels} show, globally and separated by subsample, the EWs of the galaxies in our sample as a function of their absolute magnitude in the $B$-band.

We find that our EW values are distributed over almost two orders of magnitude, with a median value of $27^{+33}_{-20}\,\rm\AA$. The different subsamples are consistent with this global result showing median values that range between $21^{+37}_{-6}\,\rm\AA$ (SINGS Large) and $31^{+47}_{-21}\,\rm\AA$ (SINGS Small). 

The relation between EW and $M(B)$ is basically flat, which indicates that, in most cases, an enhanced rate of star formation comes with a similar  increase in the continuum emission produced by evolved stars. This behavior has been suspected for almost three decades (see, for example \citealt{Kennicutt1983}) and confirmed in recent studies such as \cite{Gavazzi1998} (their Figure 5), \citet{James2004} and \citet{Knapen2009}, the latter two using H$\alpha$GS data. They also detected (see Figure 3 in \citealt{Knapen2009}) a significant number of very luminous galaxies with low EW. Their hypothesis that these galaxies are, in their majority, massive elliptical and early-type objects with almost no star formation is also supported by \citet{Lee2007} and \citet{Lee2009}, who showed that this result is also valid when comparing the EW with the rotation velocity of the galaxies. The downturn in EW is not as conspicuous in our sample as in these studies, but it seems to exist for $M(B)<-17$ mag, a region populated mostly by field and Virgo galaxies. 

Four out of the ten highest EW values are field galaxies, the top one being PGC$\,$43211, a spheroidal galaxy with an EW of 880$\,\rm\AA$. Almost no SINGS galaxies are found among the galaxies with the highest EW values. This indicates that, while these other galaxies contain SFRs usually above the mean for field galaxies, their masses of old stars are proportionally high. We also find that two of the lowest EW measurements, with EWs below $5\,\rm\AA$, are intermediate-type (Sab and Sb) galaxies from the SINGS large subsample. NGC\,4470, of Sa class and $\rm EW\sim0.1\rm\,\AA$, is the galaxy with the lowest EW in our sample.

Figure \ref{fig:EWvsT} (which shows the EW plotted as a function of the Hubble stage) does not show any significant correlation between EW and morphological type except, maybe, a certain increase of the EW values for galaxies of Sb type and later.

%
%
%%%%%%%%%%%%%%%%%%  OTHER STUDIES  %%%%%%%%%%%%%%%%%%%%%%%
%
%

\section{Other studies}
\label{sec:OtherStudies}

Our current study is one of many surveys based on H$\alpha$ as a tracer of massive stars. In the last decade alone, many samples have been observed and reduced using techniques similar to those presented here (e.g., \citealt{Gavazzi2002, Boselli2002, Gavazzi2003, HunterElmegreen2004, James2004, Meyer2004, Verdes-Montenegro2005, Gavazzi2006, Kennicutt2008a}). Notwithstanding the common goal of characterizing massive star formation in the local Universe, these samples have used different selection criteria and thus all suffer from different biases. Some of the most representative SFR studies in recent years are addressed below and their results compared with some of ours.

The GOLDMine project\footnote{http://goldmine.mib.infn.it/} has observed and collected data for over 3000 galaxies belonging to, among others, the Coma supercluser and the Virgo, Cancer and Hercules clusters. This data includes \HInoSp, IR and optical continuum measurements, as well as spectral observations. Two large sets of original H$\alpha$ data have been published by Gavazzi and collaborators. \citet{Gavazzi2002} observed 369 galaxies with types later than S0a in the Virgo and Coma/A1367 clusters. \citet{Gavazzi2006} extended the sample with some 273 galaxies in the Virgo and Cancer clusters and in the Coma/A1367 and Hercules superclusters.

We have already introduced the SINGS survey \citep{Kennicutt2003} in Section \ref{sec:Sample}. Aimed to characterize the infrared emission in the local Universe, it includes 75 galaxies within 30 Mpc from elliptical to irregular and with a large range of SFRs. The SINGS team has obtained H$\alpha$ and continuum broad-band imaging for the whole sample but, although the images have been made public, no catalogue of galaxy-integrated SFRs or EWs has been published.

Our survey has also made extensive use of the H$\alpha$GS work \citep{James2004}, which observed 334 galaxies extracted from the Uppsala Galaxy Catalogue \citep{Nilson1973} of types later than S0a and with $\rm m(B)\lesssim 14.5$ in the Palomar observatory plates. The targets were selected to include galaxies with recession velocities from 0 to $\rm 3000\,km\,s^{-1}$. As in our case, they found a clear trend between SFRs and absolute magnitudes and an apparent lack of correlation of EW with luminosity and morphological type (see also \citealt{Knapen2009}). 

Although it is generally accepted that most of the massive star formation in galaxies occurs in the spiral arms of late-type galaxies, some authors have questioned the necessity of a spiral density wave to achieve an important amount of star formation (see, for example, \citealt{Hunter1982}, \citealt{Elmegreen1986,Elmegreen2011} and references therein). \citet[hereafter HE04]{HunterElmegreen2004} focused on observing a sample of 142 mainly irregular galaxies. Some Sm and blue compact dwarfs (BCD) were observed for comparison. The sample of irregular galaxies was compiled from the \citet{Fisher1975} \HI atlas of galaxies and thus includes only galaxies containing gas. They found H$\alpha$ emission in over 80 of their 96 irregular galaxies and in all the Sm and dwarf (BCD) galaxies except one. The SFRs are low, with a mean value of $0.1\rm\,M_\odot\,yr^{-1}$ and a maximum of $0.5\,M_\odot\rm\,yr^{-1}$.

The Survey for Ionization in Neutral-Gas Galaxies (SINGG) aims to observe 468 galaxies selected from the  \HI Parkes All Sky Survey (HIPASS, \citealt{Meyer2004}) with peak flux densities above $0.05\,\rm Jy$. The galaxies have been selected to cover a range of $\rm \log(M_{\,H\,{\mathsc i}}/{\it M}_\odot)$ from 8 to 10.5. The most recent data release we have been able to find is the one published by \citet{Hanish2006} which contains observations and derived values for 111 galaxies. Among those, every galaxy with $\rm M_{\,H\,{\mathsc i}}>3\times10^7\,{\it M}_\odot$ has also been detected in H$\alpha$, thus constraining the amount of gas necessary to trigger global star formation in a galaxy.

\citet{Kennicutt2008a} (hereafter RCK08) have made, along with H$\alpha$GS, one of the first attempts to observe a volume-limited sample of galaxies using H$\alpha$. Two main factors make this a difficult task: firstly, the mapping of galaxies (especially irregular ones) in the local universe is currently not complete; and secondly, even for the known galaxies, the high uncertainties in the distances make it difficult to define the set of galaxies within a certain radius. While these problems cannot be solved at the present moment, RCK08 managed to observe a sample of 261 S and Irr galaxies within 11 Mpc with galactic latitudes above $20^\circ$ and ${\rm m}(B)\le15$ magnitudes, and an additional set of 175 galaxies with lower galactic latitudes or fainter. This sample is shared with other legacy surveys to be carried out using \textit{Galex} and \textit{Spitzer} which will provide interesting complementary information in the near and far UV and in the IR (3.6 to 160$\,\rm\mu m$).

Finally, \citet{Karachentsev2010} presented a study of 52 galaxies observed in H$\alpha$ as part of a larger survey aimed to study a sample of galaxies in the Local Volume (207 galaxies within 10\,Mpc). An important number of the galaxies observed belong to the Maffei 2/IC\,342 complex. The sample contains galaxies of all morphological types but with prevalence of irregular galaxies.

Figure \ref{fig:HistT} shows a series of histograms comparing the number of galaxies versus the morphological class (in the form of Hubble stages) for our sample and the different samples presented above. Most of these studies lack a good representation (or any representation at all) of early-type galaxies. While this is a natural thing in surveys like HE04, aimed to study irregular galaxies specifically, it indicates that a general bias towards late-type (S0 and later) galaxies dominates the literature of H$\alpha$ surveys. This comparison also shows that \HInoSp-selected samples, as SINGG and our own sample, span a wider range of morphological classes.

Figures \ref{fig:HistDist} to \ref{fig:HistEW} present similar histograms for distance, SFR and EW respectively. SINGS does not provide measurements of H$\alpha$ fluxes (and, consequently, $\rm L(H\alpha)$ and SFR) or EW. EWs for HE04 and RCK08 are not available in the public catalogues. The EWs from RCK08 we use to complete our data (see Section \ref{sec:SFRandEW}) have been taken directly from the headers of the images, but they are not included in the catalogue published in CDS.

The distance histograms show that most of the galaxies observed to date with resolved H$\alpha$ imaging are closer than 50 Mpc even when, among the surveys discussed here, only RCK08 and H$\alpha$GS include the distance as a selection criterion. The exceptions are SINGG, which observed galaxies as far as 80 Mpc, and \citet{Gavazzi2002,Gavazzi2006}, most of whose galaxies are located at the specific distances of the clusters observed ($\sim100\,\rm Mpc$\footnote{From NED} for the Coma Cluster, 16.4 to 22.9$\rm\,Mpc$ for the Virgo Cluster --\citealt{Mei2007}, and $>150\rm\,Mpc^5$ for the Hercules Cluster). The galaxies in our sample have a mean distance of 15.8 Mpc, similar to those in the rest of the surveys, except for RCK08 with 6.8 Mpc and \citet{Gavazzi2002,Gavazzi2006} with $\sim50\,\rm Mpc$.

All the SFR histograms shown in Figure \ref{fig:HistSFR} tell a similar story, with most galaxies hosting small amounts of star formation (less than $0.5\,M_\odot\rm\,yr^{-1}$). The number of galaxies with rates of star formation higher than $\sim1\,M_\odot\rm\,yr^{-1}$ drops dramatically in every case. \citet{Gavazzi2002} show a significant number of galaxies with high SFRs (up to $\sim6\,M_\odot\rm\,yr^{-1}$). This may be due to a systematic overestimation of their H$\alpha$ fluxes, since the distribution of morphological types (see Figure \ref{fig:HistT}) does not indicate an overwhelming presence of highly efficient star forming galaxies. Eleven of the SINGG galaxies present SFRs higher than $10\,M_\odot\rm\,yr^{-1}$ with the extreme case of M\,83 with a SFR of $145\,M_\odot\rm\,yr^{-1}$. Those cases are suspect, as such behaviour is not found in any of the other surveys. \citet{HunterElmegreen2004} and \citet{Karachentsev2010} show virtually no SFRs above $0.5\,M_\odot\rm\,yr^{-1}$, which can be attributed to the fact that their samples are mostly or totally composed of irregular galaxies with very low SFRs.

A similar behaviour is found (Figure \ref{fig:HistEW}) for the EW of the SINGG galaxies. The large number of galaxies with very high EW ($\gtrsim140\,\rm\AA$, with a maximum value of $450\,\rm\AA$ for the peculiar galaxy NGC\,5253) is surprising. These exceptions notwithstanding, the comparison between the different histograms show a similar pattern, with mean EWs between 20 and 35$\,\rm\AA$ (in the case of SINGG the mean is 50.7$\,\rm\AA$ but this is mainly due to some extremely high values). 

Figure \ref{fig:LitSFR} shows the comparison between our SFR data and the values found in the literature. Except for a global offset, which can be explained by the different internal absorption used in this work (see Section \ref{sec:SFR}), the global agreement is good, with a fit of slope 0.90 and a correlation coefficient $R^2=0.82$. This good agreement is also valid if we compare our data with each individual study. We find a slope of 0.95 and $R^2=0.98$ for H$\alpha$GS, 1.01 and 0.88 for RCK08, 1.35 and 0.92 for GM and 0.86 and 0.58 for SINGS. The excellent fit in the case of H$\alpha$GS is especially important as our aperture selection and flux measurement methods are essentially the same. As we have already commented, the SINGS team has not published SFR measurements based on their H$\alpha$ images. \citet{Kennicutt2003} gives the SFR values from the IRAS catalogue \citep{Fullmer1989}, based not on H$\alpha$ observations but on IR measurements from 60 to 100$\,\rm\mu m$. This explains that the fit using SINGS values is worse than those using the other points.  In any case, these comparisons show that the differences between our SFR values and those in the literature lie well within the typical errors for this kind of measurement, whatever the methods applied and the specific luminosity-to-SFR constant used in each study.

%
%
% %%%%%%%%%%%%%%%%%%  CONCLUSIONS  %%%%%%%%%%%%%%%%%%%%%%%
%
%

\section{Conclusions and future work}
\label{sec:Conclusions}

We have presented new H$\alpha$ and $R$-band observations of 72 out of the 156 galaxies sampled in the JCMT Nearby Galaxies Legacy Survey. This sample is partly \HInoSp-selected and aims to include galaxies within 25$\,$Mpc with morphological types from E to Irr. The sample is largely formed by field galaxies but also includes targets from the SINGS survey and a fair amount of Virgo cluster galaxies. 

For the remaining 84 galaxies we have compiled images from the literature. We processed all these data using well-defined methods and criteria and present H$\alpha$ fluxes, SFR and EW measurements for the whole sample. With this article, we make our new data publicly available.

We analyzed  the data reaching, among others, the following conclusions:

\begin{itemize}
    \item[(i)] The SFR shows a clear increasing trend when plotted against $M(B)$. This trend is dependent on the subsample considered, with a better fit for the field and SINGS Small galaxies than for those in the SINGS Large or Virgo subsamples.
    \item[(ii)] We find a median SFR across our sample of 0.2$\,M_\odot\rm\,yr^{-1}$, while in the case of the galaxies in the SINGS Large subsample the median SFR is 1$\,M_\odot\rm\,yr^{-1}$.
    \item[(iii)] The SFR values range across almost three orders of magnitude, except in the case of the SINGS Small galaxies where the values are confined to a range of one order of magnitude.
    \item[(iv)] The SFR of our sample peaks for Sb/Sb-c galaxies ($T\sim4$) with a secondary peak for late-type Sc galaxies ($T\sim7$). 
    \item[(v)] We plotted EW versus $M(B)$ confirming that the relation between EW and luminosity is practically flat. This is also valid when comparing the EWs with the morphological type, although a trend seems to exist leading to slightly higher EW for late-type and irregular galaxies.
    \item[(vi)] The comparison of our SFR values with those found in the literature shows an excellent agreement for SFR, well within the typical margins of error. 
\end{itemize}

This paper has presented our H$\alpha$ line observations, as well as a general study of the star forming properties of the sample and a comparison with previous similar studies. However, this is only the first step towards the more ambitious goal of comparing the star formation in the galaxies of the sample with their gas content, both atomic and molecular. $^{12}$CO $J=3-2$ maps of the whole sample are available as a part of the NGLS observations carried out with HARP-B in the JCMT, thus allowing a global comparison between $\rm L(H\alpha)$ and L(CO) for all our galaxies. This will be the largest study correlating the star formation with the amount of warm, dense molecular hydrogen present in a galaxy and may be key to understanding the physics involved in large-scale star formation in the nearby universe.

Furthermore, the rich variety of literature data available for a large part of our sample, from IR data obtained with \emph{Spitzer} to FUV from \emph{GALEX} as well as \HI from a variety of sources (THINGS: \citealt{Walter2008}; WHISP: \citealt{vanderHulst2002}; etc.) and CO $J=1-0$ and $J=2-1$ (e.g., HERACLES: \citealt{Leroy2009}) allows us to study the behavior of the different gas and dust components over the complete morphological range.

Finally, the JCMT NGLS has been awarded guaranteed time to observe the whole sample using SCUBA-2 to obtain continuum imaging in 450 and 850$\,\rm \mu m$ and study the cold dust within galaxies with an unprecedented sensitivity and resolution. These data will also be compared to the current H$\alpha$ and other data in forthcoming papers. 

% {\bf
% \begin{itemize}
%    \item Complete and modify the bullet points according with the changes in the results section.
% \end{itemize}
% }

%
%
%%%%%%%%%%%%%%%%%%  ACKNOWLEDGMENTS  %%%%%%%%%%%%%%%%%%%%%%%
%
%

\section*{Acknowledgments}

We thank Elias Brinks, Jen Golding, Frank Israel, Robert C. Kennicutt and Paul van der Werf for their comments and useful suggestions during the preparation of this manuscript. We also thank Eduardo Gonz\'{a}lez for his help during the initial steps of the INT image reduction. Finally, we thank Santiago Erroz Ferrer and Elena Mohd-Noh Velast\'{i}n for the observation and reduction of the images of NGC\,4504 with the NOT. The James Clerk Maxwell Telescope is operated by the Joint Astronomy Centre on behalf of the Science and Technology Facilities Council of the United Kingdom, the Netherlands Organisation for Scientific Research, and the National Research Council of Canada. Based on observations obtained with the WHT, the INT, and the NOT, operated on the island of La Palma by the Isaac Newton Group of Telescopes, and jointly by Denmark, Finland, Iceland, Norway and Sweden, in the Spanish Observatorio del Roque de los Muchachos of the Instituto de Astrof\'{i}sica de Canarias. We acknowledge the use of the Hiltner telescope at the MDM Observatory operated by the University of Michigan, Dartmouth College, the Ohio State University, Columbia University, and Ohio University. This research has made use of the NASA/IPAC Extragalactic Database (NED) which is operated by the Jet Propulsion Laboratory, California Institute of Technology, under contract with the National Aeronautics and Space Administration.\\

\nocite{James2004}
\nocite{Kennicutt2003}
\nocite{Knapen2004}
\nocite{Kennicutt2008a}
\nocite{Gavazzi2002}
\nocite{Gavazzi2006}
\nocite{Boselli2002}
\nocite{Paturel2003}
\nocite{Wilson2012}

\bibliographystyle{mn2e}
\bibliography{Papers}

%
%
%%%%%%%%%%%%%%%%%%  Long tables  %%%%%%%%%%%%%%%%%%%%%%%
%
%
\onecolumn
\begin{landscape}
    \begin{center}
        \begin{longtable}{lccccccccccc}
            \caption{Observations log}\\
            \hline\hline
            Galaxy & Subsample & Telescope & Instrument & Date of obs. & \multicolumn{2}{c}{Filters$\rm ^a$} & 
            \multicolumn{2}{c}{Exp. time} & Pixel scale & Seeing & Ref.$\rm ^b$\\
                   &  &  &  & (YY-MM-DD) &  H$\alpha\ \rm (\AA/\AA)$  &  Cont. $\rm (\AA/\AA)$ &  H$\alpha$ (s)  &  Cont (s)  & (arcsec/pix) & (arcsec) & \\
            \hline
            \endfirsthead
            \caption{continued.}\\
            \hline\hline
            Galaxy & Subsample & Telescope & Instrument & Date of obs. & \multicolumn{2}{c}{Filters$\rm ^a$} & 
            \multicolumn{2}{c}{Exp. time} & Pixel scale & Seeing & Ref.$\rm ^b$\\
                   &  &  &  & (YY-MM-DD) &  H$\alpha$ $\rm (\AA/\AA)$  &  Cont. $\rm (\AA/\AA)$ &  H$\alpha$ (s)  &  Cont (s)  & (arcsec/pix) & (arcsec) & \\
            \hline
            \endhead
            \hline
            \endfoot
            
            \hline
            \multicolumn{11}{l}{$\rm ^a$ Given for the H$\alpha$ and the continuum filters in the format central wavelength/FWHM.}\\
            \multicolumn{11}{l}{$\rm ^b$ (1)~This work; (2)~Gavazzi~et~al.~(2002); (3)~Boselli~et~al.~(2002); (4)~Kennicutt~et~al.~(2003); 
            (5)~James~et~al.~(2004); (6)~Knapen~et~al.~(2004);}\\ 
            \multicolumn{11}{l}{$\rm ^{\ \ }$(7)~Gavazzi~et~al.~(2006); (8)~Kennicutt~et~al.~(2008)}\\
            \endlastfoot 
            
            \input{observationsTable.txt}

        \end{longtable}
    \end{center}

\end{landscape}

\begin{landscape}
% \tiny
\centering
\begin{longtable}{llrrrrrrrrrrrr}
\caption{Results}\\
\hline\hline
Galaxy & Class & \multicolumn{1}{c}{$v_{\rm rec}$} & \multicolumn{1}{c}{Dist.} & \multicolumn{1}{c}{a/b} & \multicolumn{1}{c}{$M(B)$} & \multicolumn{1}{c}{k} & \multicolumn{1}{c}{$\eta$} & \multicolumn{1}{c}{$F\rm\left(H\alpha+[\NII]\right)$} & $A_{\rm T}$ & $\xi_{obs}^{\rm a}$ & \multicolumn{1}{c}{$L\rm(H\alpha)^b$} & \multicolumn{1}{c}{SFR$\rm{}^b$} & \multicolumn{1}{c}{EW} \\
 &       & \multicolumn{1}{c}{$\rm km\,s^{-1}$} & \multicolumn{1}{c}{Mpc} & \multicolumn{1}{c}{arcmin} & \multicolumn{1}{c}{mag} &  & \multicolumn{1}{c}{\%} & \multicolumn{1}{c}{$\rm 10^{-13}\,erg\,cm^{-2}\,s^{-1}$} & \multicolumn{1}{c}{mag} & \multicolumn{1}{c}{} & \multicolumn{1}{c}{$\rm 10^{40}\,erg\,s^{-1}$} & \multicolumn{1}{c}{$M_{\odot}\rm\,yr^{-1}$} & \multicolumn{1}{c}{\AA} \\
\hline
\endfirsthead

\caption{continued.}\\
\hline\hline
Galaxy & Class & \multicolumn{1}{c}{$v_{\rm rec}$} & \multicolumn{1}{c}{Dist.} & \multicolumn{1}{c}{a/b} & \multicolumn{1}{c}{$M(B)$} & \multicolumn{1}{c}{k} & \multicolumn{1}{c}{$\eta$} & \multicolumn{1}{c}{$F\rm\left(H\alpha+[\NII]\right)$} & $A_{\rm T}$ & $\xi_{obs}^{\rm a}$ & \multicolumn{1}{c}{$L\rm(H\alpha)^b$} & \multicolumn{1}{c}{SFR$\rm{}^b$} & \multicolumn{1}{c}{EW} \\
 &       & \multicolumn{1}{c}{$\rm km\,s^{-1}$} & \multicolumn{1}{c}{Mpc} & \multicolumn{1}{c}{arcmin} & \multicolumn{1}{c}{mag} &  & \multicolumn{1}{c}{\%} & \multicolumn{1}{c}{$\rm 10^{-13}\,erg\,cm^{-2}\,s^{-1}$} & \multicolumn{1}{c}{mag} & \multicolumn{1}{c}{} & \multicolumn{1}{c}{$\rm 10^{40}\,erg\,s^{-1}$} & \multicolumn{1}{c}{$M_{\odot}\rm\,yr^{-1}$} & \multicolumn{1}{c}{\AA} \\
\hline
\endhead

\hline
\endfoot

\hline
 \multicolumn{13}{l}{$\rm ^a$ $\rm [\NII]/H\alpha$ ratio corrected for the transmission profile of the H$\alpha$ filter used (see Section \ref{sec:NII}.).}\\
 \multicolumn{13}{l}{$\rm ^b$ Corrected for Galactic foreground absorption and internal absorption using $A_{\rm T}=A(R) + A({\rm H\alpha})$ (see Section \ref{sec:SFRandEW}).}\\
 \multicolumn{13}{l}{$\rm ^c$ From NED.}\\
 \multicolumn{13}{l}{$\rm ^d$ From LEDA (Paturel et al. 2003).}\\
\endlastfoot

\input{resultsTable.txt}

% \hline
\end{longtable}
\end{landscape}

% \ \newpage
%%%%%%%%%%%%%%%%%%  Figures  %%%%%%%%%%%%%%%%%%%%%%%
%
%

% \ \newpage

\begin{figure*}
  \centering
  \includegraphics[width=17cm]{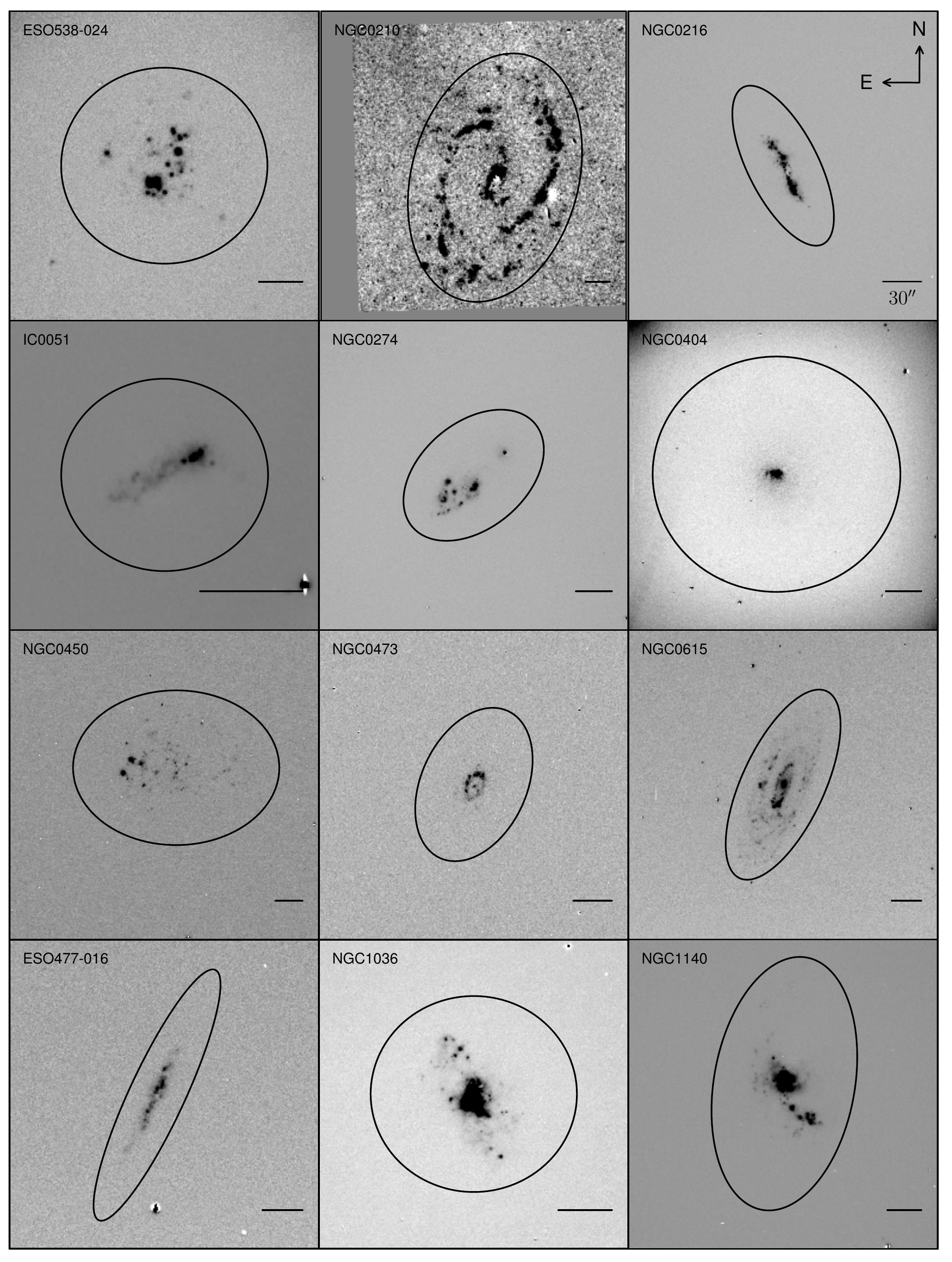}
  \caption{H$\alpha$ continuum-subtracted images of the field galaxies of our sample. The ellipse shows the aperture used to perform the photometry (see Section \ref{sec:RedFlux}). All the images are oriented North up, East to the left. The right bottom line is always $30\,\rm arcsec$.}
  \label{fig:ImagesField}
\end{figure*}
% \ \newpage

\begin{figure*}
  \ContinuedFloat 
  \centering
  \includegraphics[width=17cm]{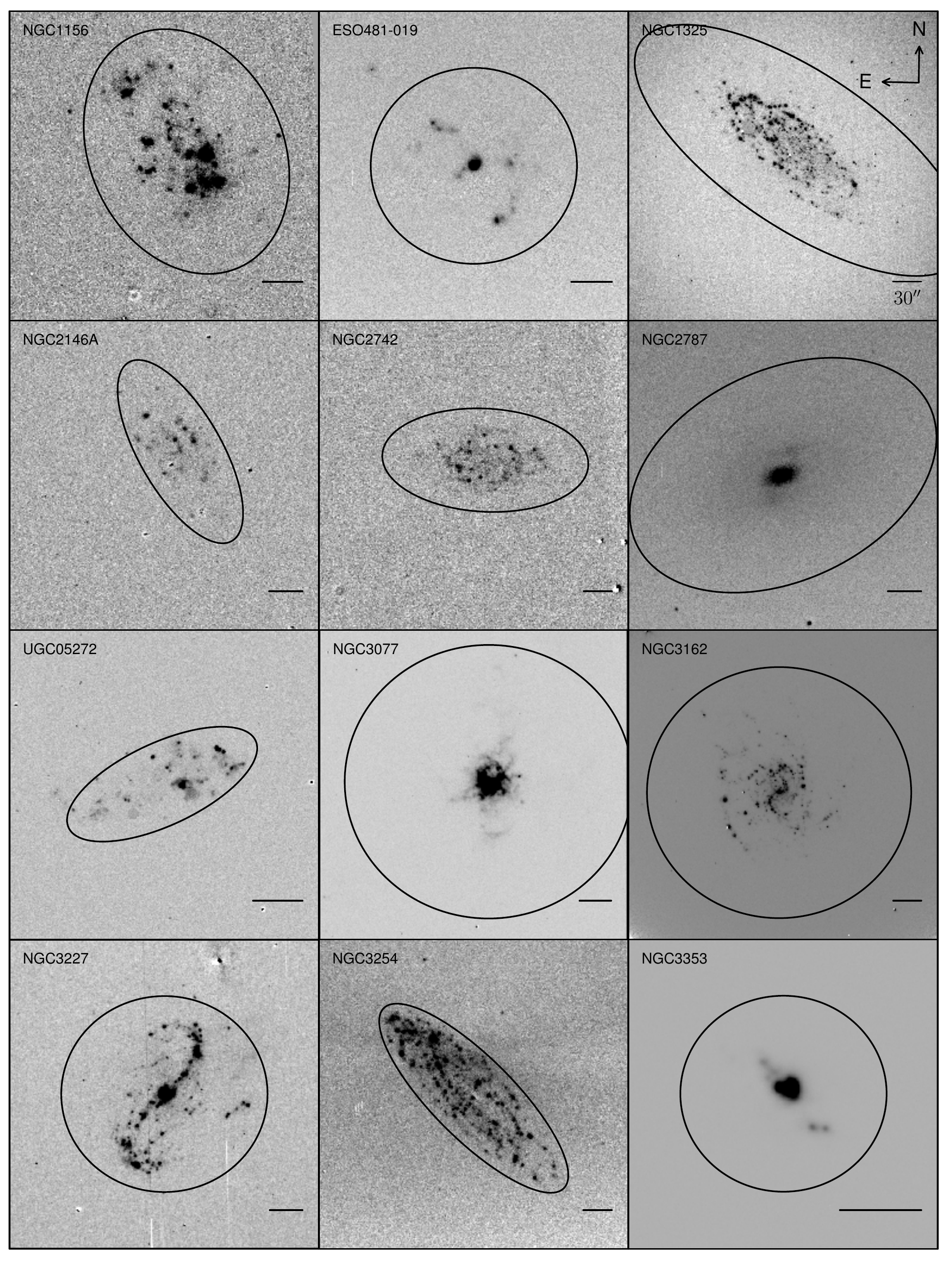}
  \caption{Continued}
\end{figure*}
% \ \newpage

\begin{figure*}
  \ContinuedFloat 
  \centering
  \includegraphics[width=17cm]{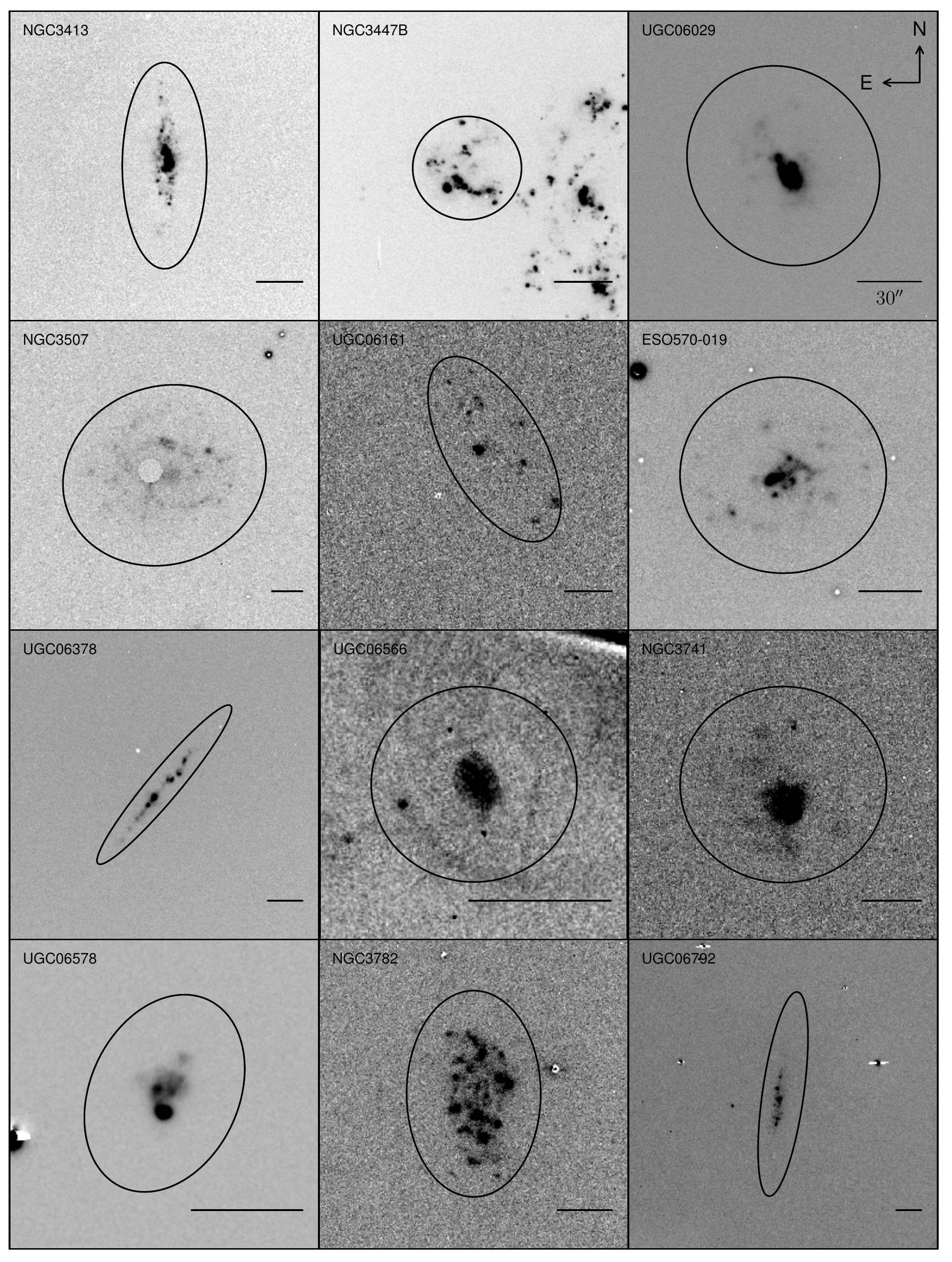}
  \caption{Continued}
\end{figure*}
% \ \newpage

\begin{figure*}
  \ContinuedFloat 
  \centering
  \includegraphics[width=17cm]{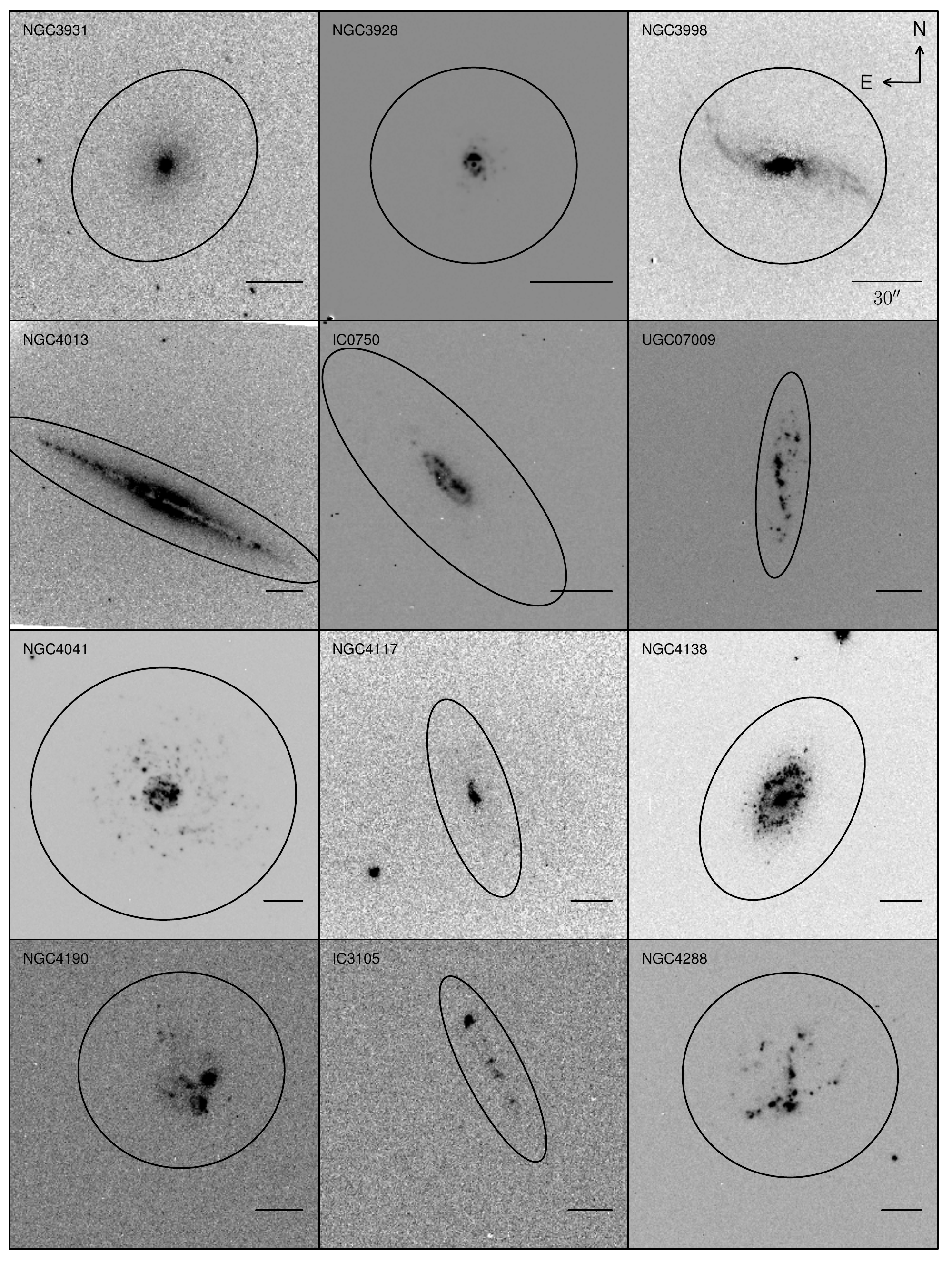}
  \caption{Continued}
\end{figure*}
% \ \newpage

\begin{figure*}
  \ContinuedFloat 
  \centering
  \includegraphics[width=17cm]{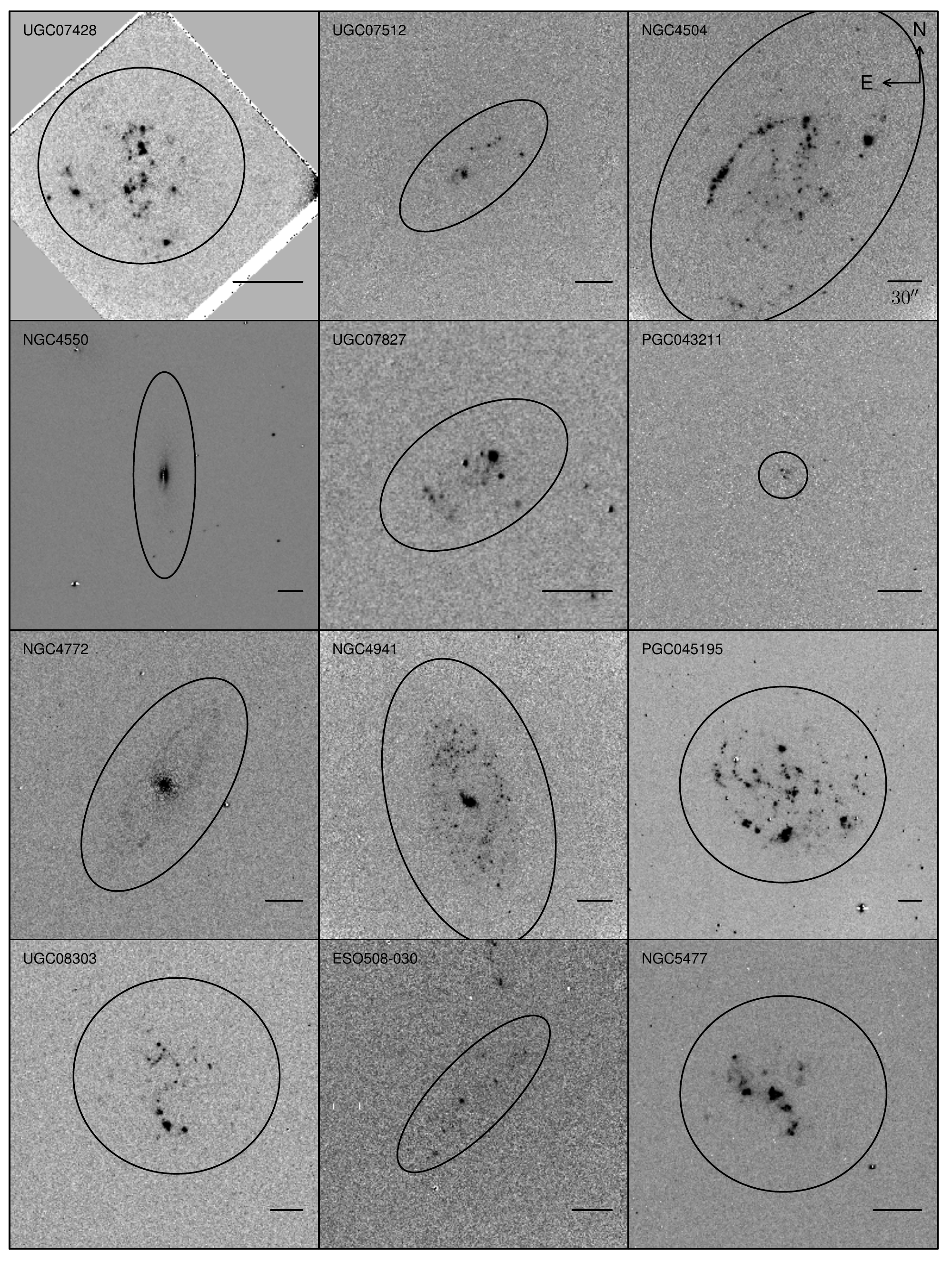}
  \caption{Continued}
\end{figure*}
% \ \newpage

\begin{figure*}
  \ContinuedFloat 
  \centering
  \includegraphics[width=17cm]{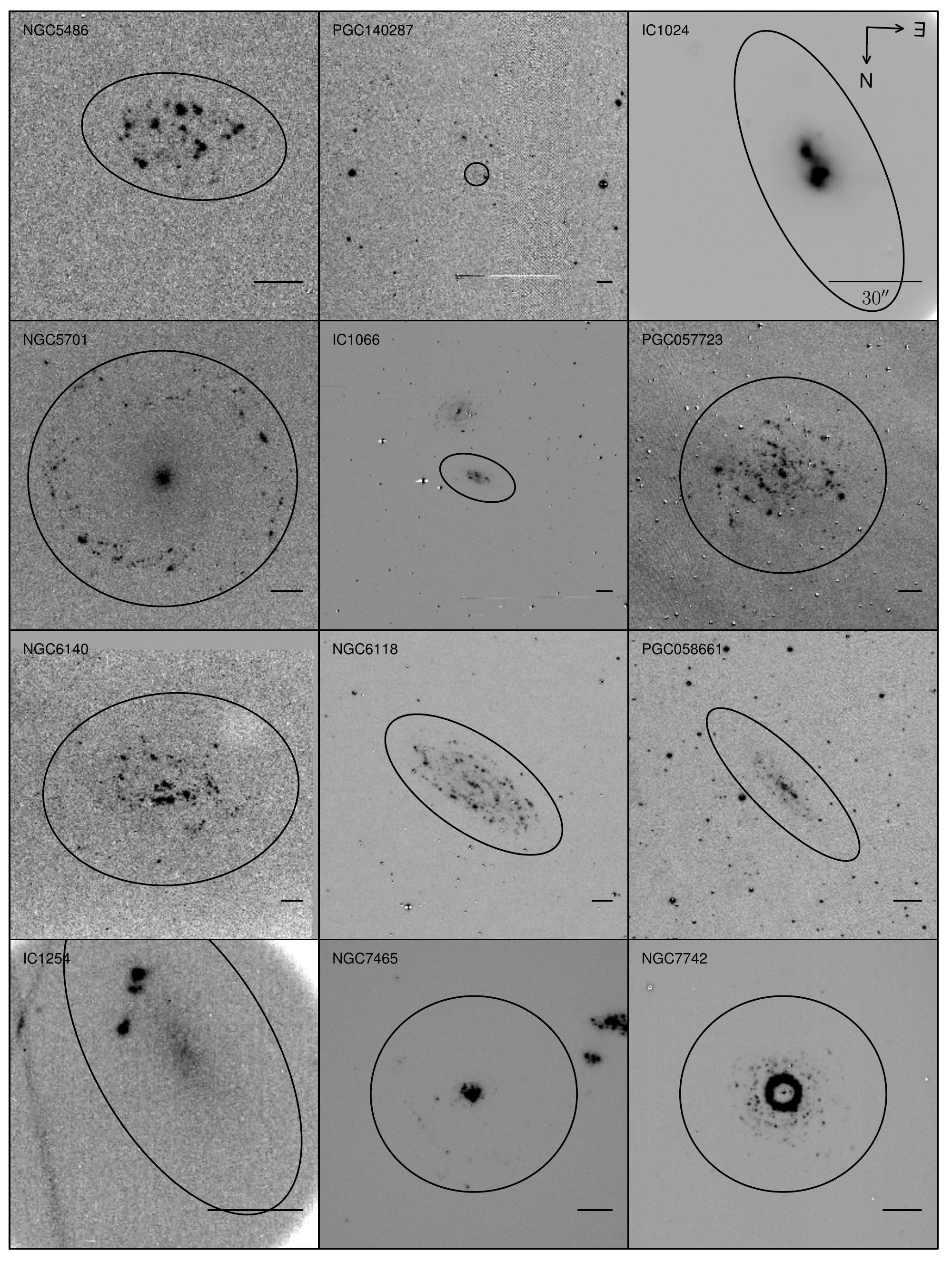}
  \caption{Continued}
\end{figure*}
% \ \newpage

\begin{figure*}
  \centering
  \includegraphics[width=17cm]{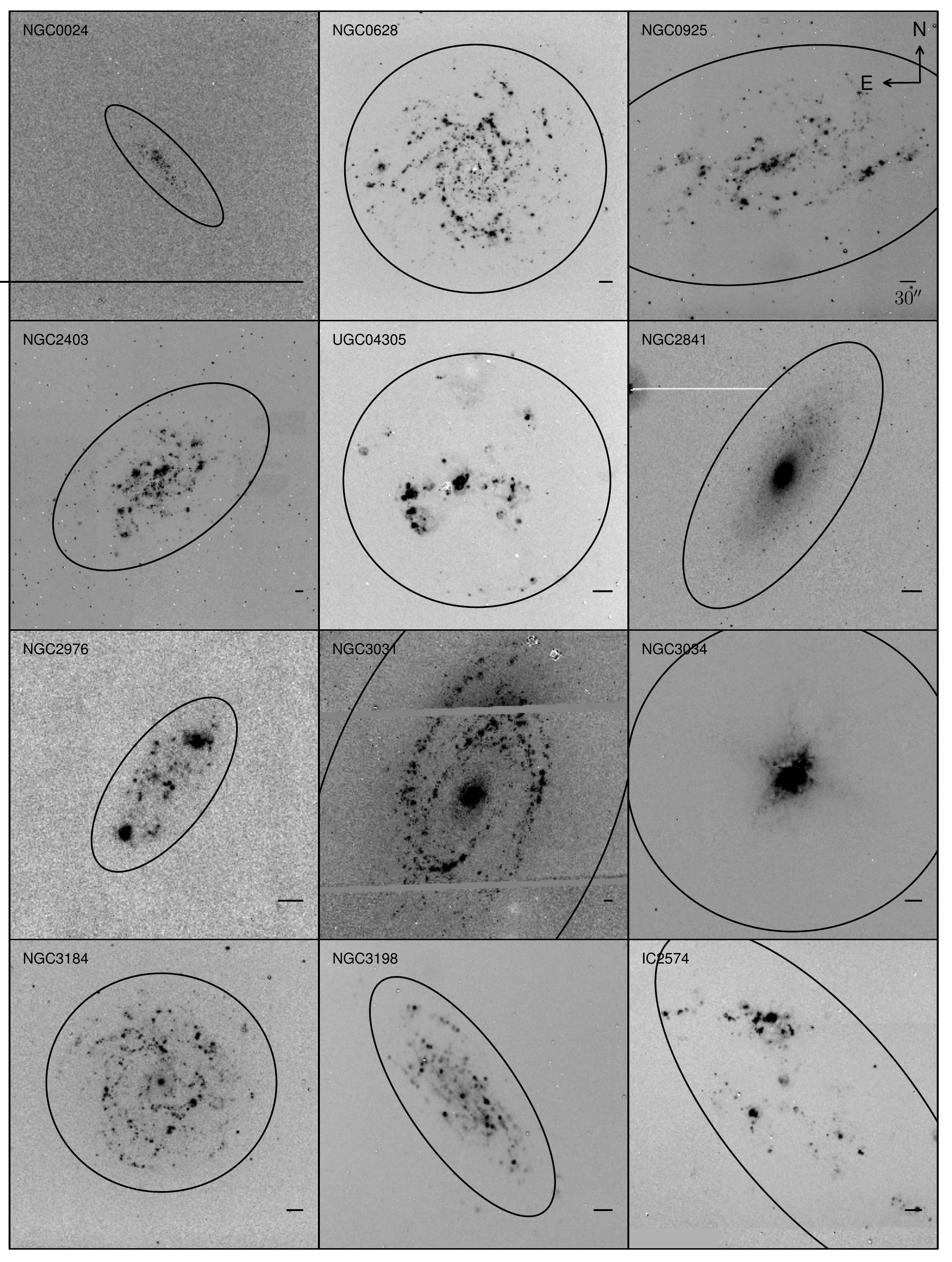}
  \caption{As Figure \ref{fig:ImagesField} but now for the SINGS Large galaxies in our sample.}
  \label{fig:ImagesSINGSLarge}
\end{figure*}
% \ \newpage

\begin{figure*}
\ContinuedFloat 
  \centering
  \includegraphics[width=17cm]{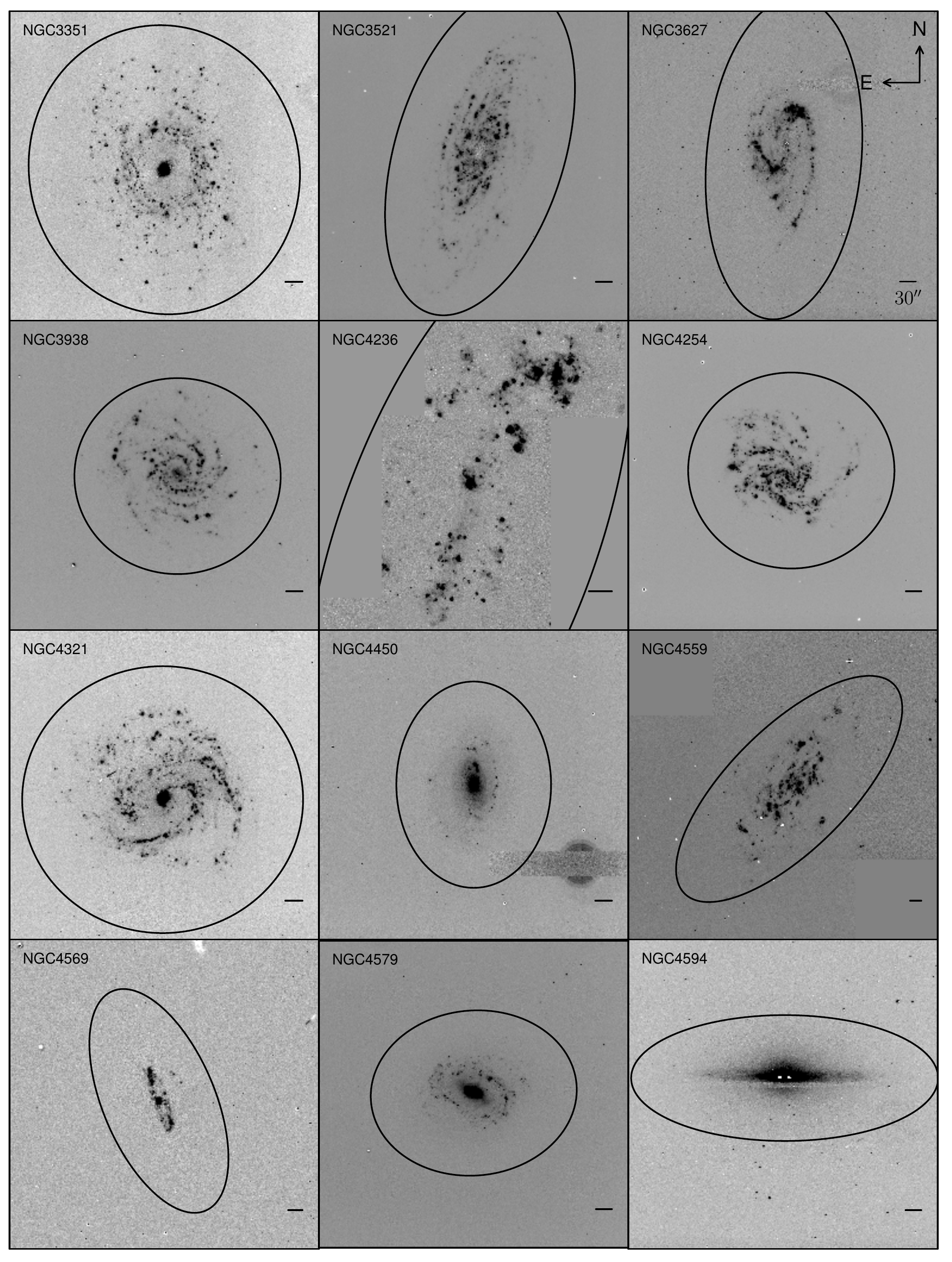}
  \caption{Continued}
\end{figure*}

\begin{figure*}
\ContinuedFloat 
  \centering
  \includegraphics[width=17cm]{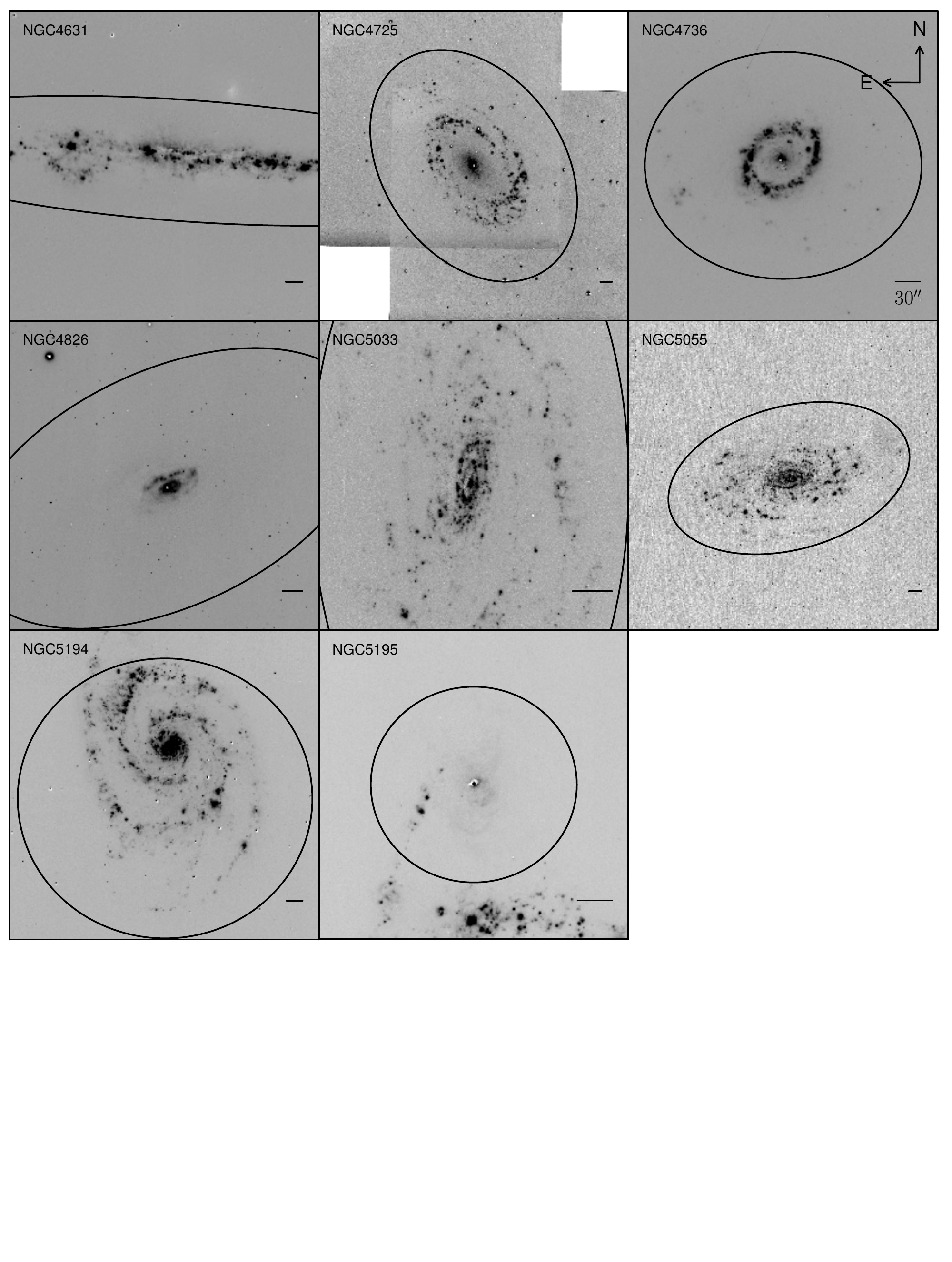}
  \caption{Continued}
\end{figure*}
% \ \newpage

\begin{figure*}
  \centering
  \includegraphics[width=17cm]{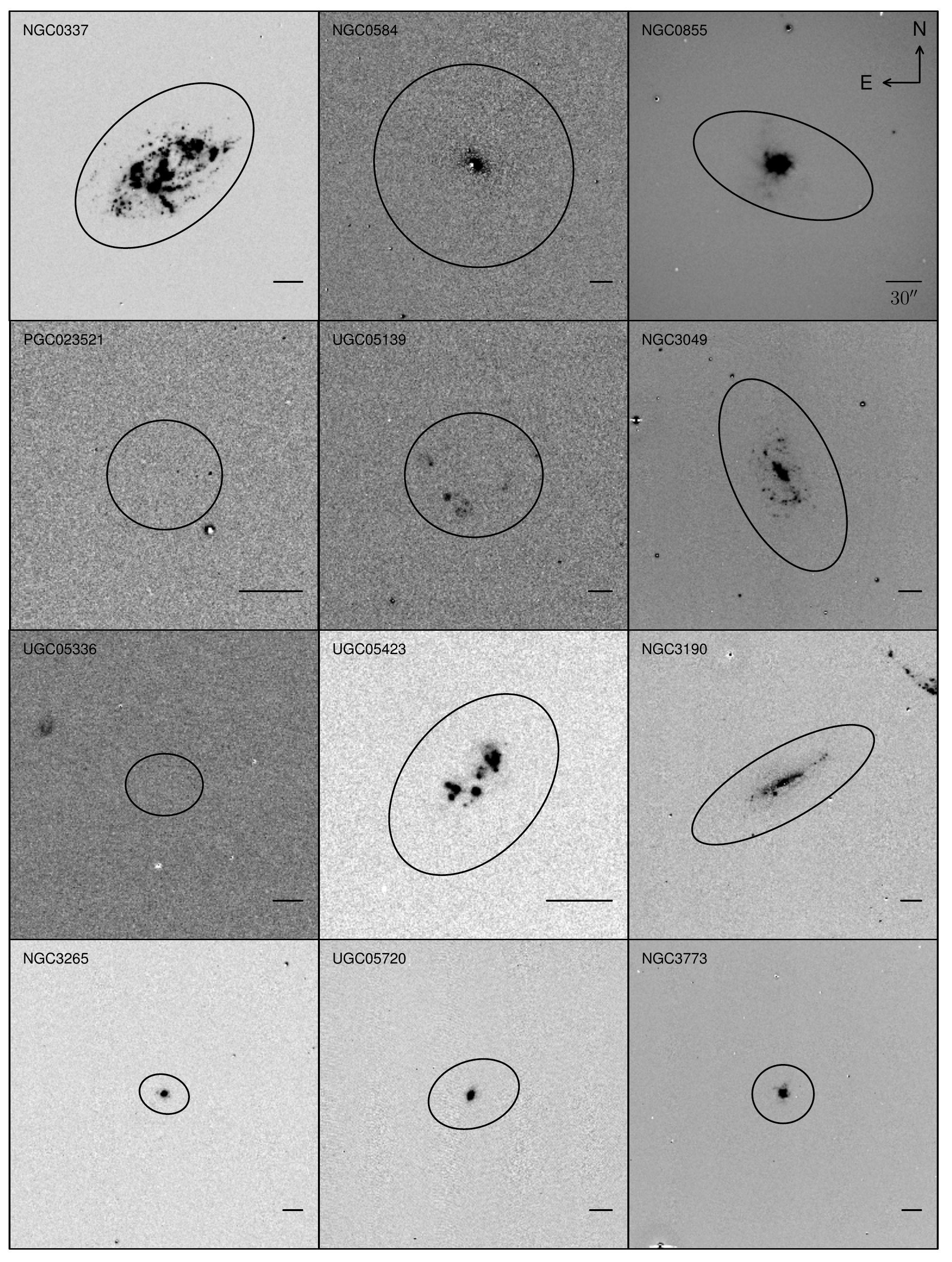}
  \caption{As Figure \ref{fig:ImagesField} but now for the SINGS Small galaxies in our sample.}
  \label{fig:ImagesSINGSSmall}
\end{figure*}
% \ \newpage

\begin{figure*}
\ContinuedFloat 
  \centering
  \includegraphics[width=17cm]{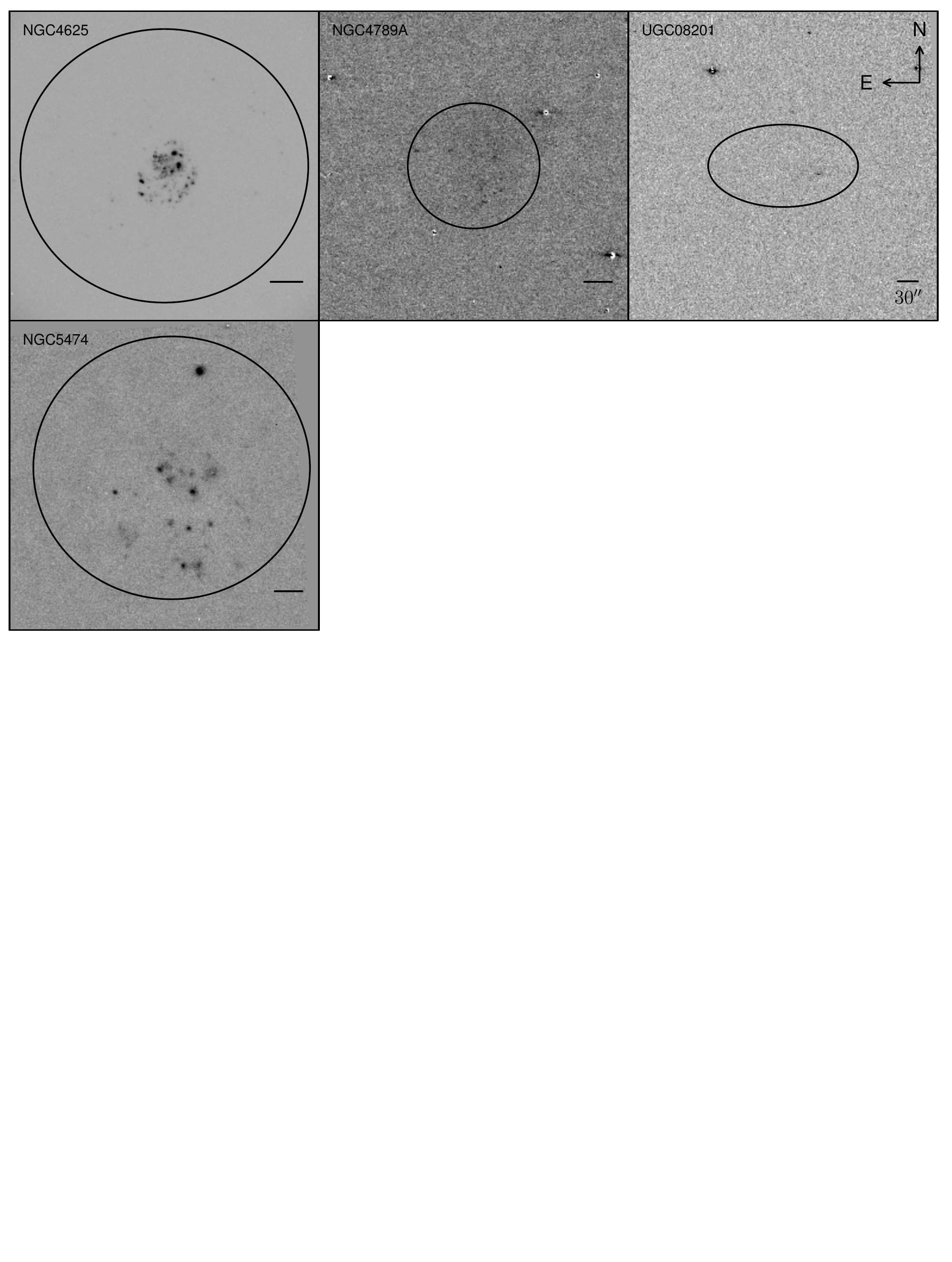}
  \caption{Continued}
\end{figure*}
% \ \newpage

\begin{figure*}
  \centering
  \includegraphics[width=17cm]{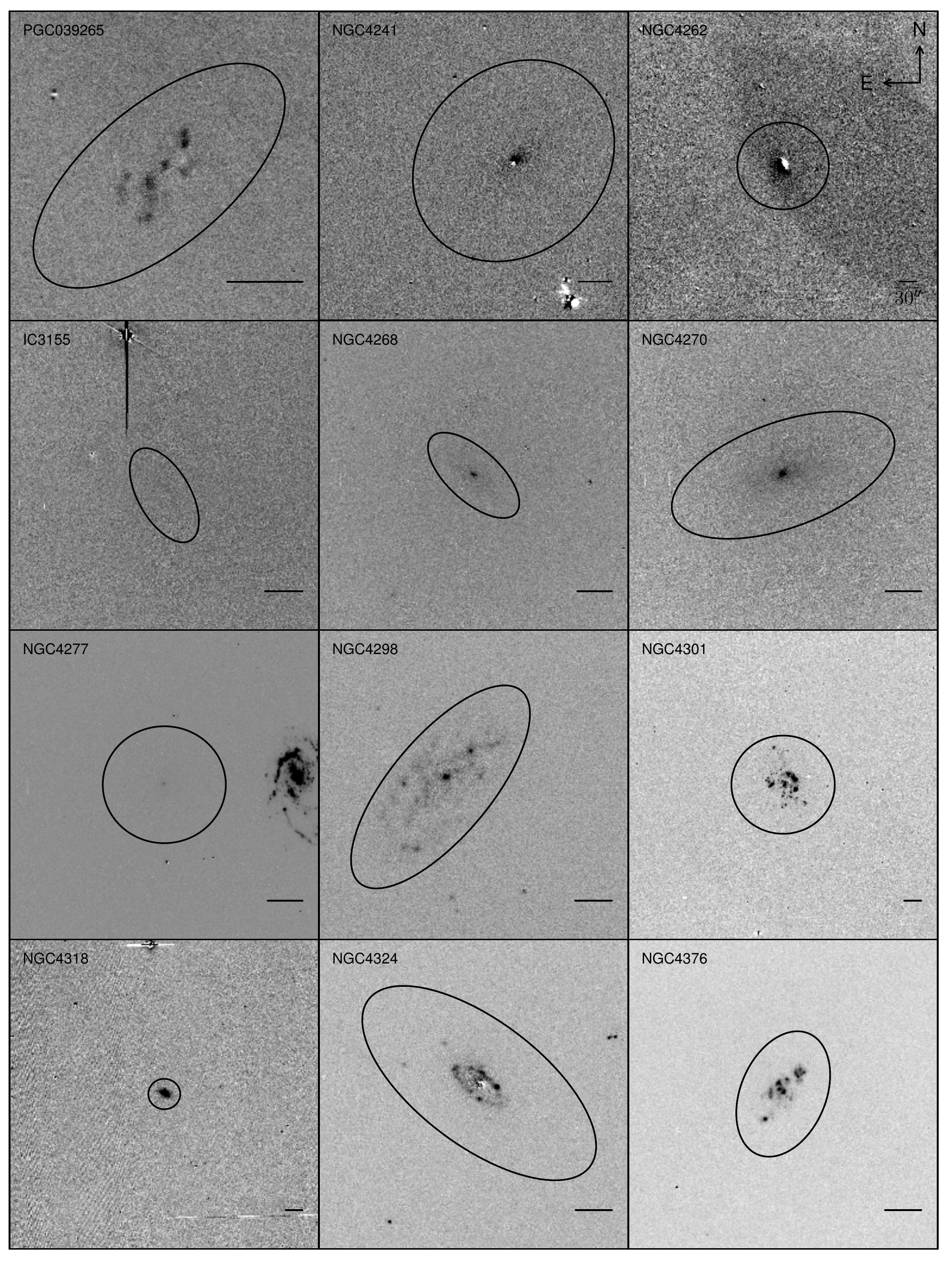}
  \caption{As Figure \ref{fig:ImagesField} but now for the Virgo Small galaxies in our sample.}
  \label{fig:ImagesVirgoSmall}
\end{figure*}

\begin{figure*}
\ContinuedFloat 
  \centering
  \includegraphics[width=17cm]{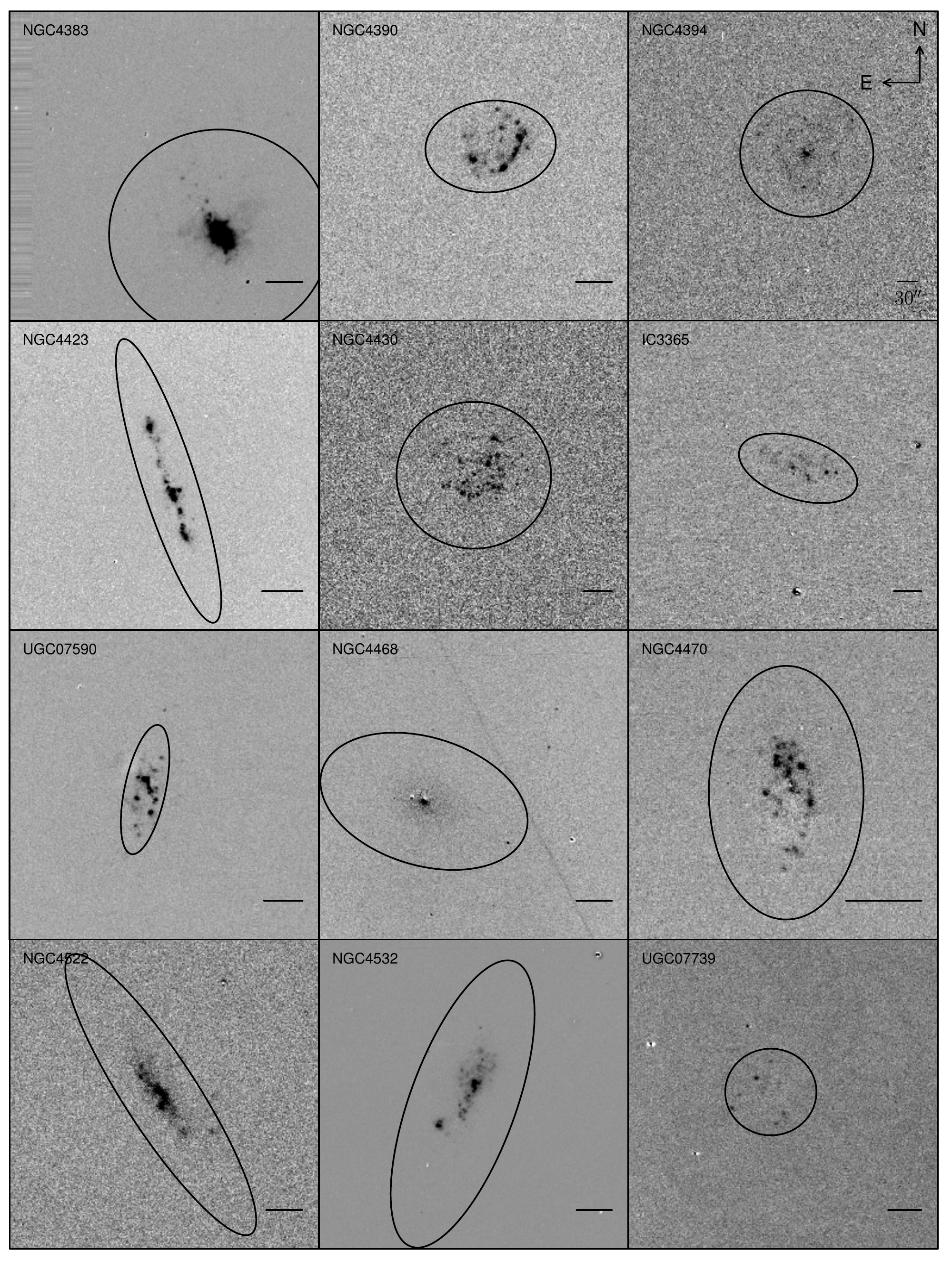}
  \caption{Continued}
\end{figure*}
% \ \newpage

\begin{figure*}
\ContinuedFloat 
  \centering
  \includegraphics[width=17cm]{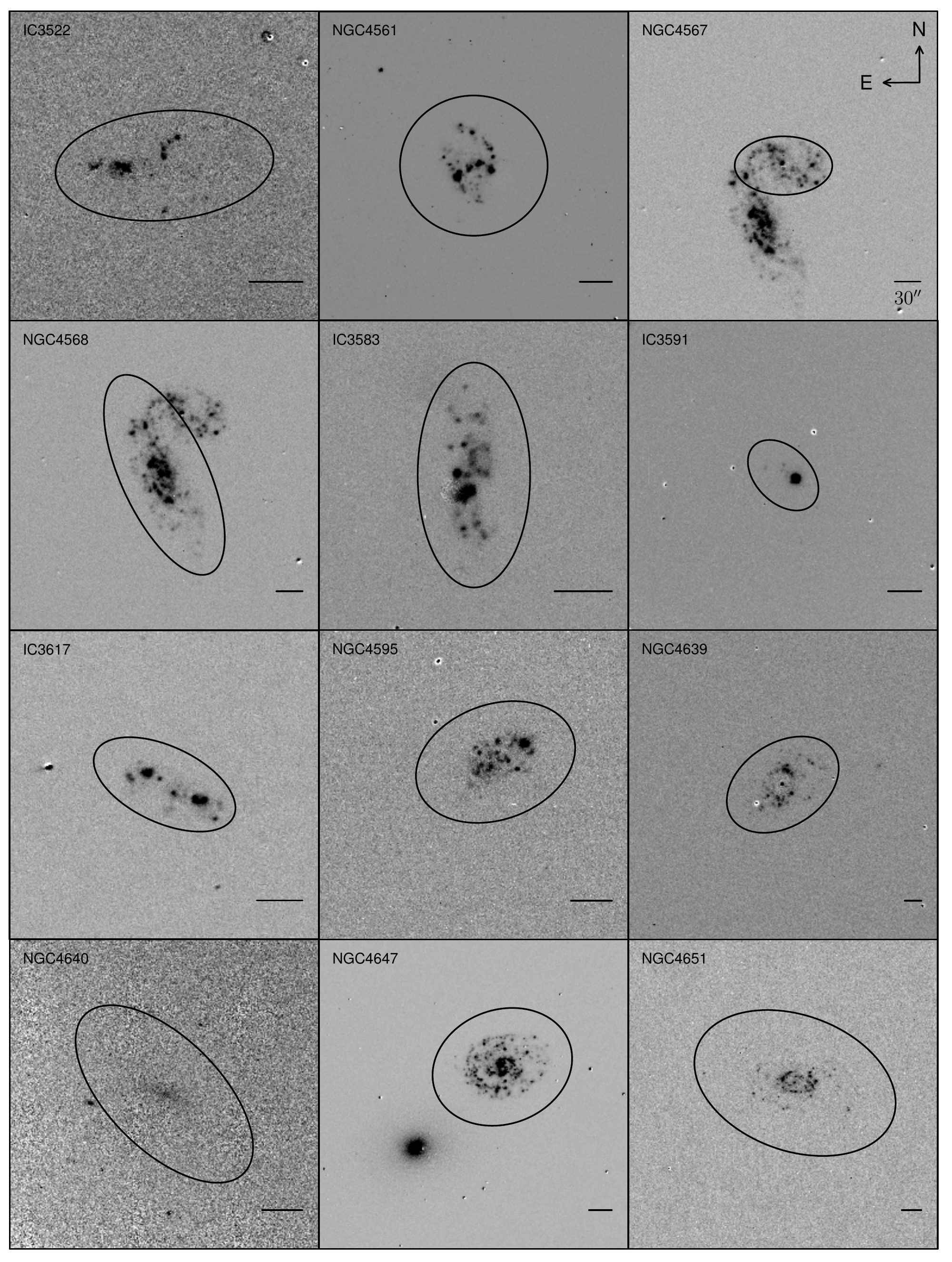}
  \caption{Continued}
\end{figure*}
\ \newpage

\begin{figure*}
  \centering
  \includegraphics[width=17cm]{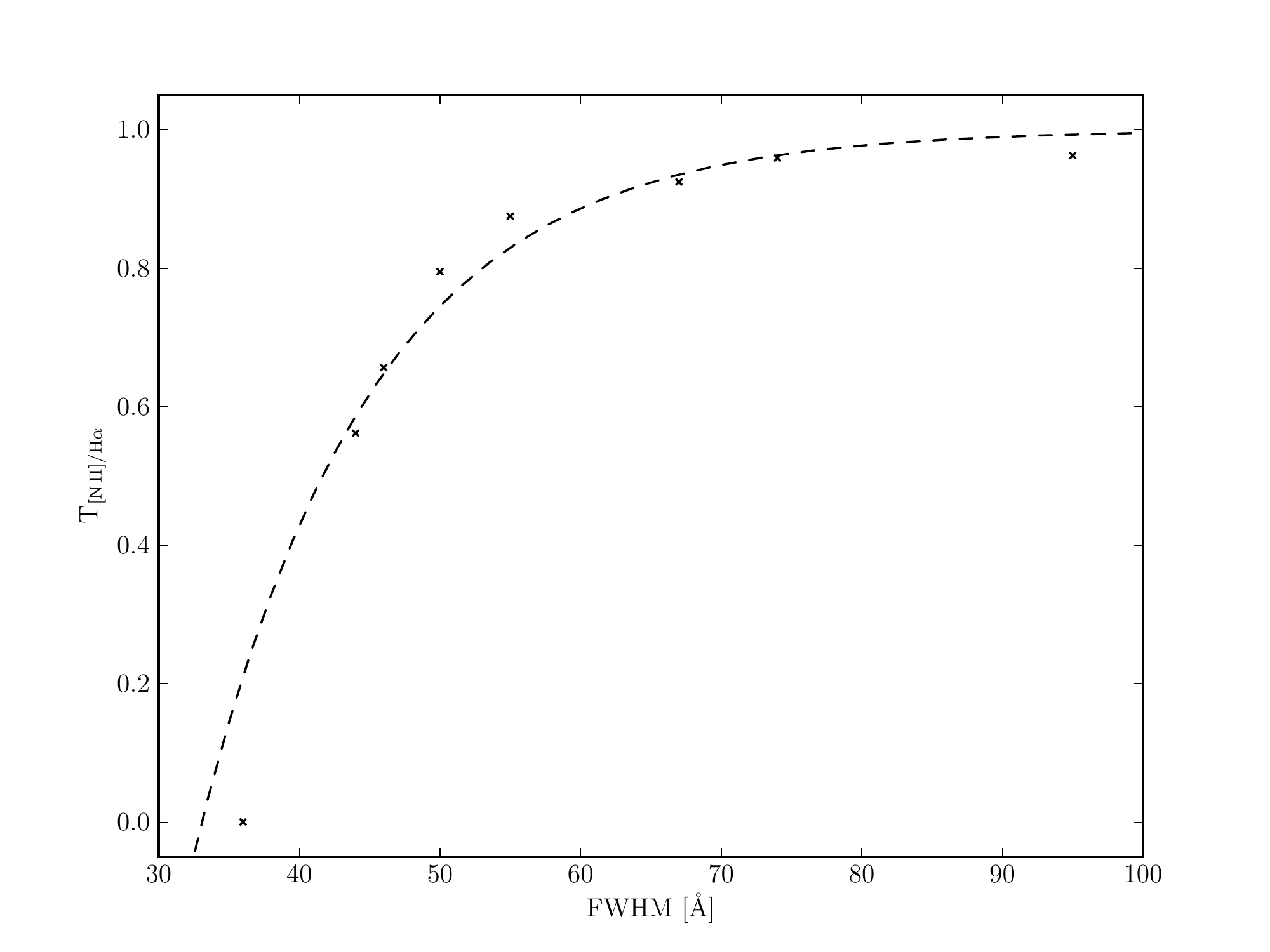}
  \caption{Relation between the transmission of the H$\alpha$ and \NII~lines for different narrow filters used in this work, as a function of the filter width. The dashed line indicates the exponential fit obtained (see Section \ref{sec:NII}).}
  \label{fig:NII}
\end{figure*}
% \ \newpage

\begin{figure*}
  \centering
  \includegraphics[width=17cm]{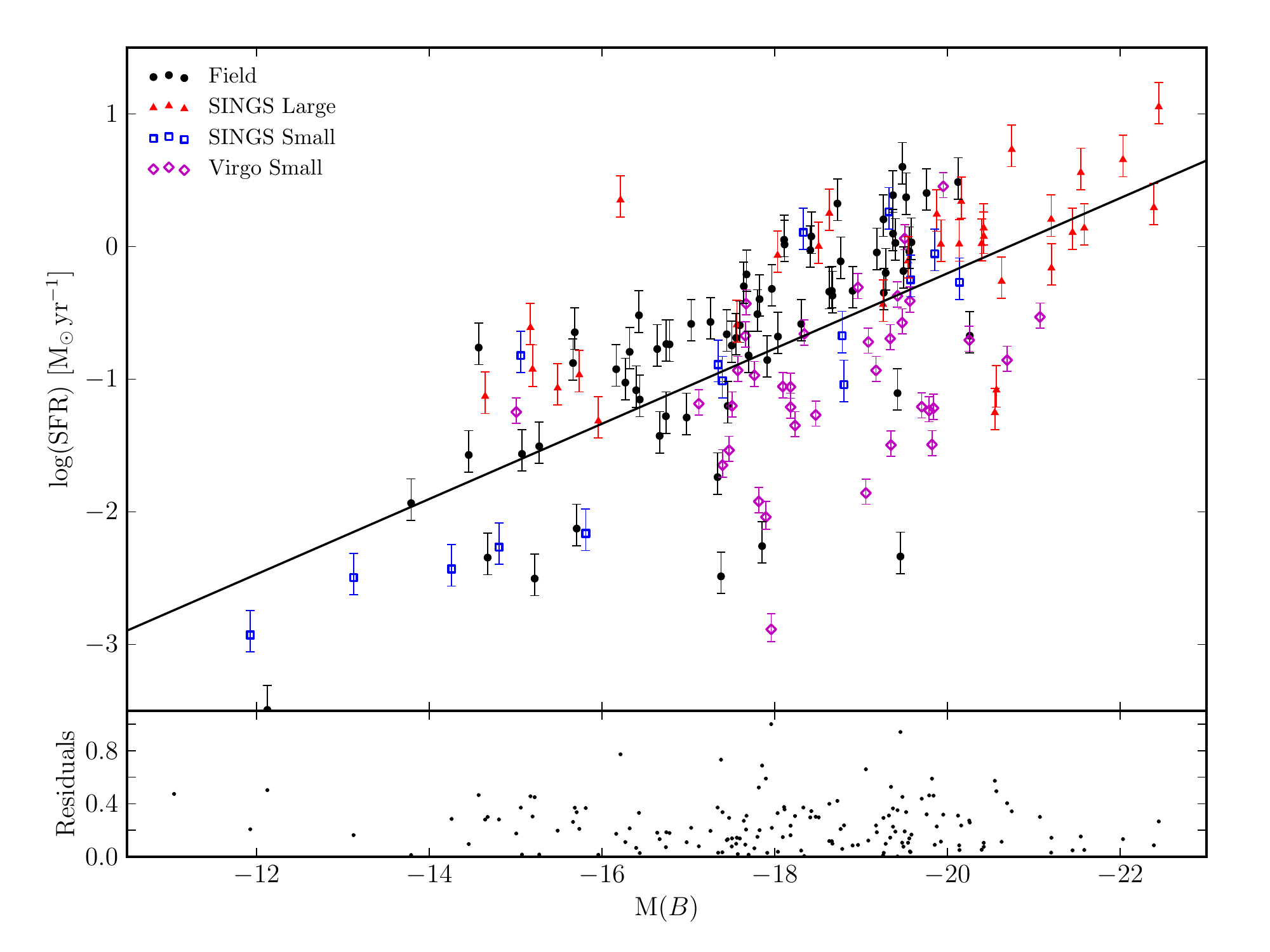}
  \caption{Logarithm of the SFR (in units of solar masses per year) versus the absolute \emph{B} magnitude of the galaxies in our sample. The four subsamples which constitute the complete list are marked with different colors and symbols. The black solid line indicates the least squares regression fit for the plotted data. The lower plot shows the normalized root mean square of each galaxy to its fitted value.}
  \label{fig:SFRvsMB}
\end{figure*}
% \ \newpage

\begin{figure*}
  \centering
  \includegraphics[width=17cm]{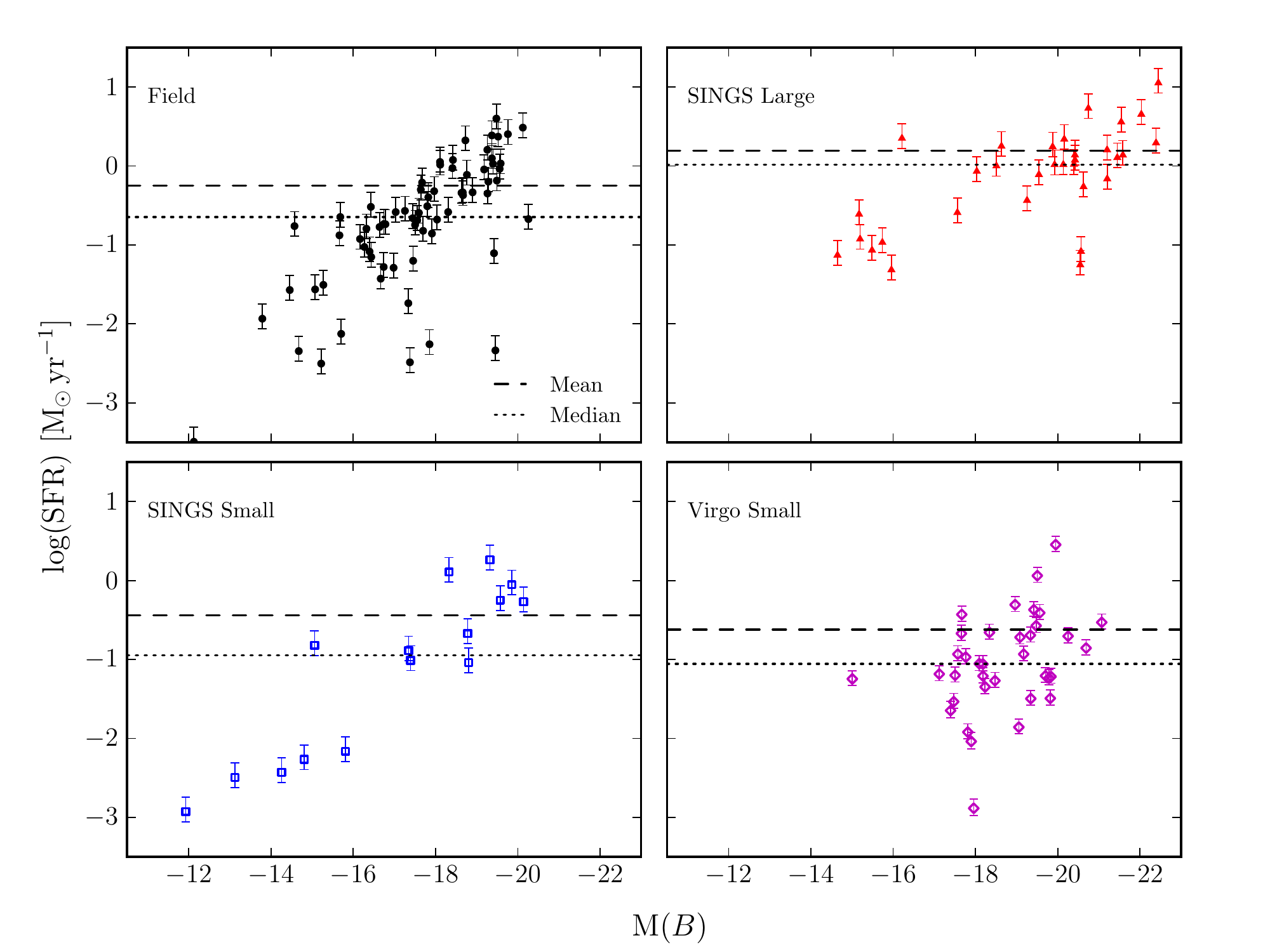}
  \caption{As Figure \ref{fig:SFRvsMB} but now with the subsamples divided into different panels. The dashed lines indicate the mean values of $\rm log(SFR)$ for each subsample while the dotted ones indicate the median values.}
  \label{fig:SFRvsMB_Panels}
\end{figure*}
% \ \newpage

\begin{figure*}
  \centering
  \includegraphics[width=17cm]{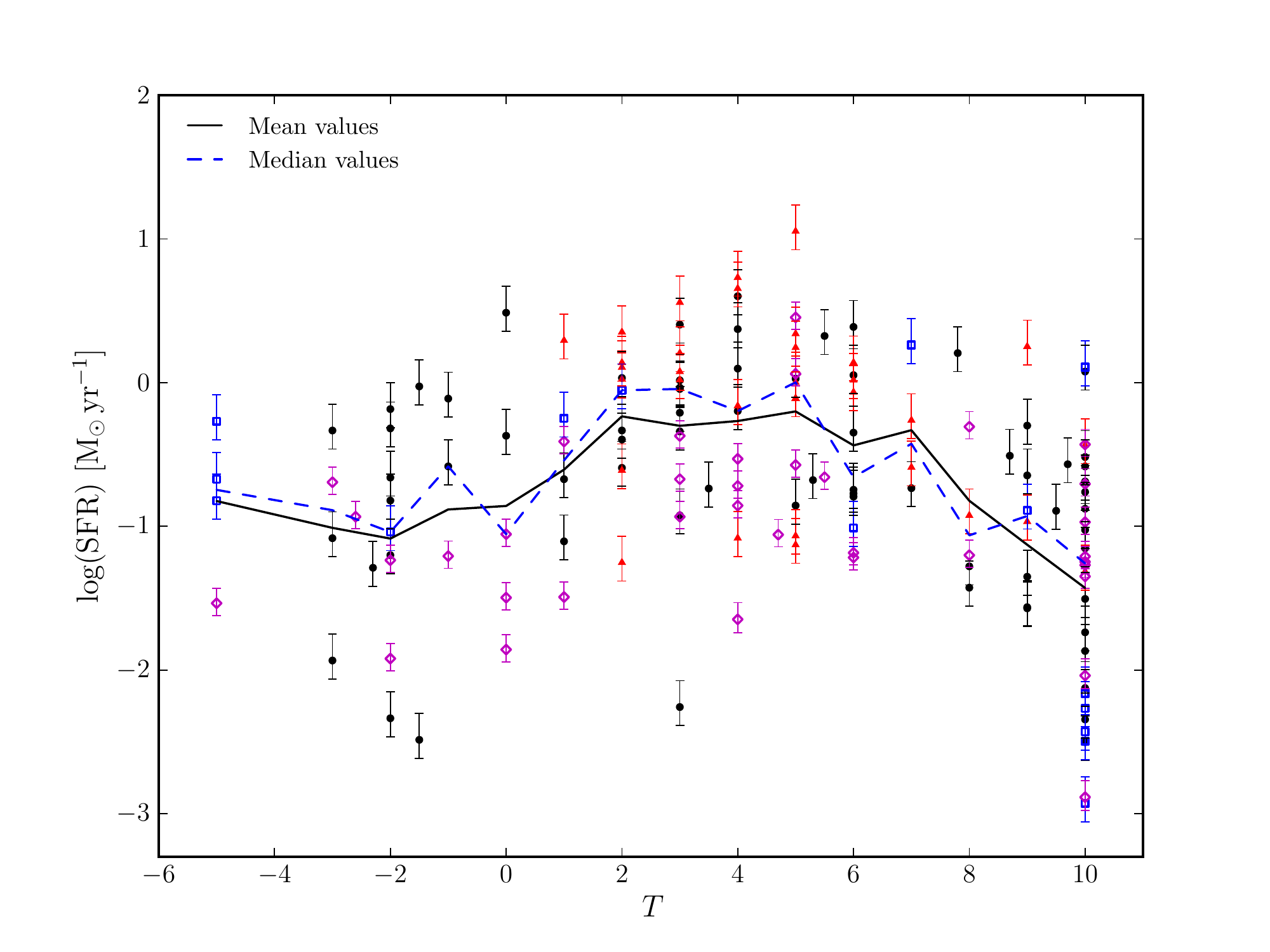}
  \caption{Logarithm of the SFR (in units of solar masses per year) versus the Hubble stage of the galaxies in the sample. The different subsamples in our complete list are plotted separately. The solid black line shows the mean values of the data binned to $\Delta \rm T=1$. The dashed blue line shows the median values calculated with the same procedure.}
  \label{fig:SFRvsT}
\end{figure*}
% \ \newpage

\begin{figure*}
  \centering
  \includegraphics[width=17cm]{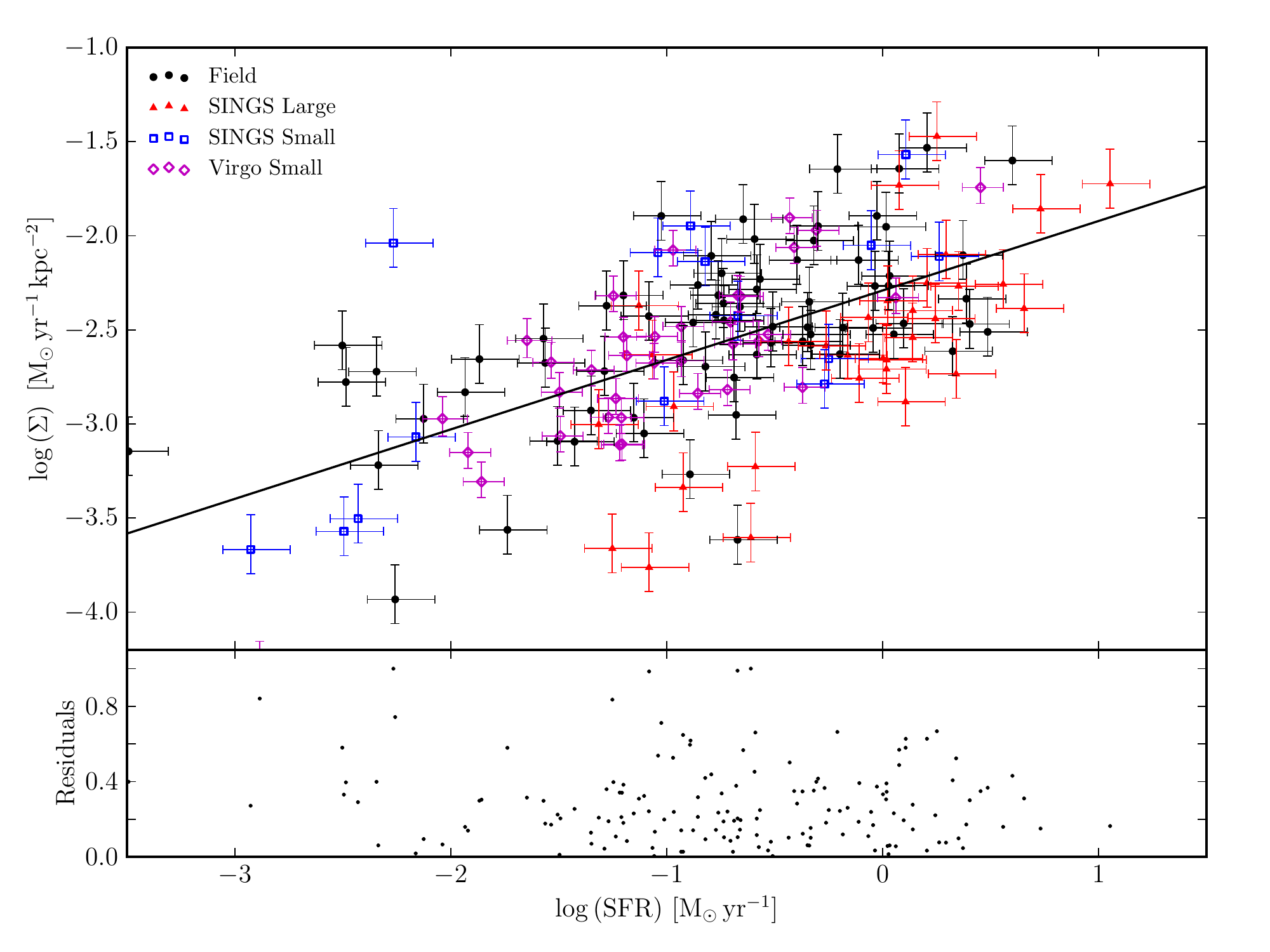}
  \caption{As Figure \ref{fig:SFRvsMB} but now plotting the logarithm of $\Sigma$ in units of solar masses per year and per square kiloparsec versus the SFR in units of solar masses per year.}
  \label{fig:SigmavsSFR}
\end{figure*}
% \ \newpage

\begin{figure*}
  \centering
  \includegraphics[width=17cm]{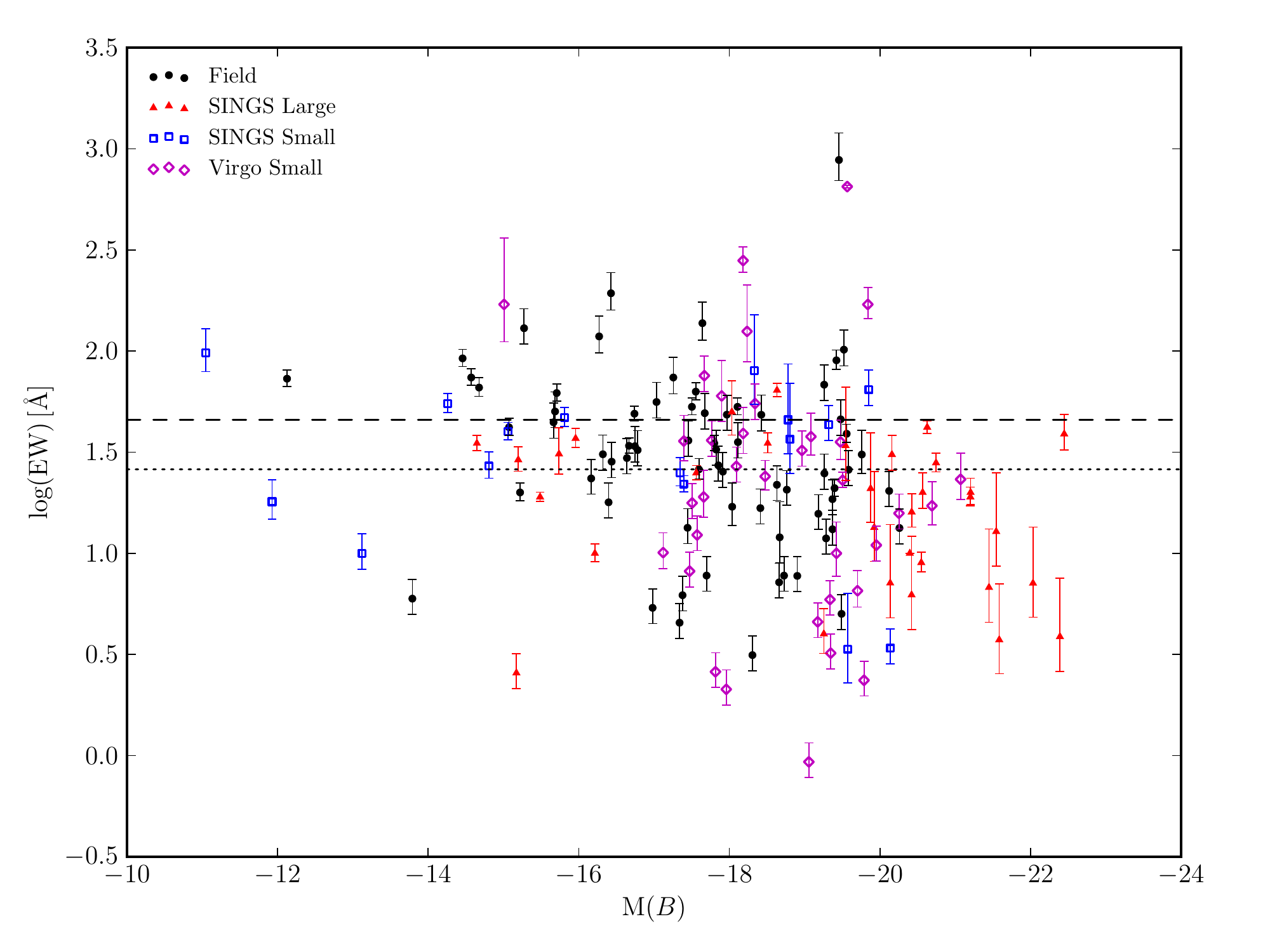}
  \caption{As Figure \ref{fig:SFRvsMB} but now with the logarithm of the equivalent width in \AA{}ngstr\"om. The dashed line indicates the mean value of $\rm log(EW)$ for the full sample while the dotted one indicates the median value.}
  \label{fig:EWvsMB}
\end{figure*}
% \ \newpage

\begin{figure*}
  \centering
  \includegraphics[width=17cm]{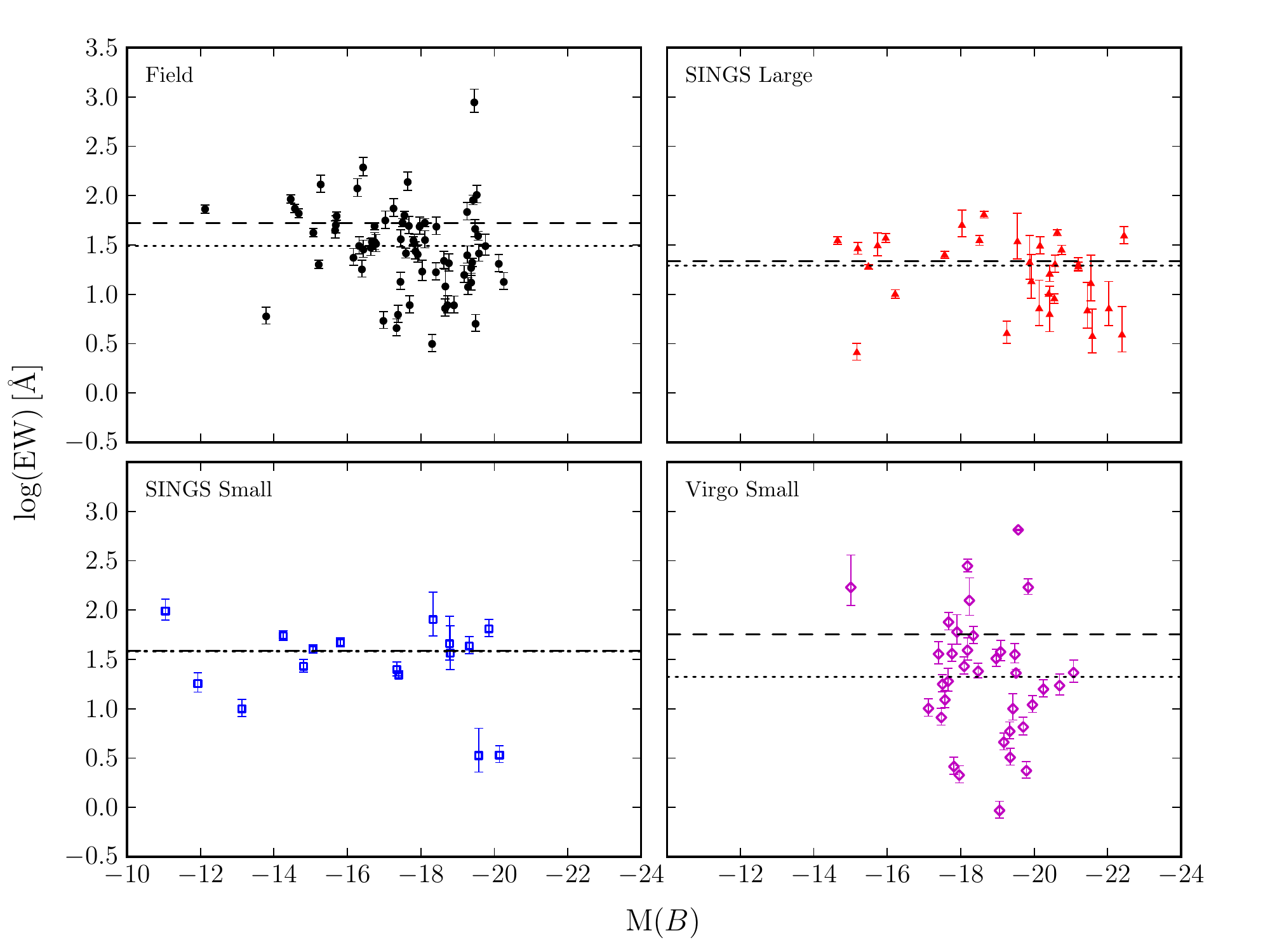}
  \caption{As Figure \ref{fig:EWvsMB} but now with the subsamples divided into different panels.}
  \label{fig:EWvsMB_Panels}
\end{figure*}
% \ \newpage

\begin{figure*}
  \centering
  \includegraphics[width=17cm]{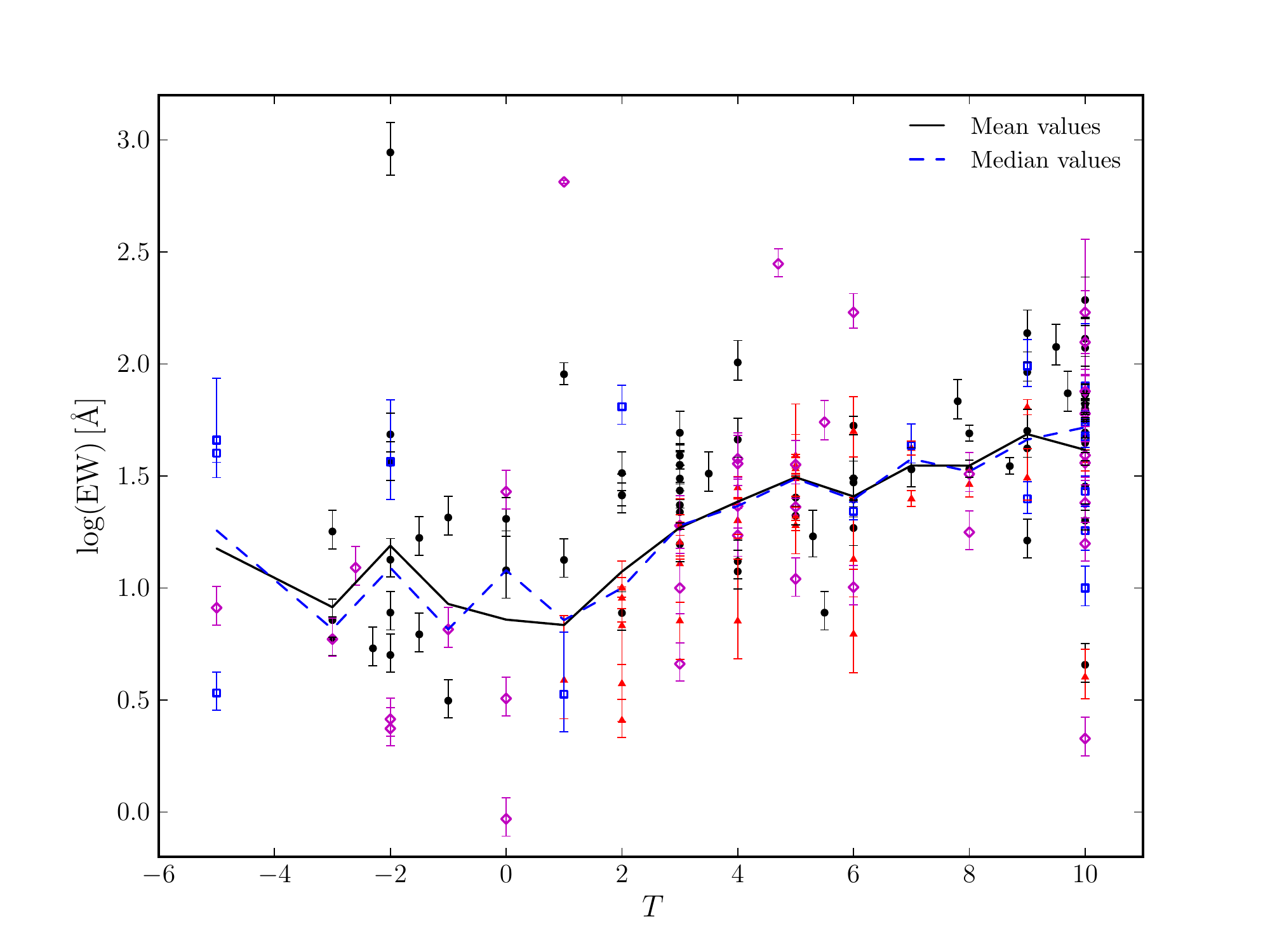}
  \caption{As Figure \ref{fig:SFRvsT} but now with the logarithm of the equivalent width in \AA{}ngstr\"om.}
  \label{fig:EWvsT}
\end{figure*}
% \ \newpage

\begin{figure*}
  \centering
  \vspace{-0.4cm}
  \includegraphics[width=17cm]{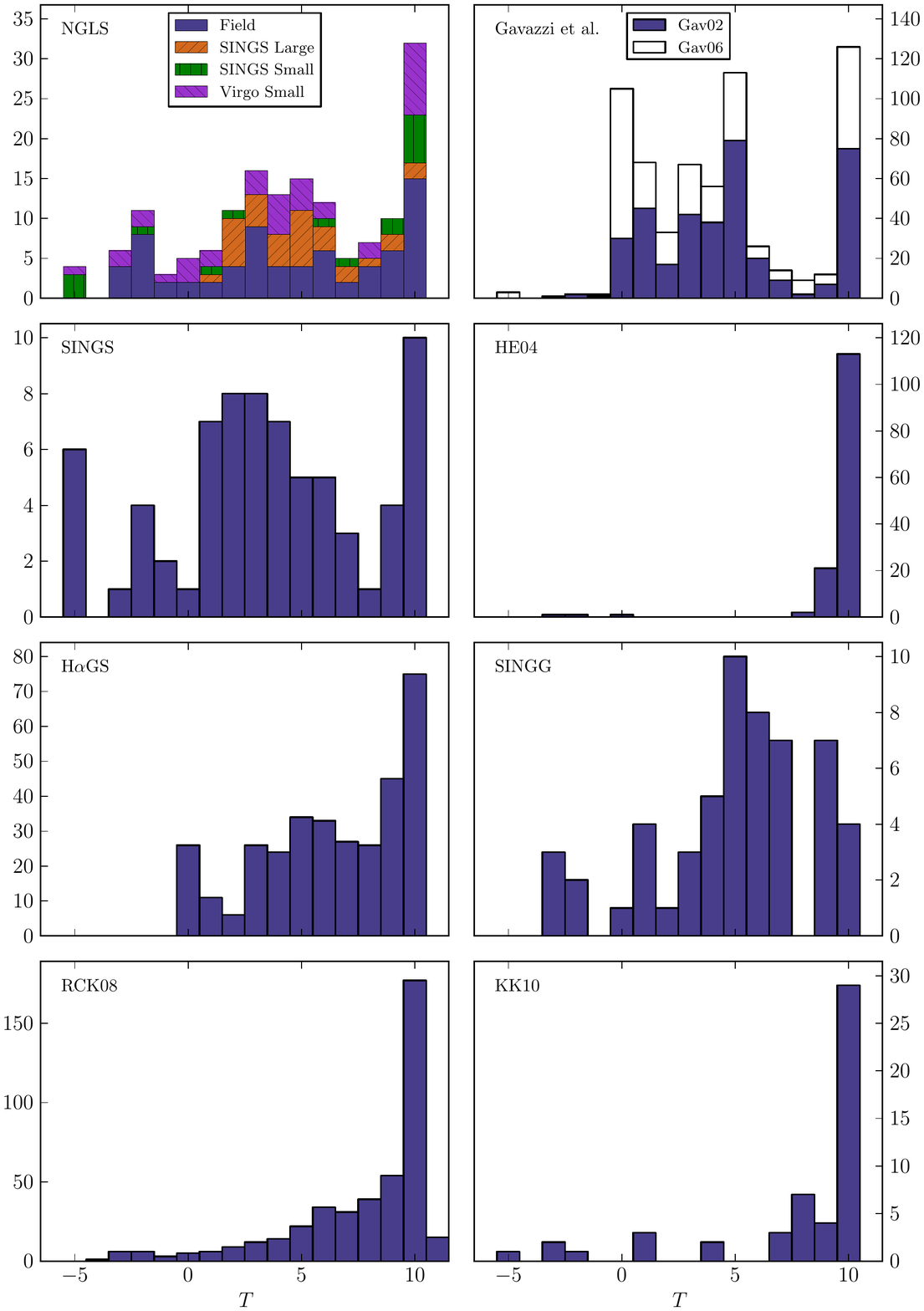}
  \vspace{-1.5cm}
  \caption{Histograms comparing the RC3 Hubble stage distribution, $T$, of our sample (top left panel) with several samples found in the literature: the \citet{Gavazzi2002,Gavazzi2006} survey of nearby galaxy clusters, the SINGS sample \citep{Kennicutt2003}, the sample of irregular and Sm galaxies observed by \citet{HunterElmegreen2004}, the H$\alpha$GS survey \citep{James2004}, the SINGG survey \citep{Meurer2006},  the volume-limited sample within $11\,\rm Mpc$ studied by \citet{Kennicutt2008a} and the 52 galaxies observed in H$\alpha$ by \citet{Karachentsev2010} as part of a larger Local Volume survey. Note that the last bin in each histogram includes the galaxies with values of $T$ larger than the upper limit in the plot. Further details are given in the text of Section \ref{sec:OtherStudies}.}
  \label{fig:HistT}
\end{figure*}
% \ \newpage

\begin{figure*}
  \centering
  \includegraphics[width=17cm]{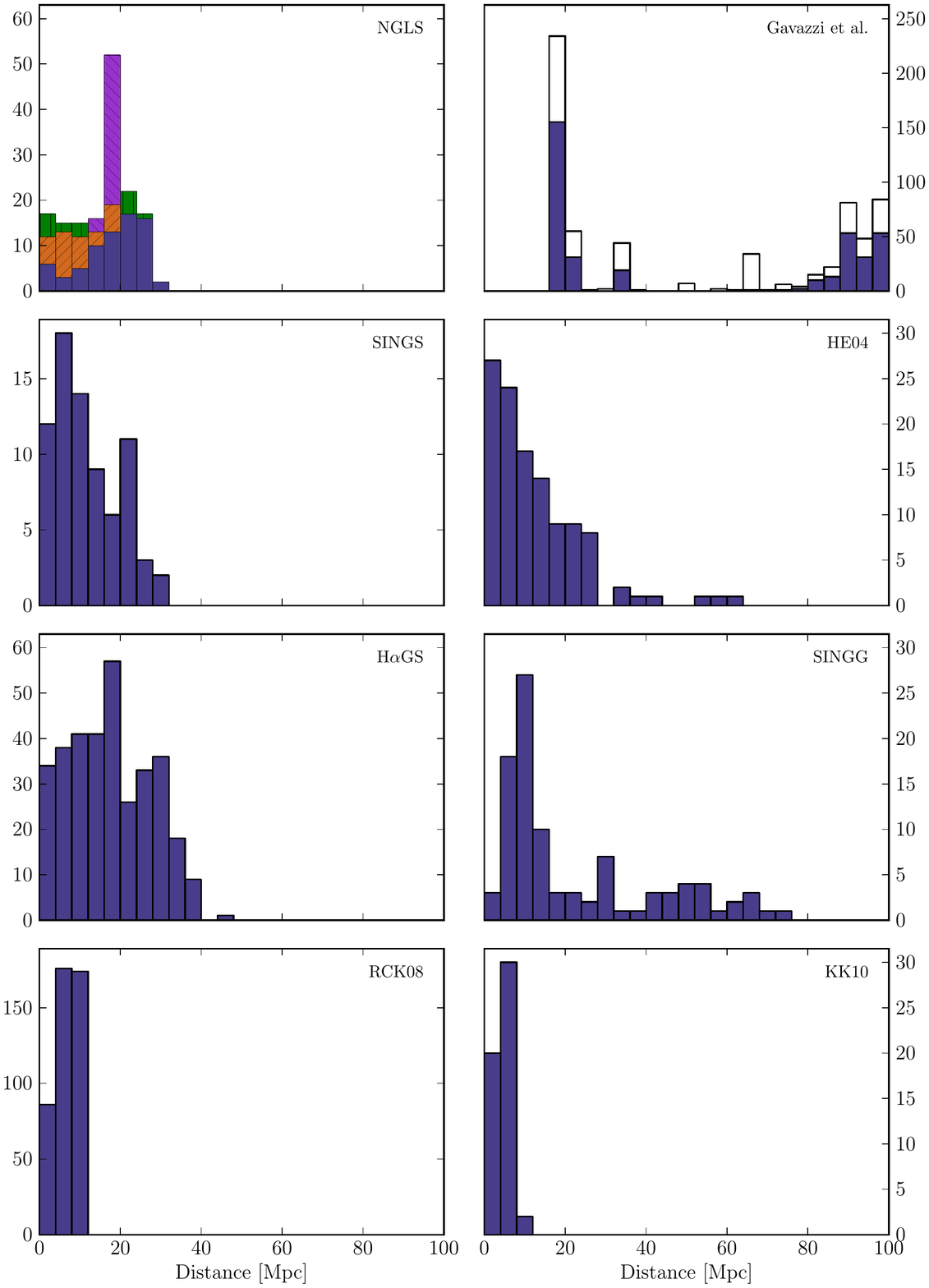}
  \vspace{-1.5cm}
  \caption{As Figure \ref{fig:HistT} but now showing the histograms of the distance in units of Mpc.}
  \label{fig:HistDist}
\end{figure*}
% \ \newpage

\begin{figure*}
  \centering
  \includegraphics[width=17cm]{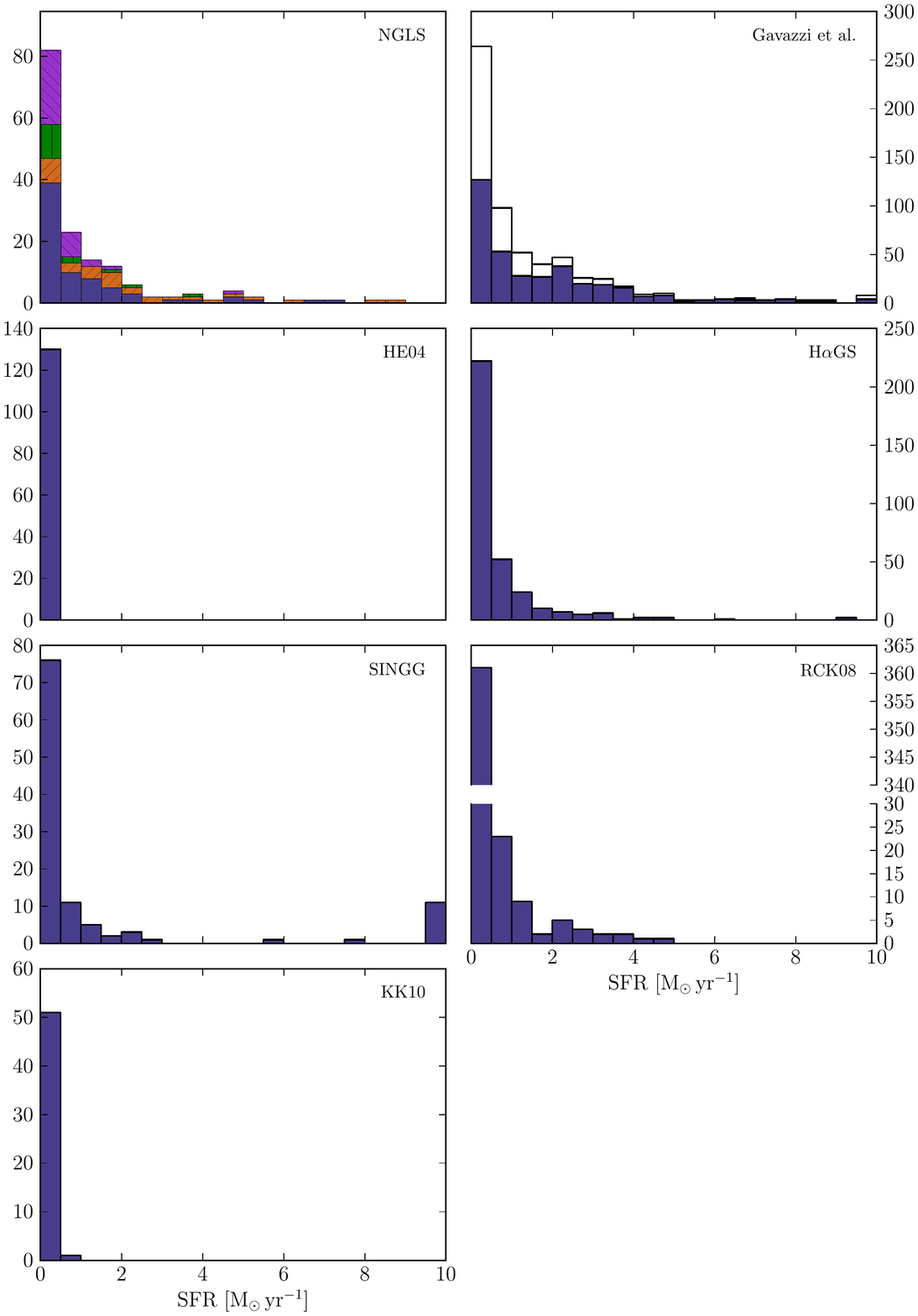}
  \vspace{-1.5cm}
  \caption{As Figure \ref{fig:HistT} but now showing the histograms of the SFR in units of $M_\odot\rm\,yr^{-1}$. The values plotted here have been calculated directly from the $F \rm (H\alpha)$ measurements taken from the catalogues using the same extinction and $F\rm (H\alpha)$-to-SFR values as in Sections \ref{sec:RedFlux} and \ref{sec:SFRandEW}.}
  \label{fig:HistSFR}
\end{figure*}
% \ \newpage

\begin{figure*}
  \centering
  \includegraphics[width=17cm]{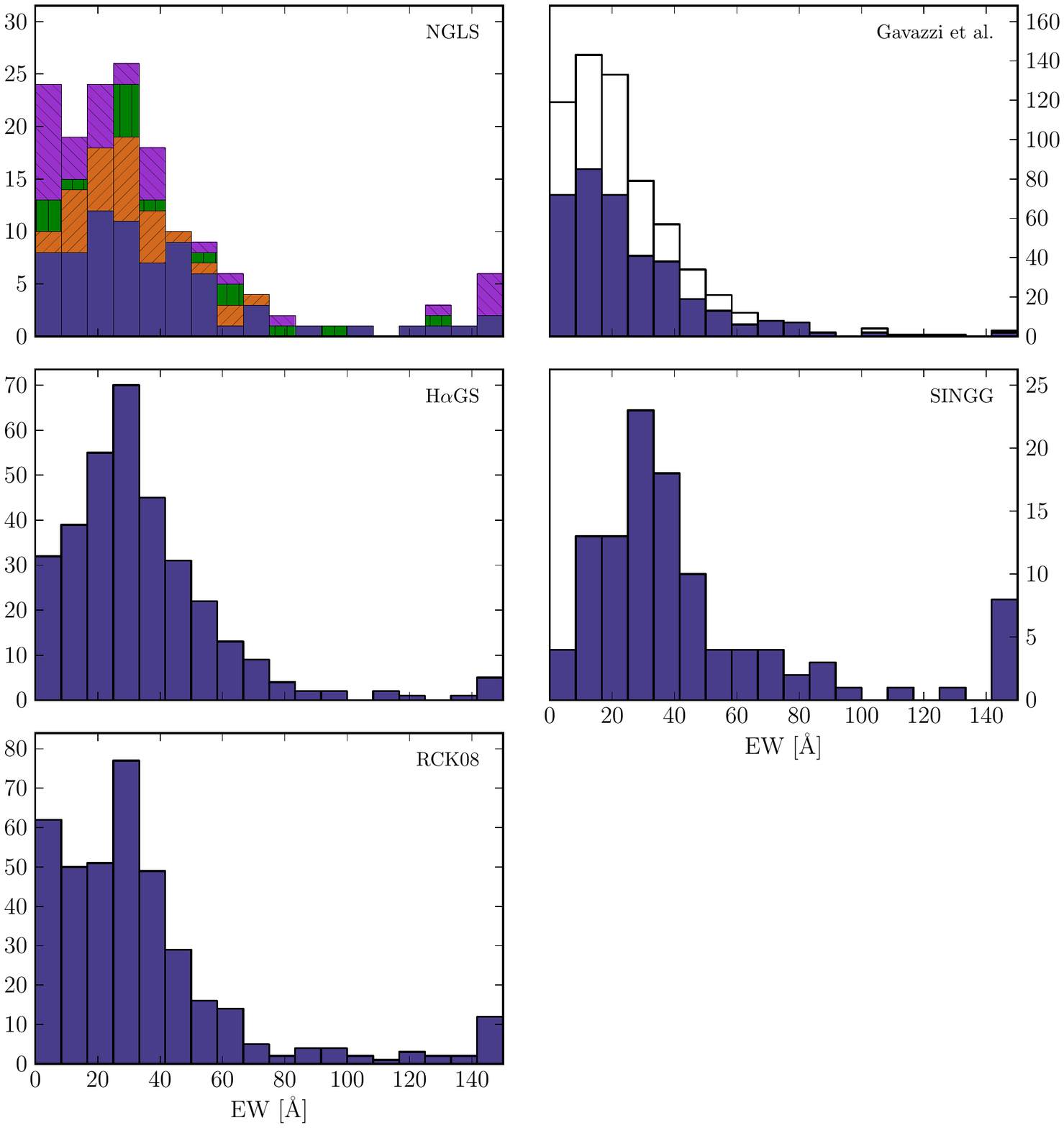}
  \vspace{-6cm}
  \caption{As Figure \ref{fig:HistT} but now showing the histograms of the EW in \r{A}ngstr\"{o}ms.}
  \label{fig:HistEW}
\end{figure*}
% \ \newpage

\begin{figure*}
  \centering
  \includegraphics[width=17cm]{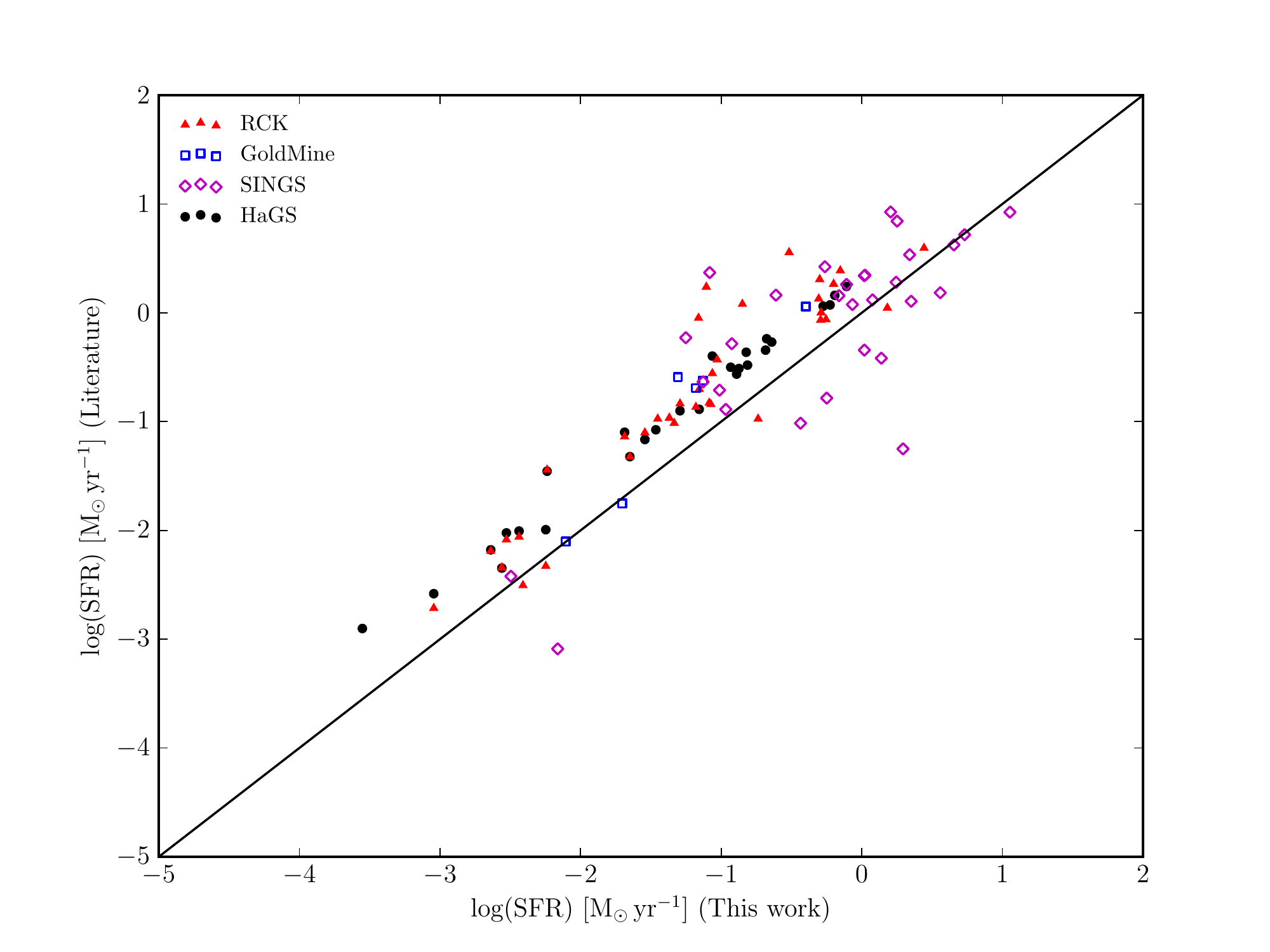}
  \caption{Comparison of SFR values calculated in this work with those obtained from the literature. Except in the case of the SINGS values, all the SFRs have been derived from the same images used in the literature. The solid line indicates $y=x$. Further details can be found in the text.}
  \label{fig:LitSFR}
\end{figure*}
\ %\newpage

\bsp

\label{lastpage}

\end{document}

%% file: observationsTable.txt
NGC0024 & SINGS Large & CTIO 1.5m & CFCCD & 01-10-19 & 6568/20 & 6425/1500 & 600 & 180 & 0.43 & 1.3 & (4)\\
ESO538-024 & Field & WHT & ACAM & 09-07-31 & 6589/15 & 6228/1322 & 720 & 360 & 0.25 & 2.2 & (1)\\
NGC0210 & Field & JKT & SITe2 CCD & 00-01-01 & 6594/44 & 6470/115 & 2400 & 300 & 0.24 & 1.3 & (6)\\
NGC0216 & Field & WHT & ACAM & 09-10-08 & 6589/15 & 6228/1322 & 720 & 360 & 0.25 & 0.9 & (1)\\
IC0051 & Field & WHT & ACAM & 09-10-07 & 6589/15 & 6228/1322 & 720 & 360 & 0.26 & 0.9 & (1)\\
NGC0274 & Field & WHT & ACAM & 09-10-07 & 6589/15 & 6228/1322 & 720 & 360 & 0.25 & 0.9 & (1)\\
NGC0337 & SINGS Small & CTIO 1.5m & CFCCD & 01-10-19 & 6602/20 & 6425/1500 & 600 & 180 & 0.43 & 1.3 & (4)\\
NGC0404 & Field & NOT & ALFOSC & 09-01-13 & 6562/46 & 6500/1300 & 1800 & 300 & 0.19 & 0.9 & (1)\\
NGC0450 & Field & JKT & JAG-CCD & 01-10-19 & 6594/44 & 6373/1491 & 1200 & 300 & 0.33 & 1.0 & (5)\\
NGC0473 & Field & JKT & JAG-CCD & 01-10-23 & 6626/44 & 6373/1491 & 1200 & 300 & 0.33 & 1.2 & (5)\\
NGC0584 & SINGS Small & CTIO 1.5m & CFCCD & 01-10-18 & 6602/18 & 6425/1500 & 601 & 121 & 0.43 & 1.1 & (4)\\
NGC0615 & Field & NOT & ALFOSC & 09-01-13 & 6611/18 & 6500/1300 & 1800 & 300 & 0.19 & 1.1 & (1)\\
NGC0628 & SINGS Large & CTIO 1.5m & CFCCD & 01-10-21 & 6586/20 & 6425/1500 & 600 & 180 & 0.43 & 2.3 & (4)\\
ESO477-016 & Field & NOT & ALFOSC & 09-01-13 & 6598/18 & 6500/1300 & 1800 & 300 & 0.19 & 1.6 & (1)\\
NGC0855 & SINGS Small & WHT & ACAM & 09-07-30 & 6577/15 & 6228/1322 & 720 & 360 & 0.25 & 1.8 & (1)\\
NGC0925 & SINGS Large & Kitt Peak 2.1m & T2KA & 01-11-09 & 6573/67 & 6513/1510 & 600 & 180 & 0.30 & 1.6 & (4)\\
NGC1036 & Field & NOT & ALFOSC & 09-01-13 & 6577/18 & 6500/1300 & 1800 & 300 & 0.19 & 0.9 & (1)\\
NGC1140 & Field & WHT & ACAM & 10-09-03 & 6589/15 & 6228/1322 & 540 & 60 & 0.25 & 1.6 & (1)\\
NGC1156 & Field & JKT & JAG-CCD & 00-11-17 & 6570/55 & 6373/1491 & 1200 & 300 & 0.33 & 1.2 & (5)\\
ESO481-019 & Field & NOT & ALFOSC & 09-01-13 & 6598/18 & 6500/1300 & 1800 & 300 & 0.19 & 2.3 & (1)\\
NGC1325 & Field & WHT & ACAM & 10-09-03 & 6589/15 & 6228/1322 & 540 & 60 & 0.42 & 2.7 & (1)\\
NGC2146A & Field & JKT & JAG-CCD & 01-01-20 & 6594/44 & 6373/1491 & 1200 & 300 & 0.33 & 2.4 & (5)\\
NGC2403 & SINGS Large & Kitt Peak 2.1m & T2KA & 01-11-16 & 6573/67 & 6513/1511 & 600 & 180 & 0.30 & 2.2 & (4)\\
UGC04305 & SINGS Large & Kitt Peak 2.1m & CFIM & 01-03-29 & 6573/67 & 6513/1511 & 900 & 210 & 0.30 & 1.7 & (4)\\
PGC023521 & SINGS Small & Bok 2.3m & CCD21 & 01-05-01 & 6580/69 & 6580/1539 & 1001 & 201 & 0.43 & 2.0 & (8)\\
NGC2742 & Field & JKT & JAG-CCD & 01-02-11 & 6594/44 & 6471/115 & 1200 & 600 & 0.33 & 2.2 & (5)\\
NGC2787 & Field & NOT & ALFOSC & 09-01-13 & 6611/18 & 6500/1300 & 1800 & 600 & 0.19 & 4.1 & (1)\\
NGC2841 & SINGS Large & Kitt Peak 2.1m & T2KA & 02-03-06 & 6573/67 & 6513/1511 & 200 & 420 & 0.30 & 1.4 & (4)\\
UGC05139 & SINGS Small & JKT & JAG-CCD & 01-03-28 & 6570/55 & 6373/1491 & 1200 & 300 & 0.33 & 1.6 & (5)\\
NGC2976 & SINGS Large & JKT & JAG-CCD & 00-02-23 & 6570/55 & 6390/1540 & 900 & 300 & 0.33 & 2.1 & (5)\\
UGC05272 & Field & Bok 2.3m & CCD21 & 01-04-01 & 6580/69 & 6580/1380 & 1000 & 200 & 0.43 & 1.3 & (8)\\
NGC3049 & SINGS Small & Kitt Peak 2.1m & T2KA & 02-04-11 & 6618/74 & 6513/1511 & 900 & 420 & 0.30 & 1.6 & (4)\\
NGC3031 & SINGS Large & INT & WFC & 08-03-15 & 6568/95 & 6380/1520 & 480 & 120 & 0.33 & 6.1 & (1)\\
NGC3034 & SINGS Large & Kitt Peak 2.1m & T2KA & 02-03-06 & 6573/67 & 6513/1511 & 200 & 200 & 0.30 & 1.5 & (4)\\
UGC05336 & SINGS Small & JKT & JAG-CCD & 01-03-29 & 6570/55 & 6373/1491 & 1200 & 300 & 0.33 & 1.7 & (5)\\
NGC3077 & Field & Bok 2.3m & CCD21 & 01-04-01 & 6580/69 & 6580/1539 & 1000 & 200 & 0.43 & 1.6 & (8)\\
UGC05423 & SINGS Small & NOT & ALFOSC & 10-03-25 & 6564/33 & 6500/1300 & 600 & 180 & 0.19 & 1.3 & (1)\\
NGC3162 & Field & NOT & ALFOSC & 08-01-03 & 6583/36 & 6500/1300 & 2700 & 540 & 0.19 & 1.2 & (1)\\
NGC3190 & SINGS Small & Kitt Peak 2.1m & T2KA & 02-04-09 & 6618/74 & 6513/1511 & 900 & 420 & 0.30 & 1.6 & (4)\\
NGC3184 & SINGS Large & Kitt Peak 2.1m & T2KA & 02-04-15 & 6573/67 & 6513/1511 & 900 & 420 & 0.30 & 1.5 & (4)\\
NGC3198 & SINGS Large & Kitt Peak 2.1m & T2KA & 02-03-09 & 6573/67 & 6513/1511 & 900 & 420 & 0.30 & 3.4 & (4)\\
NGC3227 & Field & JKT & SITe2 CCD & 00-01-01 & 6594/44 & 6580/1380 & 4800 & 900 & 0.24 & 1.4 & (6)\\
IC2574 & SINGS Large & Kitt Peak 2.1m & CFIM & 01-03-29 & 6573/67 & 6513/1511 & 900 & 360 & 0.30 & 1.9 & (4)\\
NGC3254 & Field & NOT & ALFOSC & 09-01-13 & 6598/18 & 6500/1300 & 1800 & 600 & 0.19 & 1.7 & (1)\\
NGC3265 & SINGS Small & Kitt Peak 2.1m & T2KA & 02-04-10 & 6618/74 & 6513/1511 & 900 & 420 & 0.30 & 1.6 & (4)\\
UGC05720 & SINGS Small & Kitt Peak 2.1m & T2KA & 00-01-01 & 6618/74 & 6513/1511 & 240 & 420 & 0.30 & 1.1 & (4)\\
NGC3351 & SINGS Large & Kitt Peak 2.1m & T2KA & 01-03-29 & 6573/67 & 6513/1511 & 900 & 360 & 0.30 & 1.6 & (4)\\
NGC3353 & Field & NOT & ALFOSC & 09-01-13 & 6577/18 & 6500/1300 & 1800 & 600 & 0.19 & 2.1 & (1)\\
NGC3413 & Field & NOT & ALFOSC & 10-03-25 & 6583/36 & 6500/1300 & 600 & 180 & 0.19 & 1.4 & (1)\\
NGC3447B & Field & NOT & ALFOSC & 08-01-02 & 6583/36 & 6500/1500 & 2700 & 270 & 0.19 & 1.1 & (1)\\
UGC06029 & Field & NOT & ALFOSC & 09-01-13 & 6598/18 & 6500/1300 & 1800 & 600 & 0.19 & 1.9 & (1)\\
NGC3507 & Field & JKT & JAG-CCD & 00-02-29 & 6570/55 & 6373/1491 & 1200 & 300 & 0.33 & 1.3 & (5)\\
NGC3521 & SINGS Large & Kitt Peak 2.1m & T2KA & 02-03-10 & 6573/67 & 6513/1511 & 900 & 120 & 0.30 & 1.4 & (4)\\
UGC06161 & Field & JKT & JAG-CCD & 01-02-14 & 6570/55 & 6471/115 & 1200 & 600 & 0.33 & 1.6 & (5)\\
ESO570-019 & Field & NOT & ALFOSC & 09-01-13 & 6598/18 & 6500/1300 & 1800 & 600 & 0.19 & 2.1 & (1)\\
NGC3627 & SINGS Large & Kitt Peak 2.1m & T2KA & 02-04-14 & 6618/74 & 6513/1511 & 900 & 420 & 0.30 & 1.5 & (4)\\
UGC06378 & Field & NOT & ALFOSC & 09-01-13 & 6598/18 & 6500/1300 & 1800 & 600 & 0.19 & 1.6 & (1)\\
UGC06566 & Field & NOT & ALFOSC & 09-01-13 & 6598/18 & 6500/1300 & 1200 & 600 & 0.19 & 1.9 & (1)\\
NGC3741 & Field & JKT & JAG-CCD & 01-01-20 & 6570/55 & 6373/1491 & 1200 & 300 & 0.33 & 1.0 & (5)\\
UGC06578 & Field & INT & WFC & 08-03-14 & 6568/95 & 6380/1520 & 480 & 140 & 0.33 & 1.9 & (1)\\
NGC3773 & SINGS Small & Kitt Peak 2.1m & T2KA & 02-04-11 & 6573/67 & 6513/1511 & 900 & 420 & 0.30 & 1.2 & (4)\\
NGC3782 & Field & JKT & JAG-CCD & 01-02-14 & 6570/55 & 6373/1491 & 1200 & 300 & 0.33 & 1.6 & (5)\\
UGC06792 & Field & INT & WFC & 09-05-04 & 6568/95 & 6380/1520 & 1800 & 540 & 0.33 & 1.8 & (1)\\
NGC3931 & Field & NOT & ALFOSC & 10-03-25 & 6583/36 & 6500/1300 & 480 & 180 & 0.19 & 1.3 & (1)\\
NGC3928 & Field & NOT & ALFOSC & 10-03-25 & 6583/36 & 6500/1300 & 480 & 120 & 0.19 & 1.6 & (1)\\
NGC3938 & SINGS Large & Kitt Peak 2.1m & T2KA & 02-04-12 & 6573/67 & 6513/1511 & 900 & 420 & 0.30 & 2.6 & (4)\\
NGC3998 & Field & NOT & ALFOSC & 10-03-26 & 6583/36 & 6500/1300 & 480 & 120 & 0.19 & 0.9 & (1)\\
NGC4013 & Field & NOT & ALFOSC & 10-03-26 & 6583/36 & 6500/1300 & 360 & 100 & 0.19 & 1.0 & (1)\\
IC0750 & Field & JKT & JAG-CCD & 99-02-11 & 6570/55 & 6373/1491 & 1200 & 750 & 0.33 & 1.5 & (5)\\
UGC07009 & Field & NOT & ALFOSC & 09-01-13 & 6583/36 & 6500/1300 & 1800 & 600 & 0.19 & 1.7 & (1)\\
NGC4041 & Field & NOT & ALFOSC & 09-01-13 & 6583/36 & 6500/1300 & 1800 & 600 & 0.19 & 1.1 & (1)\\
NGC4117 & Field & NOT & ALFOSC & 10-03-26 & 6583/36 & 6500/1300 & 480 & 120 & 0.19 & 0.8 & (1)\\
NGC4138 & Field & NOT & ALFOSC & 10-03-26 & 6583/36 & 6500/1300 & 480 & 90 & 0.19 & 0.8 & (1)\\
NGC4190 & Field & JKT & JAG-CCD & 01-01-20 & 6570/55 & 6373/1491 & 1200 & 300 & 0.33 & 1.1 & (5)\\
PGC039265 & Virgo Small & SPM 2.1m & SIT3 & 04-03-17 & 6603/100 & 6550/900 & 900 & 240 & 0.31 & 1.8 & (7)\\
NGC4236 & SINGS Large & NOT & ALFOSC & 10-03-25 & 6564/33 & 6500/1300 & 480 & 120 & 0.19 & 0.9 & (1)\\
IC3105 & Field & JKT & JAG-CCD & 01-05-18 & 6570/55 & 6373/1491 & 1200 & 300 & 0.33 & 1.5 & (5)\\
NGC4241 & Virgo Small & Calar Alto 1.23m & SITe & 99-04-19 & 6569/113 & 6744/97 & 600 & 600 & 0.50 & 1.7 & (3)\\
NGC4254 & SINGS Large & Kitt Peak 2.1m & T2KA & 02-04-12 & 6618/74 & 6513/1511 & 900 & 420 & 0.30 & 2.5 & (4)\\
NGC4262 & Virgo Small & INT & WFC & 08-03-14 & 6568/95 & 6380/1520 & 480 & 120 & 0.33 & 1.6 & (1)\\
IC3155 & Virgo Small & NOT & ALFOSC & 10-03-26 & 6611/18 & 6500/1300 & 480 & 108 & 0.19 & 1.0 & (1)\\
NGC4268 & Virgo Small & NOT & ALFOSC & 10-03-26 & 6611/18 & 6500/1300 & 480 & 120 & 0.19 & 1.1 & (1)\\
NGC4270 & Virgo Small & NOT & ALFOSC & 10-03-26 & 6611/18 & 6500/1300 & 480 & 120 & 0.19 & 0.9 & (1)\\
NGC4277 & Virgo Small & NOT & ALFOSC & 10-03-26 & 6611/18 & 6500/1300 & 480 & 120 & 0.19 & 0.8 & (1)\\
NGC4288 & Field & Bok 2.3m & CCD21 & 01-05-01 & 6580/69 & 6580/1539 & 1000 & 200 & 0.43 & 1.2 & (8)\\
NGC4298 & Virgo Small & Calar Alto 1.23m & SITe & 99-04-17 & 6569/110 & 6744/97 & 600 & 600 & 0.50 & 3.0 & (3)\\
UGC07428 & Field & WHT & TEK2 & 07-04-03 & 6594/44 & 6391/1455 & 360 & 300 & 0.22 & 15.0 & (1)\\
NGC4301 & Virgo Small & INT & WFC & 08-03-15 & 6568/95 & 6380/1520 & 480 & 120 & 0.33 & 1.9 & (1)\\
NGC4318 & Virgo Small & INT & WFC & 08-03-15 & 6568/95 & 6380/1520 & 480 & 120 & 0.33 & 1.3 & (1)\\
NGC4321 & SINGS Large & Kitt Peak 2.1m & CFIM & 01-03-28 & 6573/67 & 6513/1511 & 900 & 420 & 0.30 & 1.3 & (4)\\
NGC4324 & Virgo Small & SPM 2.1m & TEK & 00-04-01 & 6603/90 & 6723/90 & 420 & 420 & 0.31 & 2.3 & (2)\\
NGC4376 & Virgo Small & INT & WFC & 09-05-06 & 6568/95 & 6380/1520 & 1800 & 540 & 0.33 & 1.6 & (1)\\
NGC4383 & Virgo Small & SPM 2.1m & SITe & 04-03-11 & 6603/90 & 6550/900 & 1200 & 300 & 0.31 & 2.1 & (7)\\
UGC07512 & Field & NOT & ALFOSC & 10-03-26 & 6598/18 & 6500/1300 & 480 & 120 & 0.19 & 1.0 & (1)\\
NGC4390 & Virgo Small & SPM 2.1m & SITe & 04-03-11 & 6603/90 & 6550/900 & 1200 & 240 & 0.31 & 1.9 & (7)\\
NGC4394 & Virgo Small & JKT & JAG-CCD & 01-02-17 & 6570/55 & 6373/1491 & 1200 & 300 & 0.33 & 1.2 & (5)\\
NGC4423 & Virgo Small & NOT & ALFOSC & 10-03-26 & 6583/36 & 6500/1300 & 360 & 90 & 0.19 & 1.5 & (1)\\
NGC4430 & Virgo Small & OHP 1.2m & TEK & 93-03-07 & 6612/55 & 6000/1500 & 600 & 200 & 0.41 & 2.0 & (3)\\
IC3365 & Virgo Small & JKT & JAG-CCD & 01-01-21 & 6607/50 & 6373/1491 & 1200 & 300 & 0.33 & 1.6 & (5)\\
UGC07590 & Virgo Small & SPM 2.1m & Thomson & 01-04-25 & 6603/90 & 6723/90 & 300 & 300 & 0.39 & 1.7 & (2)\\
NGC4450 & SINGS Large & Kitt Peak 2.1m & T2KA & 02-04-09 & 6618/74 & 6513/1511 & 900 & 420 & 0.30 & 1.4 & (4)\\
NGC4468 & Virgo Small & MDM 2.4m & MDM8K & 08-12-30 & 6563/100 & 6470/1330 & 2700 & 540 & 0.35 & 2.2 & (1)\\
NGC4470 & Virgo Small & Calar Alto 1.23m & SITe & 99-04-18 & 6569/113 & 6744/97 & 600 & 607 & 0.51 & 1.9 & (3)\\
NGC4504 & Field & NOT & ALFOSC & 11-07-24 & 6583/36 & 6500/1300 & 900 & 300 & 0.19 & 1.1 & (1)\\
NGC4522 & Virgo Small & SPM 2.1m & SIT3 & 04-03-11 & 6603/90 & 6550/900 & 900 & 240 & 0.31 & 2.5 & (7)\\
NGC4532 & Virgo Small & MDM 2.4m & MDM8K & 08-12-30 & 6563/100 & 6470/1330 & 2700 & 540 & 0.35 & 1.8 & (1)\\
UGC07739 & Virgo Small & SPM 2.1m & Thomson & 01-04-23 & 6603/90 & 6723/90 & 600 & 600 & 0.39 & 2.9 & (2)\\
IC3522 & Virgo Small & MDM 2.4m & MDM8K & 08-12-31 & 6563/100 & 6470/1330 & 1800 & 540 & 0.35 & 1.2 & (1)\\
NGC4550 & Field & INT & WFC & 08-03-16 & 6568/95 & 6380/1520 & 480 & 120 & 0.33 & 1.7 & (1)\\
NGC4559 & SINGS Large & Kitt Peak 2.1m & T2KA & 02-03-09 & 6673/67 & 6513/1511 & 900 & 420 & 0.30 & 2.0 & (4)\\
NGC4561 & Virgo Small & INT & WFC & 09-06-13 & 6568/95 & 6380/1520 & 1199 & 540 & 0.33 & 1.8 & (1)\\
NGC4567 & Virgo Small & OHP 1.2m & TEK & 98-02-28 & 6612/55 & 6561/48 & 1800 & 1800 & 0.69 & 2.8 & (3)\\
NGC4568 & Virgo Small & OHP 1.2m & TEK & 98-02-28 & 6612/55 & 6561/48 & 1800 & 1800 & 0.69 & 2.8 & (3)\\
IC3583 & Virgo Small & SPM 2.1m & TEK & 99-04-17 & 6603/90 & 6723/90 & 420 & 420 & 0.30 & 3.3 & (2)\\
NGC4569 & SINGS Large & OHP 1.2m & TEK & 98-02-28 & 6561/48 & 6600/55 & 1800 & 1800 & 0.68 & 3.2 & (7)\\
IC3591 & Virgo Small & NOT & ALFOSC & 10-02-09 & 6583/36 & 6500/1300 & 1800 & 540 & 0.19 & 1.8 & (1)\\
NGC4579 & SINGS Large & Kitt Peak 2.1m & T2KA & 02-03-10 & 6618/74 & 6513/1511 & 900 & 420 & 0.30 & 1.3 & (4)\\
IC3617 & Virgo Small & SPM 2.1m & Thomson & 01-04-25 & 6603/90 & 6723/90 & 300 & 300 & 0.39 & 2.5 & (2)\\
UGC07827 & Field & INT & WFC & 09-05-05 & 6568/95 & 6380/1520 & 1800 & 540 & 0.33 & 1.5 & (1)\\
NGC4595 & Virgo Small & JKT & JAG-CCD & 01-03-27 & 6570/55 & 6373/1491 & 1200 & 300 & 0.33 & 2.0 & (5)\\
NGC4594 & SINGS Large & Kitt Peak 2.1m & T2KA & 02-03-10 & 6573/67 & 6513/1511 & 900 & 420 & 0.30 & 1.2 & (4)\\
NGC4625 & SINGS Small & NOT & ALFOSC & 10-02-09 & 6583/36 & 6500/1300 & 1800 & 540 & 0.19 & 1.3 & (1)\\
NGC4631 & SINGS Large & Kitt Peak 2.1m & T2KA & 02-03-11 & 6573/67 & 6513/1511 & 900 & 420 & 0.30 & 2.5 & (4)\\
NGC4639 & Virgo Small & OHP 1.20m & TEK & 00-03-09 & 6612/55 & 6450/45 & 900 & 900 & 0.69 & 3.3 & (3)\\
NGC4640 & Virgo Small & NOT & ALFOSC & 10-03-27 & 6610/50 & 6500/1300 & 480 & 120 & 0.19 & 0.9 & (1)\\
NGC4647 & Virgo Small & INT & WFC & 09-06-13 & 6568/95 & 6380/1520 & 2399 & 540 & 0.33 & 2.0 & (1)\\
NGC4651 & Virgo Small & JKT & JAG-CCD & 00-04-23 & 6570/55 & 6373/1491 & 1200 & 300 & 0.33 & 1.7 & (5)\\
PGC043211 & Field & NOT & ALFOSC & 10-03-27 & 6583/36 & 6500/1300 & 480 & 120 & 0.19 & 0.8 & (1)\\
NGC4725 & SINGS Large & Kitt Peak 2.1m & T2KA & 02-03-11 & 6573/67 & 6513/1511 & 900 & 420 & 0.30 & 2.2 & (4)\\
NGC4736 & SINGS Large & JKT & JAG CCD & 02-01-15 & 6570/55 & 6580/1380 & 3600 & 1620 & 0.24 & 1.4 & (6)\\
NGC4772 & Field & SPM 2.1m & SIT3 & 03-02-28 & 6603/90 & 6550/900 & 900 & 240 & 0.31 & 2.0 & (7)\\
NGC4789A & SINGS Small & JKT & JAG-CCD & 00-04-26 & 6570/55 & 6471/115 & 1200 & 600 & 0.33 & 1.6 & (5)\\
NGC4826 & SINGS Large & Kitt Peak 0.9m & T2KA & 99-04-22 & 6573/67 & 6513/1511 & 360 & 20 & 0.78 & 1.8 & (4)\\
NGC4941 & Field & NOT & ALFOSC & 10-03-27 & 6583/36 & 6500/1300 & 480 & 120 & 0.19 & 0.8 & (1)\\
PGC045195 & Field & INT & WFC & 08-03-16 & 6568/95 & 6380/1520 & 528 & 130 & 0.33 & 1.3 & (1)\\
UGC08201 & SINGS Small & JKT & JAG-CCD & 01-03-31 & 6570/55 & 6373/1491 & 1200 & 300 & 0.33 & 2.0 & (5)\\
UGC08303 & Field & JKT & JAG-CCD & 01-01-23 & 6570/55 & 6373/1491 & 1200 & 300 & 0.33 & 1.4 & (5)\\
NGC5033 & SINGS Large & NOT & ALFOSC & 10-03-27 & 6583/36 & 6500/1300 & 360 & 83 & 0.19 & 0.8 & (1)\\
ESO508-030 & Field & NOT & ALFOSC & 10-03-27 & 6598/18 & 6500/1300 & 480 & 120 & 0.19 & 0.9 & (1)\\
NGC5055 & SINGS Large & Kitt Peak 0.9m & T2KA & 99-04-20 & 6573/67 & -/- & 420 & - & 0.38 & - & (4)\\
NGC5194 & SINGS Large & Kitt Peak 2.1m & CFIM & 01-03-28 & 6573/67 & 6425/1500 & 900 & 372 & 0.30 & 1.4 & (4)\\
NGC5195 & SINGS Large & Kitt Peak 2.1m & CFIM & 01-03-28 & 6573/67 & 6425/1500 & 900 & 372 & 0.30 & 1.4 & (4)\\
NGC5474 & SINGS Small & JKT & JAG-CCD & 00-02-29 & 6570/55 & 6471/115 & 1200 & 1200 & 0.33 & 1.4 & (5)\\
NGC5477 & Field & JKT & JAG-CCD & 00-07-04 & 6570/55 & 6373/1491 & 1200 & 300 & 0.33 & 0.9 & (5)\\
NGC5486 & Field & JKT & JAG-CCD & 01-05-13 & 6594/44 & 1471/115 & 1200 & 1200 & 0.33 & 1.3 & (5)\\
PGC140287 & Field & INT & WFC & 08-03-17 & 6568/95 & 6380/1520 & 450 & 135 & 0.33 & 2.2 & (1)\\
IC1024 & Field & WHT & TEK2 & 07-04-28 & 6594/44 & 6391/1455 & 360 & 300 & 0.22 & 1.9 & (1)\\
NGC5701 & Field & NOT & ALFOSC & 10-03-27 & 6598/18 & 6500/1300 & 480 & 120 & 0.19 & 0.8 & (1)\\
IC1066 & Field & INT & WFC & 08-03-14 & 6568/95 & 6380/1520 & 600 & 285 & 0.33 & 1.9 & (1)\\
PGC057723 & Field & INT & WFC & 08-03-16 & 6568/95 & 6380/1520 & 480 & 120 & 0.33 & 1.7 & (1)\\
NGC6140 & Field & INT & PFCU & 96-05-14 & 6594/44 & 6580/1380 & 3600 & 1800 & 0.24 & 1.4 & (6)\\
NGC6118 & Field & INT & WFC & 08-03-14 & 6568/95 & 6380/1520 & 600 & 180 & 0.33 & 1.8 & (1)\\
PGC058661 & Field & INT & WFC & 08-03-16 & 6568/95 & 6380/1520 & 510 & 135 & 0.33 & 1.3 & (1)\\
IC1254 & Field & WHT & TEK2 & 07-04-28 & 6594/44 & 6391/1455 & 360 & 300 & 0.21 & 1.3 & (1)\\
NGC7465 & Field & WHT & ACAM & 09-07-30 & 6613/15 & 6228/1322 & 720 & 360 & 0.25 & 1.5 & (1)\\
NGC7742 & Field & WHT & ACAM & 10-09-03 & 6589/15 & 6228/1322 & 900 & 60 & 0.30 & 1.1 & (1)\\

%% file: resultsTable.txt
NGC0024 & SA(s)c & 554 & 7.6\,(1.1) & 5.8/1.3 & -15.49$\rm ^{~}$ & 0.011 & 1.1 & 18.0\,(3.1) & 0.3 & 0.03 & 1.6\,(0.6) & 0.09\,(0.03) & 19\,(1) [3]\\
ESO538-024 & IB(s)m & 1551 & 21.4\,(3.2) & 1.4/1.2 & -16.44$\rm ^c$ & 0.014 & 1.1 & 1.8\,(0.3) & 0.4 & 0.02 & 1.3\,(0.5) & 0.07\,(0.02) & 28\,(6) [1]\\
NGC0210 & SAB(s)b & 1663 & 23.1\,(3.5) & 5.0/3.3 & -19.76$\rm ^{~}$ & 0.516 & 0.0 & 52.6\,(8.9) & 0.8 & 0.3 & 47.8\,(16.5) & 2.53\,(0.87) & 31\,(7) [1]\\
NGC0216 & S0 & 1541 & 21.0\,(3.1) & 2.0/0.7 & -17.44$\rm ^{~}$ & 0.015 & 1.1 & 5.1\,(0.9) & 0.5 & 0.03 & 4.1\,(1.4) & 0.22\,(0.08) & 13\,(3) [1]\\
IC0051 & S0 & 1753 & 23.8\,(3.6) & 1.3/1.2 & -17.70$\rm ^c$ & 0.015 & 1.1 & 2.7\,(0.5) & 0.5 & 0.03 & 2.8\,(1.0) & 0.15\,(0.05) & 8\,(2) [1]\\
NGC0274 & SAB(r) & 1773 & 24.6\,(3.7) & 2.3/1.5 & -18.66$\rm ^c$ & 0.012 & 1.1 & 6.4\,(1.1) & 0.7 & 0.04 & 8.8\,(3.0) & 0.46\,(0.16) & 7\,(1) [1]\\
NGC0337 & SB(s)d & 1648 & 23.1\,(3.5) & 2.9/1.8 & -19.32$\rm ^{~}$ & 0.009 & 1.1 & 27.2\,(4.6) & 0.9 & 0.13 & 34.5\,(11.9) & 1.83\,(0.63) & 43\,(9) [1]\\
NGC0404 & SA(s)0 & -48 & 2.8\,(0.4) & 3.5/3.5 & -13.79$\rm ^{~}$ & 0.035 & 3.5 & 18.1\,(3.1) & 0.3 & 0.03 & 0.2\,(0.1) & 0.01\,(0.00) & 6\,(1) [1]\\
NGC0450 & SAB(s)cd & 1761 & 25.0\,(3.7) & 3.1/2.3 & -18.11$\rm ^c$ & - & 2.0 & 19.7\,(3.4) & 0.6 & 0.16 & 21.3\,(7.3) & 1.13\,(0.39) & 53\,(5) [3]\\
NGC0473 & SAB(r)0/a & 2222 & 31.3\,(4.7) & 1.7/1.1 & -18.67$\rm ^{~}$ & - & 2.0 & 3.4\,(0.6) & 1.0 & 0.2 & 8.0\,(2.8) & 0.43\,(0.15) & 12\,(4) [3]\\
NGC0584 & E4 & 1802 & 20.1\,(3.0) & 4.2/2.3 & -20.14$\rm ^{~}$ & 0.010 & 1.1 & 7.2\,(1.2) & 1.3 & 0.14 & 10.2\,(3.5) & 0.54\,(0.19) & 3\,(1) [1]\\
NGC0615 & SA(rs)b & 1948 & 25.6\,(3.8) & 3.6/1.4 & -19.18$\rm ^{~}$ & 0.010 & 1.5 & 12.9\,(2.2) & 0.7 & 0.1 & 17.0\,(5.9) & 0.90\,(0.31) & 16\,(3) [1]\\
NGC0628 & SA(s)c & 657 & 7.3\,(1.1) & 10.5/9.5 & -18.51$\rm ^{~}$ & - & 1.1 & 146.1\,(24.8) & 0.9 & 0.09 & 18.9\,(6.5) & 1.00\,(0.35) & 35\,(4) [3]\\
ESO477-016 & Sb & 1646 & 24.2\,(3.6) & 2.8/0.3 & -16.78$\rm ^c$ & 0.018 & 1.5 & 3.6\,(0.6) & 0.4 & 0.04 & 3.5\,(1.2) & 0.18\,(0.06) & 32\,(6) [1]\\
NGC0855 & E & 575 & 9.7\,(1.5) & 2.6/1.0 & -15.06$\rm ^{~}$ & 0.014 & 1.1 & 7.8\,(1.3) & 1.3 & 0.01 & 2.8\,(1.0) & 0.15\,(0.05) & 40\,(4) [3]\\
NGC0925 & SAB(s)d & 553 & 9.1\,(1.4) & 10.5/5.9 & -17.56$\rm ^{~}$ & 0.046 & 4.2 & 26.9\,(4.6) & 0.8 & 0.13 & 4.9\,(1.7) & 0.26\,(0.09) & 25\,(2) [3]\\
NGC1036 & Pec? & 802 & 12.5\,(1.9) & 1.4/1.0 & -15.69$\rm ^{~}$ & 0.013 & 3.0 & 13.9\,(2.4) & 0.6 & 0.02 & 4.3\,(1.5) & 0.23\,(0.08) & 50\,(10) [1]\\
NGC1140 & IBm & 1501 & 20.1\,(3.0) & 1.7/0.9 & -18.42$\rm ^{~}$ & 0.017 & 1.1 & 30.0\,(5.1) & 0.5 & 0.04 & 22.5\,(7.8) & 1.19\,(0.41) & 48\,(10) [1]\\
NGC1156 & IB(s)m & 375 & 7.2\,(1.1) & 3.3/2.5 & -14.57$\rm ^{~}$ & - & 2.0 & 26.5\,(4.5) & 0.8 & 0.04 & 3.3\,(1.1) & 0.17\,(0.06) & 74\,(7) [3]\\
ESO481-019 & IB(s)m & 1480 & 19.8\,(3.0) & 2.0/1.7 & -16.43$\rm ^c$ & 0.012 & 1.5 & 9.3\,(1.6) & 0.3 & 0.03 & 5.7\,(2.0) & 0.30\,(0.10) & 193\,(41) [1]\\
NGC1325 & SA(s)bc & 1646 & 20.6\,(3.1) & 4.7/1.6 & -19.29$\rm ^{~}$ & 0.012 & 1.1 & 12.7\,(2.2) & 0.7 & 0.06 & 12.0\,(4.1) & 0.63\,(0.22) & 12\,(2) [1]\\
NGC2146A & SAB(s)c & 1386 & 25.9\,(3.9) & 3.0/1.1 & -18.04$\rm ^{~}$ & - & 1.8 & 2.7\,(0.5) & 0.8 & 0.15 & 4.0\,(1.4) & 0.21\,(0.07) & 17\,(4) [3]\\
NGC2403 & SAB(s)cd & 131 & 3.2\,(0.5) & 21.9/12.3 & -18.03$\rm ^{~}$ & - & 4.2 & 791.6\,(134.6) & 0.7 & 0.16 & 16.2\,(5.6) & 0.86\,(0.30) & 50\,(15) [3]\\
UGC04305 & Im & 142 & 3.4\,(0.5) & 7.9/6.3 & -15.96$\rm ^{~}$ & 0.012 & 4.2 & 48.2\,(8.2) & 0.4 & 0.07 & 0.9\,(0.3) & 0.05\,(0.02) & 37\,(4) [3]\\
PGC023521 & I & 113 & 3.6\,(0.5) & 1.3/0.7 & -11.05$\rm ^c$ & 0.240 & 4.5 & 0.2\,(0.0) & 0.1 & 0.01 & 0.0\,(0.0) & 0.00\,(0.00) & 98\,(23) [1]\\
NGC2742 & SA(s)c & 1285 & 22.7\,(3.4) & 3.0/1.5 & -19.40$\rm ^{~}$ & - & 0.0 & 18.2\,(3.1) & 1.0 & 0.26 & 20.1\,(6.9) & 1.06\,(0.37) & 21\,(2) [3]\\
NGC2787 & SB(r)0+ & 696 & 14.4\,(2.2) & 3.2/2.0 & -18.31$\rm ^{~}$ & 0.011 & 1.5 & 8.0\,(1.4) & 1.1 & 0.07 & 4.9\,(1.7) & 0.26\,(0.09) & 3\,(1) [1]\\
NGC2841 & SA(r)b & 638 & 14.1\,(2.1) & 8.1/3.5 & -20.14$\rm ^{~}$ & - & 4.2 & 38.2\,(6.5) & 1.4 & 0.38 & 19.7\,(6.8) & 1.04\,(0.36) & 7\,(3) [2]\\
UGC05139 & IAB(s)m & 139 & 3.6\,(0.5) & 3.6/3.0 & -14.26$\rm ^{~}$ & - & 2.0 & 3.6\,(0.6) & 0.3 & 0.04 & 0.1\,(0.0) & 0.00\,(0.00) & 55\,(6) [3]\\
NGC2976 & SAc & 3 & 3.6\,(0.5) & 5.9/2.7 & -14.65$\rm ^{~}$ & - & 2.0 & 62.0\,(10.5) & 0.5 & 0.04 & 1.4\,(0.5) & 0.07\,(0.03) & 35\,(3) [3]\\
UGC05272 & ImIII & 524 & 7.0\,(1.1) & 2.1/0.8 & -15.71$\rm ^{~}$ & 0.250 & 4.5 & 1.8\,(0.3) & 0.4 & 0.06 & 0.1\,(0.0) & 0.01\,(0.00) & 62\,(6) [3]\\
NGC3049 & SB(rs)ab & 1455 & 22.7\,(3.4) & 2.1/1.1 & -19.85$\rm ^c$ & 0.044 & 4.5 & 18.1\,(3.1) & 0.9 & 0.34 & 16.7\,(5.8) & 0.88\,(0.30) & 65\,(13) [1]\\
NGC3031 & SA(s)ab & -34 & 3.6\,(0.5) & 26.9/14.1 & -16.21$\rm ^{~}$ & 0.060 & 3.0 & 1849.0\,(314.3) & 0.5 & 0.08 & 42.4\,(14.6) & 2.25\,(0.77) & 10\,(1) [3]\\
NGC3034 & I0 & 203 & 3.6\,(0.5) & 11.2/4.3 & -18.63$\rm ^{~}$ & 0.041 & 4.2 & 766.9\,(130.4) & 1.4 & 0.21 & 33.6\,(11.6) & 1.78\,(0.61) & 64\,(5) [3]\\
UGC05336 & Im & 46 & 3.6\,(0.5) & 2.5/2.0 & -11.93$\rm ^{~}$ & - & 3.0 & 1.1\,(0.2) & 0.3 & 0.01 & 0.0\,(0.0) & 0.00\,(0.00) & 18\,(4) [3]\\
NGC3077 & I0 & 14 & 2.5\,(0.4) & 5.4/4.5 & -15.07$\rm ^{~}$ & 0.217 & 4.5 & 51.5\,(8.8) & 0.4 & 0.05 & 0.5\,(0.2) & 0.03\,(0.01) & 42\,(4) [3]\\
UGC05423 & Im & 347 & 3.6\,(0.5) & 0.9/0.6 & -14.81$\rm ^c$ & 0.028 & 3.0 & 4.7\,(0.8) & 0.4 & 0.04 & 0.1\,(0.0) & 0.01\,(0.00) & 27\,(4) [3]\\
NGC3162 & SAB(rs)bc & 1384 & 21.7\,(3.3) & 3.0/2.5 & -19.52$\rm ^{~}$ & 0.022 & 2.5 & 75.8\,(12.9) & 0.4 & 0.25 & 44.4\,(15.3) & 2.36\,(0.81) & 102\,(21) [1]\\
NGC3190 & SA(s)a & 1271 & 21.3\,(3.2) & 4.4/1.5 & -19.57$\rm ^{~}$ & - & 4.5 & 8.5\,(1.5) & 1.3 & 0.3 & 10.6\,(3.7) & 0.56\,(0.19) & 3\,(2) [2]\\
NGC3184 & SAB(rs)cd & 592 & 11.1\,(1.7) & 7.4/6.9 & -19.92$\rm ^{~}$ & - & 4.2 & 69.9\,(11.9) & 1.2 & 0.35 & 19.7\,(6.8) & 1.04\,(0.36) & 13\,(6) [2]\\
NGC3198 & SB(rs)c & 663 & 13.7\,(2.1) & 8.5/3.3 & -19.54$\rm ^{~}$ & - & 4.2 & 38.4\,(6.5) & 1.0 & 0.3 & 14.7\,(5.1) & 0.78\,(0.27) & 34\,(17) [2]\\
NGC3227 & SAB(s) & 1157 & 23.1\,(3.5) & 5.4/3.6 & -20.26$\rm ^{~}$ & 0.020 & 3.0 & 4.5\,(0.8) & 0.9 & 0.37 & 4.0\,(1.4) & 0.21\,(0.07) & 13\,(3) [1]\\
IC2574 & SAB(s)m & 57 & 3.8\,(0.6) & 13.2/5.4 & -15.74$\rm ^{~}$ & 0.047 & 4.2 & 89.8\,(15.3) & 0.4 & 0.07 & 2.0\,(0.7) & 0.11\,(0.04) & 31\,(8) [3]\\
NGC3254 & SA(s)bc & 1262 & 23.2\,(3.5) & 5.0/1.6 & -19.37$\rm ^{~}$ & 0.014 & 1.5 & 16.1\,(2.7) & 1.0 & 0.11 & 23.6\,(8.1) & 1.25\,(0.43) & 13\,(3) [1]\\
NGC3265 & E & 1319 & 22.7\,(3.4) & 1.3/1.0 & -18.78$\rm ^c$ & - & 4.5 & 5.0\,(0.9) & 0.6 & 0.22 & 4.0\,(1.4) & 0.21\,(0.07) & 46\,(22) [2]\\
UGC05720 & Im & 1430 & 25.0\,(3.8) & 1.1/0.8 & -18.33$\rm ^c$ & - & 4.5 & 21.3\,(3.6) & 0.7 & 0.18 & 24.1\,(8.3) & 1.28\,(0.44) & 80\,(38) [2]\\
NGC3351 & SB(r)b & 778 & 9.3\,(1.4) & 3.1/2.9 & -20.42$\rm ^{~}$ & 0.047 & 4.2 & 101.5\,(17.3) & 1.4 & 0.42 & 22.5\,(7.8) & 1.19\,(0.41) & 16\,(3) [3]\\
NGC3353 & BCD/Irr & 944 & 18.0\,(2.7) & 1.2/0.8 & -17.67$\rm ^{~}$ & 0.018 & 1.5 & 19.3\,(3.3) & 0.5 & 0.05 & 11.6\,(4.0) & 0.62\,(0.21) & 49\,(10) [1]\\
NGC3413 & S0 & 667 & 8.8\,(1.3) & 2.2/0.9 & -17.46$\rm ^c$ & 0.029 & 3.0 & 9.4\,(1.6) & 0.5 & 0.11 & 1.2\,(0.4) & 0.06\,(0.02) & 36\,(7) [1]\\
NGC3447B & IB(s)m & 1023 & 17.0\,(2.6) & 4.2/2.3 & -$\rm ^{~}$ & 0.028 & 2.5 & 4.9\,(0.8) & 0.4 & 0.0 & 2.4\,(0.8) & 0.13\,(0.04) & 119\,(25) [1]\\
UGC06029 & Pec & 1403 & 24.2\,(3.6) & 1.0/0.9 & -17.64$\rm ^c$ & 0.019 & 1.5 & 9.6\,(1.6) & 0.4 & 0.05 & 9.5\,(3.3) & 0.50\,(0.17) & 137\,(29) [1]\\
NGC3507 & SB(s)b & 967 & 14.3\,(2.1) & 3.4/2.9 & -19.56$\rm ^{~}$ & - & 2.0 & 41.0\,(7.0) & 1.0 & 0.29 & 17.4\,(6.0) & 0.92\,(0.32) & 39\,(4) [3]\\
NGC3521 & SAB(rs)bc & 801 & 7.9\,(1.2) & 11.0/5.1 & -21.20$\rm ^{~}$ & 0.047 & 4.2 & 126.8\,(21.6) & 1.2 & 0.53 & 13.0\,(4.5) & 0.69\,(0.24) & 20\,(3) [3]\\
UGC06161 & SBdm & 804 & 13.3\,(2.0) & 2.6/1.2 & -16.67$\rm ^c$ & - & 2.0 & 2.6\,(0.4) & 0.4 & 0.09 & 0.7\,(0.2) & 0.04\,(0.01) & 34\,(3) [3]\\
ESO570-019 & (R\')SAc & 1249 & 19.4\,(2.9) & 1.6/0.5 & -17.91$\rm ^c$ & 0.016 & 1.5 & 3.7\,(0.6) & 0.6 & 0.06 & 2.6\,(0.9) & 0.14\,(0.05) & 25\,(5) [1]\\
NGC3627 & SAB(s)b & 727 & 9.4\,(1.4) & 9.1/4.2 & -21.20$\rm ^{~}$ & 0.049 & 4.5 & 125.5\,(21.3) & 1.7 & 0.54 & 30.3\,(10.4) & 1.60\,(0.55) & 19\,(2) [3]\\
UGC06378 & Sd & 1309 & 23.8\,(3.6) & 2.7/0.4 & -16.75$\rm ^c$ & 0.019 & 1.5 & 3.6\,(0.6) & 0.4 & 0.04 & 3.5\,(1.2) & 0.18\,(0.06) & 34\,(7) [1]\\
UGC06566 & SBm & 1249 & 22.5\,(3.4) & 1.1/0.8 & -$\rm ^{~}$ & 0.011 & 1.5 & 1.0\,(0.2) & 0.3 & 0.0 & 0.8\,(0.3) & 0.04\,(0.02) & 16\,(3) [1]\\
NGC3741 & ImIII/BCD & 223 & 3.6\,(0.5) & 2.0/1.1 & -14.68$\rm ^{~}$ & - & 2.0 & 4.5\,(0.8) & 0.3 & 0.04 & 0.1\,(0.0) & 0.00\,(0.00) & 66\,(7) [3]\\
UGC06578 & WR/BCD & 1178 & 15.8\,(2.4) & 0.7/0.5 & -16.27$\rm ^c$ & 0.076 & 6.0 & 4.6\,(0.8) & 0.4 & 0.08 & 1.8\,(0.6) & 0.09\,(0.03) & 118\,(24) [1]\\
NGC3773 & SA0 & 987 & 10.5\,(1.6) & 1.2/1.0 & -18.80$\rm ^c$ & - & 4.2 & 9.3\,(1.6) & 0.6 & 0.22 & 1.7\,(0.6) & 0.09\,(0.03) & 37\,(17) [2]\\
NGC3782 & SAB(s)cd & 748 & 13.4\,(2.0) & 1.7/1.1 & -17.50$\rm ^{~}$ & - & 2.0 & 11.3\,(1.9) & 0.5 & 0.13 & 3.4\,(1.2) & 0.18\,(0.06) & 53\,(5) [3]\\
UGC06792 & Scd & 832 & 14.7\,(2.2) & 2.8/0.4 & -16.32$\rm ^c$ & 0.064 & 6.0 & 7.9\,(1.3) & 0.5 & 0.08 & 3.0\,(1.0) & 0.16\,(0.06) & 31\,(6) [1]\\
NGC3931 & SA0- & 929 & 16.2\,(2.4) & 1.1/0.9 & -16.40$\rm ^c$ & 0.019 & 3.0 & 3.4\,(0.6) & 0.5 & 0.07 & 1.6\,(0.5) & 0.08\,(0.03) & 18\,(3) [1]\\
NGC3928 & SA(s)b? & 951 & 18.9\,(2.8) & 1.3/1.2 & -17.85$\rm ^{~}$ & 0.027 & 3.0 & 0.1\,(0.0) & 0.7 & 0.13 & 0.1\,(0.0) & 0.01\,(0.00) & 27\,(5) [1]\\
NGC3938 & SA(s)c & 809 & 14.7\,(2.2) & 5.4/4.9 & -19.88$\rm ^{~}$ & - & 4.2 & 71.4\,(12.1) & 1.1 & 0.34 & 33.1\,(11.4) & 1.75\,(0.60) & 21\,(10) [2]\\
NGC3998 & SA(r)0 & 1040 & 20.0\,(3.0) & 2.7/2.2 & -19.49$\rm ^{~}$ & 0.024 & 3.0 & 13.7\,(2.3) & 1.0 & 0.25 & 12.3\,(4.3) & 0.65\,(0.23) & 5\,(1) [1]\\
NGC4013 & SAb & 831 & 15.3\,(2.3) & 5.2/1.0 & -18.63$\rm ^{~}$ & 0.016 & 3.0 & 18.4\,(3.1) & 0.8 & 0.18 & 8.6\,(3.0) & 0.46\,(0.16) & 22\,(4) [1]\\
IC0750 & Sab: & 701 & 11.8\,(1.8) & 2.5/0.9 & -17.60$\rm ^{~}$ & - & 2.0 & 17.4\,(3.0) & 0.7 & 0.13 & 4.8\,(1.7) & 0.26\,(0.09) & 26\,(3) [3]\\
UGC07009 & Im & 1097 & 21.3\,(3.2) & 2.0/0.5 & -15.66$\rm ^{~}$ & 0.031 & 2.5 & 3.5\,(0.6) & 0.4 & 0.05 & 2.5\,(0.9) & 0.13\,(0.05) & 44\,(9) [1]\\
NGC4041 & SA(rs)bc: & 1221 & 22.9\,(3.4) & 2.1/1.7 & -19.48$\rm ^{~}$ & 0.029 & 3.0 & 74.4\,(12.6) & 0.8 & 0.25 & 75.3\,(26.0) & 3.99\,(1.38) & 46\,(9) [1]\\
NGC4117 & S0 & 886 & 17.7\,(2.7) & 1.7/0.6 & -16.98$\rm ^c$ & 0.024 & 3.0 & 1.7\,(0.3) & 0.6 & 0.09 & 1.0\,(0.3) & 0.05\,(0.02) & 5\,(1) [1]\\
NGC4138 & SA(r)0+ & 888 & 16.7\,(2.5) & 2.6/1.7 & -18.76$\rm ^{~}$ & 0.018 & 3.0 & 26.3\,(4.5) & 0.8 & 0.19 & 14.6\,(5.0) & 0.77\,(0.27) & 21\,(4) [1]\\
NGC4190 & Im & 192 & 2.8\,(0.4) & 1.5/1.2 & -15.22$\rm ^{~}$ & - & 2.0 & 4.6\,(0.8) & 0.4 & 0.05 & 0.1\,(0.0) & 0.00\,(0.00) & 20\,(2) [3]\\
PGC039265 & IAm? & 2175 & 13.6\,(1.1) & 1.1/0.5 & -17.90$\rm ^c$ & 0.300 & 10.0 & 0.7\,(0.1) & 0.4 & 0.16 & 0.2\,(0.0) & 0.01\,(0.00) & 60\,(20) [3]\\
NGC4236 & SB(s)dm & 0 & 4.4\,(0.7) & 21.9/7.2 & -15.20$\rm ^{~}$ & 0.011 & 3.0 & 76.4\,(13.0) & 0.3 & 0.04 & 2.2\,(0.8) & 0.12\,(0.04) & 29\,(4) [3]\\
IC3105 & Im & -159 & 2.4\,(0.4) & 1.8/0.5 & -12.13$\rm ^{~}$ & - & 2.0 & 0.7\,(0.1) & 0.2 & 0.02 & 0.0\,(0.0) & 0.00\,(0.00) & 73\,(7) [3]\\
NGC4241 & SB(s)cd & 733 & 16.7\,(1.1) & 1.2/1.0 & -17.12$\rm ^{~}$ & 0.660 & 0.0 & 3.0\,(0.5) & 0.4 & 0.11 & 1.2\,(0.3) & 0.07\,(0.01) & 10\,(2) [1]\\
NGC4254 & SA(s)c & 2407 & 16.7\,(2.5) & 5.4/4.7 & -22.45$\rm ^{~}$ & 0.051 & 4.5 & 161.2\,(27.4) & 2.3 & 0.54 & 213.4\,(73.6) & 11.31\,(3.90) & 39\,(8) [1]\\
NGC4262 & SB(s)0-? & 1363 & 16.7\,(1.1) & 1.9/1.7 & -19.34$\rm ^{~}$ & 0.067 & 6.0 & 6.7\,(1.1) & 0.9 & 0.28 & 3.8\,(0.8) & 0.20\,(0.04) & 6\,(1) [1]\\
IC3155 & S0 & 2209 & 16.7\,(1.1) & 1.2/0.6 & -17.82$\rm ^c$ & 0.011 & 1.5 & 0.4\,(0.1) & 0.7 & 0.06 & 0.2\,(0.0) & 0.01\,(0.00) & 3\,(1) [1]\\
NGC4268 & SB0/a & 2374 & 16.7\,(1.1) & 1.6/0.7 & -19.06$\rm ^{~}$ & 0.012 & 1.5 & 0.4\,(0.1) & 1.0 & 0.09 & 0.3\,(0.1) & 0.01\,(0.00) & 1\,(0) [1]\\
NGC4270 & S0 & 2357 & 16.7\,(1.1) & 2.0/0.9 & -19.79$\rm ^{~}$ & 0.011 & 1.5 & 1.6\,(0.3) & 1.0 & 0.12 & 1.1\,(0.2) & 0.06\,(0.01) & 2\,(0) [1]\\
NGC4277 & SAB(rs)0/a & 2196 & 16.7\,(1.1) & 1.1/0.8 & -19.35$\rm ^{~}$ & 0.010 & 1.5 & 1.1\,(0.2) & 0.7 & 0.1 & 0.6\,(0.1) & 0.03\,(0.01) & 3\,(1) [1]\\
NGC4288 & SB(s)dm & 556 & 8.7\,(1.3) & 1.6/1.2 & -16.74$\rm ^{~}$ & 0.247 & 4.5 & 8.9\,(1.5) & 0.3 & 0.1 & 1.0\,(0.3) & 0.05\,(0.02) & 49\,(4) [3]\\
NGC4298 & SA(rs)c & 1135 & 16.7\,(1.1) & 3.3/1.2 & -19.48$\rm ^{~}$ & 0.640 & 0.0 & 14.8\,(2.5) & 0.4 & 0.29 & 5.0\,(1.1) & 0.27\,(0.06) & 36\,(8) [1]\\
UGC07428 & Im & 1170 & 22.5\,(3.4) & 1.3/1.2 & -17.34$\rm ^{~}$ & 0.021 & 3.0 & 0.5\,(0.1) & 0.3 & 0.12 & 0.3\,(0.1) & 0.02\,(0.01) & 5\,(1) [1]\\
NGC4301 & SAB(s)cd & 1275 & 16.7\,(1.1) & 1.5/1.3 & -18.34$\rm ^{~}$ & 0.067 & 6.0 & 10.9\,(1.9) & 0.4 & 0.19 & 4.1\,(0.9) & 0.22\,(0.05) & 55\,(11) [1]\\
NGC4318 & E? & 1231 & 16.7\,(1.1) & 0.8/0.7 & -17.47$\rm ^{~}$ & 0.062 & 6.0 & 1.4\,(0.2) & 0.4 & 0.13 & 0.5\,(0.1) & 0.03\,(0.01) & 8\,(2) [1]\\
NGC4321 & SAB(s)bc & 1571 & 16.7\,(2.5) & 7.4/6.3 & -22.03$\rm ^{~}$ & - & 4.2 & 85.7\,(14.6) & 2.0 & 0.53 & 85.3\,(29.4) & 4.52\,(1.56) & 7\,(3) [2]\\
NGC4324 & SA(r)0+ & 1685 & 16.7\,(1.1) & 2.8/1.2 & -19.70$\rm ^{~}$ & 1.000 & 0.0 & 3.7\,(0.6) & 0.4 & 0.32 & 1.2\,(0.3) & 0.06\,(0.01) & 7\,(1) [1]\\
NGC4376 & Im & 1136 & 16.7\,(1.1) & 1.4/0.9 & -17.67$\rm ^{~}$ & 0.064 & 6.0 & 14.3\,(2.4) & 0.6 & 0.14 & 7.0\,(1.5) & 0.37\,(0.08) & 76\,(15) [1]\\
NGC4383 & Sa? & 1710 & 16.7\,(1.1) & 1.9/1.0 & -19.57$\rm ^{~}$ & 0.330 & 10.0 & 22.6\,(3.8) & 0.4 & 0.3 & 7.3\,(1.6) & 0.39\,(0.08) & 650\,(10) [3]\\
UGC07512 & ImIV & 1505 & 27.6\,(4.1) & 1.3/0.6 & -17.03$\rm ^c$ & 0.013 & 1.5 & 3.7\,(0.6) & 0.5 & 0.04 & 4.9\,(1.7) & 0.26\,(0.09) & 56\,(11) [1]\\
NGC4390 & Sbc(s) & 1131 & 16.7\,(1.1) & 1.6/1.1 & -18.18$\rm ^{~}$ & 0.420 & 10.0 & 4.2\,(0.7) & 0.4 & 0.17 & 1.6\,(0.4) & 0.09\,(0.02) & 280\,(40) [3]\\
NGC4394 & (R)SB(r)b & 922 & 16.7\,(1.1) & 3.6/3.2 & -19.42$\rm ^{~}$ & - & 2.0 & 12.9\,(2.2) & 1.0 & 0.28 & 8.0\,(1.7) & 0.43\,(0.09) & 10\,(3) [3]\\
NGC4423 & Sdm & 1104 & 16.7\,(1.1) & 2.3/0.4 & -17.51$\rm ^{~}$ & 0.032 & 3.0 & 2.5\,(0.4) & 0.5 & 0.11 & 1.2\,(0.3) & 0.06\,(0.01) & 18\,(3) [1]\\
NGC4430 & SB(rs)b & 1468 & 16.7\,(1.1) & 1.9/1.2 & -19.17$\rm ^{~}$ & - & 3.0 & 6.4\,(1.1) & 0.3 & 0.25 & 2.2\,(0.5) & 0.12\,(0.03) & 5\,(1) [1]\\
IC3365 & Im & 2375 & 16.7\,(1.1) & 2.1/1.0 & -18.48$\rm ^{~}$ & - & 2.0 & 1.9\,(0.3) & 0.7 & 0.19 & 1.0\,(0.2) & 0.05\,(0.01) & 24\,(4) [3]\\
UGC07590 & Sbc & 1118 & 13.6\,(1.1) & 1.3/0.4 & -17.40$\rm ^{~}$ & 1.150 & 0.0 & 1.6\,(0.3) & 0.4 & 0.13 & 0.4\,(0.1) & 0.02\,(0.01) & 36\,(9) [1]\\
NGC4450 & SA(s)ab & 1954 & 16.7\,(2.5) & 5.2/3.9 & -21.58$\rm ^{~}$ & - & 4.5 & 28.0\,(4.8) & 1.9 & 0.54 & 26.0\,(9.0) & 1.38\,(0.48) & 4\,(2) [2]\\
NGC4468 & SA0-? & 908 & 16.7\,(1.1) & 1.5/1.0 & -17.57$\rm ^{~}$ & 0.054 & 3.0 & 4.5\,(0.8) & 0.6 & 0.14 & 2.2\,(0.5) & 0.12\,(0.03) & 12\,(2) [1]\\
NGC4470 & Sa? & 2340 & 16.7\,(1.1) & 1.6/1.0 & -19.82$\rm ^{~}$ & 1.170 & 0.0 & 2.0\,(0.3) & 0.4 & 0.34 & 0.6\,(0.1) & 0.03\,(0.01) & 0\,(0) [1]\\
NGC4504 & SA(s)cd & 1000 & 11.7\,(1.8) & 4.4/2.7 & -19.26$\rm ^c$ & 0.036 & 3.0 & 35.6\,(6.0) & 0.7 & 0.23 & 8.5\,(2.9) & 0.45\,(0.15) & 25\,(5) [1]\\
NGC4522 & SB(s)cd & 2307 & 16.7\,(1.1) & 4.0/0.8 & -19.84$\rm ^{~}$ & 0.300 & 10.0 & 3.7\,(0.6) & 0.4 & 0.34 & 1.1\,(0.2) & 0.06\,(0.01) & 170\,(30) [3]\\
NGC4532 & IBm & 2012 & 16.7\,(1.1) & 2.6/0.9 & -20.25$\rm ^{~}$ & 0.054 & 3.0 & 13.3\,(2.3) & 0.4 & 0.4 & 3.7\,(0.8) & 0.20\,(0.04) & 16\,(3) [1]\\
UGC07739 & Im & 2039 & 13.6\,(1.1) & 1.3/1.2 & -17.96$\rm ^{~}$ & 1.180 & 0.0 & 0.1\,(0.0) & 0.4 & 0.16 & 0.0\,(0.0) & 0.00\,(0.00) & 2\,(0) [1]\\
IC3522 & ImIII-IV & 668 & 16.7\,(1.1) & 1.0/0.5 & -15.01$\rm ^c$ & 0.052 & 3.0 & 1.1\,(0.2) & 1.2 & 0.05 & 1.1\,(0.2) & 0.06\,(0.01) & 170\,(90) [3]\\
NGC4550 & SB0 & 409 & 2.8\,(0.4) & 3.3/0.9 & -17.38$\rm ^{~}$ & 0.062 & 6.0 & 4.5\,(0.8) & 0.6 & 0.13 & 0.1\,(0.0) & 0.00\,(0.00) & 6\,(1) [1]\\
NGC4559 & SAB(rs)cd & 816 & 9.3\,(1.4) & 10.7/4.4 & -20.42$\rm ^{~}$ & - & 4.2 & 139.2\,(23.7) & 1.2 & 0.42 & 26.0\,(9.0) & 1.38\,(0.48) & 6\,(3) [2]\\
NGC4561 & SB(rs)dm & 1407 & 16.7\,(1.1) & 1.5/1.3 & -18.96$\rm ^{~}$ & 0.070 & 6.0 & 17.4\,(3.0) & 0.8 & 0.24 & 9.3\,(2.0) & 0.49\,(0.11) & 32\,(6) [1]\\
NGC4567 & SA(rs)bc & 2274 & 16.7\,(1.1) & 2.9/1.4 & -20.69$\rm ^{~}$ & 1.130 & 0.0 & 10.2\,(1.7) & 0.4 & 0.46 & 2.6\,(0.6) & 0.14\,(0.03) & 17\,(4) [1]\\
NGC4568 & SA(rs)bc & 2255 & 16.7\,(1.1) & 4.3/1.0 & -21.07$\rm ^{~}$ & 1.130 & 0.0 & 24.4\,(4.2) & 0.4 & 0.52 & 5.6\,(1.2) & 0.29\,(0.06) & 23\,(6) [1]\\
IC3583 & SmIII & 1121 & 16.7\,(1.1) & 2.2/1.1 & -18.18$\rm ^{~}$ & 1.100 & 0.0 & 2.9\,(0.5) & 0.4 & 0.17 & 1.2\,(0.3) & 0.06\,(0.01) & 39\,(10) [1]\\
NGC4569 & SAB(rs)ab & -235 & 16.7\,(2.5) & 9.5/4.4 & -15.17$\rm ^{~}$ & - & 0.0 & 11.1\,(1.9) & 0.3 & 0.05 & 4.6\,(1.6) & 0.25\,(0.08) & 3\,(0) [1]\\
IC3591 & SBmIII & 1632 & 16.7\,(1.1) & 0.9/0.6 & -17.76$\rm ^{~}$ & 0.029 & 3.0 & 3.9\,(0.7) & 0.6 & 0.13 & 2.0\,(0.4) & 0.11\,(0.02) & 36\,(7) [1]\\
NGC4579 & SAB(rs)b & 1519 & 16.7\,(2.5) & 5.9/4.7 & -21.54$\rm ^{~}$ & - & 4.5 & 67.8\,(11.5) & 2.0 & 0.54 & 68.2\,(23.5) & 3.62\,(1.25) & 13\,(6) [2]\\
IC3617 & SBmIII/BCD & 2079 & 16.7\,(1.1) & 1.4/0.7 & -18.24$\rm ^{~}$ & 1.130 & 0.0 & 2.2\,(0.4) & 0.4 & 0.18 & 0.8\,(0.2) & 0.04\,(0.01) & 125\,(51) [1]\\
UGC07827 & Im: & 552 & 9.3\,(1.4) & 1.2/0.7 & -$\rm ^{~}$ & 0.060 & 6.0 & 1.9\,(0.3) & 0.3 & 0.0 & 0.3\,(0.1) & 0.01\,(0.00) & 49\,(10) [1]\\
NGC4595 & SAB(rs)b? & 604 & 16.7\,(1.1) & 1.7/1.1 & -17.66$\rm ^{~}$ & - & 2.0 & 7.6\,(1.3) & 0.7 & 0.14 & 4.0\,(0.9) & 0.21\,(0.05) & 19\,(5) [3]\\
NGC4594 & SA(s)a & 1024 & 9.8\,(1.5) & 8.7/3.5 & -22.39$\rm ^{~}$ & - & 4.2 & 65.7\,(11.2) & 2.6 & 0.53 & 37.1\,(12.8) & 1.96\,(0.68) & 4\,(2) [2]\\
NGC4625 & SAB(rs)m & 609 & 9.4\,(1.4) & 1.3/1.2 & -17.35$\rm ^{~}$ & 0.027 & 3.0 & 15.1\,(2.6) & 0.6 & 0.11 & 2.4\,(0.8) & 0.13\,(0.04) & 25\,(4) [3]\\
NGC4631 & SB(s)d & 606 & 7.7\,(1.2) & 15.5/2.7 & -20.63$\rm ^{~}$ & 0.051 & 4.2 & 80.9\,(13.8) & 1.3 & 0.46 & 10.3\,(3.6) & 0.55\,(0.19) & 42\,(3) [3]\\
NGC4639 & SAB(rs)bc & 977 & 16.7\,(1.1) & 2.8/1.9 & -19.08$\rm ^{~}$ & 0.440 & 0.0 & 10.1\,(1.7) & 0.4 & 0.24 & 3.6\,(0.8) & 0.19\,(0.04) & 38\,(9) [1]\\
NGC4640 & dS0 & 2082 & 16.7\,(1.1) & 1.6/0.8 & -18.09$\rm ^c$ & 0.025 & 4.0 & 3.0\,(0.5) & 0.7 & 0.16 & 1.7\,(0.4) & 0.09\,(0.02) & 27\,(5) [1]\\
NGC4647 & SAB(rs)c & 1334 & 16.7\,(1.1) & 2.9/2.3 & -19.95$\rm ^{~}$ & 0.065 & 6.0 & 31.0\,(5.3) & 2.3 & 0.35 & 53.6\,(11.5) & 2.84\,(0.61) & 11\,(2) [1]\\
NGC4651 & SA(rs)c & 805 & 16.7\,(1.1) & 4.0/2.6 & -19.50$\rm ^{~}$ & - & 2.0 & 36.3\,(6.2) & 1.0 & 0.29 & 21.7\,(4.7) & 1.15\,(0.25) & 23\,(2) [3]\\
PGC043211 & dE & 1146 & 13.6\,(2.0) & 0.7/0.7 & -19.46$\rm ^{~}$ & 0.022 & 3.0 & 0.4\,(0.1) & 0.2 & 0.25 & 0.1\,(0.0) & 0.00\,(0.00) & 880\,(233) [1]\\
NGC4725 & SAB(r)ab & 1206 & 11.9\,(1.8) & 10.7/7.6 & -21.45$\rm ^{~}$ & - & 4.2 & 57.2\,(9.7) & 1.8 & 0.53 & 24.1\,(8.3) & 1.28\,(0.44) & 7\,(3) [2]\\
NGC4736 & (R)SA(r)ab & 308 & 5.2\,(0.8) & 11.2/9.1 & -20.40$\rm ^{~}$ & 0.043 & 3.0 & 297.3\,(50.5) & 1.4 & 0.41 & 19.9\,(6.9) & 1.05\,(0.36) & 10\,(0) [3]\\
NGC4772 & SA(s)a & 998 & 13.4\,(2.0) & 3.4/1.7 & -19.42$\rm ^{~}$ & 0.320 & 10.0 & 4.3\,(0.7) & 0.9 & 0.29 & 1.5\,(0.5) & 0.08\,(0.03) & 90\,(10) [3]\\
NGC4789A & IB(s)m & 374 & 3.8\,(0.6) & 3.0/2.2 & -15.81$\rm ^{~}$ & - & 2.0 & 6.2\,(1.0) & 0.3 & 0.07 & 0.1\,(0.0) & 0.01\,(0.00) & 47\,(5) [3]\\
NGC4826 & (R)SA(rs)ab & 408 & 7.5\,(1.1) & 10.0/5.4 & -20.55$\rm ^{~}$ & - & 4.2 & 19.3\,(3.3) & 0.4 & 0.44 & 1.1\,(0.4) & 0.06\,(0.02) & 9\,(1) [3]\\
NGC4941 & (R)SAB(r)ab & 1108 & 17.4\,(2.6) & 3.6/1.9 & -19.58$\rm ^{~}$ & 0.025 & 3.0 & 27.2\,(4.6) & 1.1 & 0.26 & 20.3\,(7.0) & 1.08\,(0.37) & 26\,(5) [1]\\
PGC045195 & SAB(s)dm & 1362 & 24.4\,(3.7) & 1.2/0.9 & -19.26$\rm ^c$ & 0.063 & 6.0 & 30.8\,(5.2) & 0.7 & 0.27 & 30.3\,(10.4) & 1.60\,(0.55) & 68\,(14) [1]\\
UGC08201 & Im & 31 & 4.6\,(0.7) & 3.5/1.9 & -13.12$\rm ^{~}$ & - & 2.0 & 2.1\,(0.3) & 0.2 & 0.02 & 0.1\,(0.0) & 0.00\,(0.00) & 10\,(2) [3]\\
UGC08303 & IAB(s)m & 929 & 18.1\,(2.7) & 2.2/1.9 & -17.55$\rm ^{~}$ & - & 2.0 & 7.2\,(1.2) & 0.5 & 0.13 & 3.9\,(1.3) & 0.21\,(0.07) & 63\,(6) [3]\\
NGC5033 & SA(s)c & 875 & 16.2\,(2.4) & 10.7/5.0 & -20.16$\rm ^{~}$ & 0.033 & 3.0 & 97.4\,(16.6) & 0.8 & 0.33 & 41.3\,(14.2) & 2.19\,(0.75) & 31\,(6) [1]\\
ESO508-030 & IB(s)m & 1510 & 23.1\,(3.5) & 1.7/0.6 & -17.26$\rm ^c$ & 0.018 & 1.5 & 4.7\,(0.8) & 0.6 & 0.05 & 5.1\,(1.8) & 0.27\,(0.09) & 74\,(15) [1]\\
NGC5055 & SA(rs)bc & 504 & 7.9\,(1.2) & 12.6/7.2 & -20.57$\rm ^{~}$ & - & 4.2 & 19.8\,(3.4) & 0.7 & 0.45 & 1.6\,(0.5) & 0.08\,(0.03) & 20\,(4) [3]\\
NGC5194 & SA(s)bc & 463 & 7.7\,(1.2) & 11.2/6.9 & -20.75$\rm ^{~}$ & - & 4.2 & 737.0\,(125.3) & 1.4 & 0.48 & 101.6\,(35.0) & 5.39\,(1.86) & 28\,(3) [3]\\
NGC5195 & SB0\_1 & 465 & 7.7\,(1.2) & 5.8/4.6 & -19.26$\rm ^{~}$ & - & 4.2 & 48.7\,(8.3) & 1.1 & 0.27 & 6.9\,(2.4) & 0.37\,(0.13) & 4\,(1) [3]\\
NGC5474 & SA(s)cd & 273 & 6.5\,(1.0) & 4.8/4.3 & -17.39$\rm ^{~}$ & - & 2.0 & 26.1\,(4.4) & 0.5 & 0.12 & 1.8\,(0.6) & 0.10\,(0.03) & 22\,(2) [3]\\
NGC5477 & SA(s)m & 304 & 7.1\,(1.1) & 1.7/1.3 & -14.46$\rm ^{~}$ & - & 3.0 & 7.0\,(1.2) & 0.2 & 0.04 & 0.5\,(0.2) & 0.03\,(0.01) & 92\,(9) [3]\\
NGC5486 & SA(s)m & 1386 & 25.6\,(3.8) & 1.7/1.0 & -17.80$\rm ^c$ & - & 1.8 & 5.0\,(0.8) & 0.6 & 0.14 & 5.8\,(2.0) & 0.31\,(0.11) & 35\,(3) [3]\\
PGC140287 & Im & 1470 & 25.5\,(3.8) & 1.0/0.7 & -15.27$\rm ^d$ & 0.025 & 6.0 & 0.3\,(0.0) & 1.2 & 0.05 & 0.6\,(0.2) & 0.03\,(0.01) & 130\,(26) [1]\\
IC1024 & S0? & 1411 & 25.0\,(3.8) & 1.6/0.6 & -17.97$\rm ^c$ & 0.018 & 3.0 & 10.0\,(1.7) & 0.4 & 0.15 & 9.1\,(3.1) & 0.48\,(0.17) & 48\,(10) [1]\\
NGC5701 & (R)SB(rs)0/a & 1524 & 25.8\,(3.9) & 4.3/4.1 & -20.12$\rm ^{~}$ & 0.012 & 1.5 & 28.6\,(4.9) & 1.2 & 0.14 & 57.8\,(19.9) & 3.07\,(1.06) & 20\,(4) [1]\\
IC1066 & Sab & 1577 & 26.5\,(4.0) & 1.3/0.7 & -17.82$\rm ^c$ & 0.061 & 6.0 & 7.2\,(1.2) & 0.4 & 0.15 & 7.6\,(2.6) & 0.40\,(0.14) & 33\,(6) [1]\\
PGC057723 & SAB(r)b? & 934 & 16.4\,(2.5) & 2.4/1.7 & -18.11$\rm ^c$ & 0.065 & 6.0 & 22.6\,(3.8) & 1.3 & 0.17 & 19.6\,(6.7) & 1.04\,(0.36) & 35\,(7) [1]\\
NGC6140 & SB(s)cd & 866 & 18.8\,(2.8) & 6.3/4.6 & -18.73$\rm ^{~}$ & 0.020 & 3.0 & 84.1\,(14.3) & 0.4 & 0.2 & 39.8\,(13.7) & 2.11\,(0.73) & 8\,(2) [1]\\
NGC6118 & SA(s)cd & 1611 & 25.7\,(3.9) & 4.7/2.0 & -19.37$\rm ^{~}$ & 0.066 & 6.0 & 27.2\,(4.6) & 1.2 & 0.28 & 46.0\,(15.9) & 2.44\,(0.84) & 19\,(4) [1]\\
PGC058661 & Sc & 1581 & 25.6\,(3.8) & 1.6/0.5 & -16.64$\rm ^c$ & 0.064 & 6.0 & 1.3\,(0.2) & 1.3 & 0.09 & 3.2\,(1.1) & 0.17\,(0.06) & 30\,(6) [1]\\
IC1254 & Sb? & 1283 & 24.0\,(3.6) & 1.6/0.7 & -16.16$\rm ^{~}$ & 0.017 & 3.0 & 2.4\,(0.4) & 0.4 & 0.07 & 2.2\,(0.8) & 0.12\,(0.04) & 23\,(5) [1]\\
NGC7465 & (R\')SB(s) & 1972 & 30.1\,(4.5) & 1.2/0.8 & -18.41$\rm ^{~}$ & 0.012 & 1.1 & 6.8\,(1.2) & 1.0 & 0.04 & 17.7\,(6.1) & 0.94\,(0.32) & 17\,(3) [1]\\
NGC7742 & SA(r)b & 1663 & 25.2\,(3.8) & 1.7/1.7 & -18.90$\rm ^c$ & 0.013 & 1.1 & 8.0\,(1.4) & 0.4 & 0.05 & 8.8\,(3.0) & 0.46\,(0.16) & 8\,(2) [1]\\